\documentclass[12pt,preprint]{aastex}
\usepackage{epsfig}
\usepackage{natbib}

\bibpunct{(}{)}{;}{a}{}{,}

\newcommand{\GS}{\lower.5ex\hbox{$\buildrel>\over\sim$}}
\newcommand{\LS}{\lower.5ex\hbox{$\buildrel<\over\sim$}}

\shorttitle{The cocoon of Hercules~A}
\shortauthors{Saxton et al.}

\received{2002 April 27}
\begin{document}

\title{Production of ring-like structure in the cocoon of Hercules~A}
\author{Curtis~J.~Saxton\altaffilmark{1,2}
\and {Geoffrey~V.~Bicknell\altaffilmark{2,1}
\and {Ralph~S.~Sutherland\altaffilmark{2}}}
}

\altaffiltext{1}{Department of Physics \& Theoretical Physics,
	Faculty of Science,
	Australian National University, ACT 0200, Australia }
\altaffiltext{2}{Research School of Astronomy \& Astrophysics,
	Mt Stromlo Observatory, Australian National University,
	Cotter Road, Weston ACT 2611, Australia }

\begin{abstract}
The radio lobes of the radio galaxy Hercules~A
contain intriguing ring-like structures concentric with the jet axis.
To investigate the occurrence of such features,
we have used hydrodynamic simulations of jets
with a range of Mach numbers (from $M=2$ to $50$)
and densities (down to a ratio of $10^{-4}$ relative to the background)
to generate ray-traced images
simulating synchrotron emission from the time-dependent shock structures.
We compare these images with observations of Hercules~A,
and consider the physical nature and temporal evolution
of the most plausible configurations.
We find that the observed ring-like structures
are well explained as nearly annular shocks
propagating in the backflow surrounding the jet.
We infer that the jet is oriented at between $30^\circ$ and $70^\circ$
to the line of sight,
consistent with radio depolarization observations of Gizani \& Leahy.
The observational lack of hot-spots at the extremities of the radio lobes,
and the possible presence of a buried hot-spot
near the base of the western lobe,
are explained in terms of the intrinsic brightness fluctuations
and dynamics of the terminal shock of
an ultra-light, low Mach number jet that surges along its axis
due to intermittent pinching and  obstruction
by turbulent backflow in the cocoon.
We conclude from the appearance of both sides of the Hercules~A,
that both jets are on the verge of becoming fully turbulent.
\end{abstract}



\clearpage

\section{Introduction}

Hercules~A (3C~348) is a luminous
radio galaxy \citep{dreher1984,mason1988}
at a redshift of $z=0.154$, in a cluster of uncertain richness
\citep{greenstein1962, owen1989, yates1989, allington-smith1993}. 
Its power density at $5 \> \mathrm{GHz}$,
$P_\mathrm{5 GHz}\approx6.9\times10^{25} \> \mathrm{W} \>
\mathrm{Hz}^{-1} \> \mathrm{sr}^{-1}$ 
for $H_0=65 \> \mathrm{km} \> \mathrm{s}^{-1} \> \mathrm{Mpc}^{-1}$
\citep{gizani1999,gizani2000}.
In optical observations,
the elliptical galaxy is found to have a double nucleus
(with a separation $\sim4''$)
and tail-like distortions of the stellar isophotes
implying a merger of galaxies
\citep{minkowski1957,sadun1993}.

The X-ray emitting cluster medium
was the subject of ASCA and ROSAT observations by
\citet{siebert1999}
and ROSAT observations by \citet{gizani1999}.
The diffuse X-ray distribution is elongated
in the direction of the radio source axis,
suggesting an interaction with the radio lobes.
The medium is otherwise well described
by a $\beta$-model profile with best-fit parameters
$\beta\approx0.63$
and core radius $r_\mathrm{c}\approx35\pm3''$
\citep{siebert1999}.
The contribution from the nucleus is $\la 3\%$ of total X-ray luminosity,
but the cluster surface brightness may be centrally peaked
\citep{gizani1999}.
However there may actually be two thermal X-ray emitting components present,
as the ASCA spectrum suggests a temperature $kT\approx4.3\mathrm{keV}$
and a central electron density of
$n_0\approx 9.1\times10^{-3} \> \mathrm{cm}^{-3}$
\citep{siebert1999},
whereas the ROSAT/PSPC spectrum suggests
$kT \approx2.4 \> \mathrm{keV}$.
ROSAT observations by
\citet{gizani2002}
determined a temperature $kT=2.45 \> \mathrm{keV}$
and $n_0=7.8 \times10^{-3} \> \mathrm{cm}^{-3}$.
\citet{trussoni2001}
presented BeppoSAX observations
indicating a temperature $4$--$5 \> \mathrm{keV}$ on scales of 
several arcminutes
with the possibility of an additional $\sim3 \> \mathrm{keV}$ component.

In VLA observations,
\citet{gizani1999} found that the cluster medium affects
the depolarization of the radio features:
the western side exhibits greater depolarization than the east.
As the more distant jet and lobe are seen through
a thicker column of the cluster medium,
they deduced  a jet inclination of $\theta\approx 50^\circ$
to the line of sight. 
Thus Hercules~A exhibits a classical Laing-Garrington effect
\citep{laing88a,garrington88a}.

The radio lobes span $\sim 3'$
and in overall shape they are symmetric about the nucleus
(see the upper panel of Figure~\ref{'fig.radio'}). 
However, the eastern lobe accounts for
$3/4$ of the total radio luminosity. 
Although the radio power of Hercules~A
is in the range of FR~2 sources \citep[See ][]{fanaroff74a}
the morphology of the radio lobes
belongs ambiguously to the FR~1 class.
Most notably, neither lobe shows a hot-spot near its outer extremity.

The western radio lobe
(on side of the receding jet)
contains a sequence of arc-like and ring-like features that are
brighter than the lobe,
and which are approximately concentric with the jet axis
\citep{dreher1984,mason1988}.
Some of the ring-like features are partially filled;
some are open; some are incomplete.
The most distinct of these features,
and the furthest from the nucleus,
is a ``semi-ring'' that is bright on only one side.
Radio emission from the rings is linearly polarised:
up to $\sim 50\%$ in the two outer rings
\citep{dreher1984}.
The radio spectral indices are flatter in the jets and rings
($\alpha\approx1.1$ where $S_\nu\propto\nu^{-\alpha}$)
than in the radio lobes generally
($\alpha\approx1.6$),
indicating that the radio emitting plasma in the rings is younger
than the background \citep{gizani1999,gizani2002}.
The eastern (approaching) jet
shows ``helical'' features,
which may be a projected superposition of rings
\citep{gizani2002}.

The radio lobes of some other galaxies
display ring-like internal structures,
which have been likened to
the rings in the western lobe of Hercules~A
\citep{morrison1996}.
The source 3C~310 also shows several circular features
in both the northern and southern lobes;
these rings are larger in proportion to the lobes
than those of Hercules~A
\citep{vanbreugel1984,morrison1996}.
Other radio galaxies with ring-like features include
$\mathrm{MG}~0248+0641$
\citep{conner1998}
and 3C~219
\citep{perley1980}.

Several explanations have been offered for
the nature and origin of the rings.
\citet{dreher1984}
suggested that the rings in Hercules~A originate as distinct
ejections from the nucleus,
or interruptions of the jet due to either intrinsic pinching instabilities
or collisions with clouds straying into the jet axis.
In that case, the bright rings are surfaces where
the ambient radio lobe or cluster medium contacts
and interacts with a cloud.

\citet{mason1988}
proposed that the jet loses collimation at the point
where the radio structure widens,
and that concentrated streams of relativistic protons
intermittently cross or self-cross,
causing energetic particle bursts
which deposit kinetic energy in the plasma at that site.
These disturbances expand spherically,
as they drift outwardly within the medium of the radio lobe.
In this scenario, the rings are the projected edges of the spheres.
This would account for the approximately linear relation between
ring radius and displacement from the core.

\citet{morrison1996}
proposed a qualitatively similar theory
in which uniformly expanding spheres
originate as acoustic disturbances carried outwardly in a galactic wind.
The subsequent onset of jet activity
then floods the region with radio-emitting material.
Radio-bright plasma interacts with the shells and tends to concentrate
on the surfaces of these structures.
One implication of this model is that the extensive wind
has an age of Gigayears.
However, this requirement would be difficult to reconcile
with the likely disturbances in the cluster environment of Hercules~A,
especially related to the merger of another galaxy with Hercules~A
\citep[as implied by the double nucleus and stellar tail,][]{sadun1993}.
Thus, models for the radio structure of Hercules~A in which the 
formation of rings is
intrinsic to the jet and lobe dynamics seem to have more appeal.

\citet{meier1991}
proposed such a model, based on extensive numerical simulations, in 
which an  {\em over-dense} jet undergoes a sudden expansion 
when it becomes highly over-pressured some
$20^{\prime\prime}$ from the core and subsequently expands conically 
giving rise to the ``ice-cream
cone'' shape of the lobe. 
\citet{meier1991} 
offered a qualitative explanation for the prominent rings in
the western lobe as being the result of weak spherical shocks. 
Our difficulty with this model, 
which is really a difficulty with all models involving dense jets, 
is that the implied jet energy flux under most circumstances is too large.
This is most likely the case for Hercules~A. 
Moreover, the implied Mach number is also very large. 
In the next section, therefore, 
we examine the case for $\eta\sim 1$ 
in some detail and in so doing we present our case for a more 
conventional range of $\eta \ll 1$.

\section{Choice of parameters}
\label{s:pars}
Useful constraints on the parameters of jets may be obtained by 
relating the energy flux to the observed radio power. 
The energy flux of a relativistic jet 
consisting of cold matter plus relativistic
particles is given by the following: Let
$\beta_{\rm jet} = v_{\rm jet}/c$, $\Gamma$ be the jet Lorentz 
factor, $p_{\rm jet}$ be the jet pressure,
$\chi = \rho c^2/4 p_{\rm jet}$ where $\rho_{\rm jet}$ is the cold 
matter rest-mass density and $A_{\rm
jet} = \pi D_{\rm jet}^2/4$ be the jet cross-sectional area for 
diameter $D_{\rm jet}$. 
Then, the energy flux, $F_E$ is given by:
\begin{equation}
F_E = 4 p_{\rm jet} c \beta_{\rm jet} A_{\rm jet} \Gamma^2 \left[ 1 + 
\frac{\Gamma-1}{\Gamma} \chi
\right]
\end{equation}
\cite{bicknell1994}.
The parameter, $\chi$ can be expressed in terms of $\eta$, the ratio
of jet to ISM pressures and the ISM temperature, 
$T_{\rm ism} = 10^7 T_7 \, \mathrm{K}$ as
\begin{equation}
\chi = \eta \, \left( \frac{p_{\rm ism}}{p_{\rm jet}} \right)  \,
\left( \frac {\mu m_pc^2}{k T_{\rm ism}} \right)
\approx 1.7
\times 10^{5} \left( \frac{p_{\rm ism}}{p_{\rm jet}} \right) \,  \eta 
\, T_7^{-1}
\end{equation}
Numerically, the energy flux is
\begin{eqnarray}
F_E &\approx& 9.0 \times 10^{42} \,
\left(  \frac {p_{\rm ism}}{10^{-11} \mathrm{dyn} \> \mathrm{cm}^{-2}} \right)
\, \left( \frac {D_{\rm jet}}{\rm kpc} \right)^2  \Gamma^2 
\beta_{\rm jet} \nonumber \\
&\times & \, \left[\frac {p_{\rm jet}}{p_{\rm ism}} + 1.7 \times 10^5 \,
  \frac {\Gamma-1}{\Gamma} \, \eta T_7^{-1}
\right]
\> \mathrm{erg} \> \mathrm{s}^{-1}
\label{e:fe_rel}
\end{eqnarray}
We have chosen a fiducial value of 1~kpc for $D_{\rm jet}$ 
since the 
$0.5^{\prime\prime}$ resolution limit
of the observations \citep[e.g. ][]{dreher1984}
implies an upper limit 
on the unresolved jet diameter of about
$1.3 \> \rm kpc$. 
On the other hand the jet diameter at $20^{\prime\prime}$ 
is probably not much less than
$0.5^{\prime\prime}$ since that corresponds to an opening angle of 
$0.6^\circ$.  
In the non-relativistic limit, for a hypersonic jet,
\begin{equation}
F_E \approx \frac {1}{2} \rho_{jet} v_{\rm jet}^3 A_{\rm jet}
\approx 1.1 \times 10^{48} \, \eta \,
\left( \frac {n_{\rm ism}}{10^{-2} \> \mathrm{cm}^{-3}} \right) \, 
\beta_{\rm jet}^3 \,
\left( \frac {D_{\rm jet}}{\rm kpc} \right)^2
\> \mathrm{erg} \> \mathrm{s}^{-1}
\end{equation}
and this expression serves as a reasonable order of magnitude 
estimate of the effects of various parameters. 
We should compare these expressions for the jet energy 
flux to estimates of this parameter
obtained from the radio power.

In making these comparisons, 
first note that the ISM parameters obtained from X-ray observations
summarised in the introduction, are reasonably well determined. 
The central electron number density density,
$n_e$, of the interstellar medium is approximately 
$7.8 \times 10^{-3} \> \mathrm{cm}^{-3}$ 
with a corresponding total number density, 
$n_{\rm ism} \approx 1.5 \times 10^{-2} \> \mathrm{cm}^{-3}$; 
the temperature is
$2.45 \> \rm keV \approx 2.8 \times 10^7 \> \mathrm{K}$; 
the  corresponding pressure, 
$p_{\rm ism} \approx 5.7 \times 10^{-11} \> \mathrm{dyn} \> \mathrm{cm}^{-2}$. 
We use these central estimates of
density and pressure within the core radius
$\approx 35^{\prime\prime}$ of the hot gas distribution. 
Second, although we have allowed for a
difference between jet and ISM pressure, 
we evaluate the above expressions for energy flux in the region
within
$20^{\prime\prime}$ of the core where both Meier~et~al. 
and we assume that the jet is confined. 
The jet is unresolved in this region so that 
we only know that $D \la 500 \> \rm pc$;
we have taken a fiducial value of 100~pc. 
Third, given that we are dealing with a supersonic jet, 
it is highly likely that the jet is at least mildly relativistic 
for the following reason. 
There currently seems to be little doubt 
that jets in radio loud objects are initially relativistic. 
Given that this is the case, 
a jet that is supersonic on the kpc scale has a velocity of 
at least $0.3 \> \rm c$
\citep{bicknell1994}. 
Using the flux density given in 
\citet{kellermann1969}, 
the total power of Hercules~A at 1.4~GHz is approximately
$3.1 \times 10^{27} \> \rm W \> Hz^{-1}$. 
For reasonable limits on the  ratio of radio power to jet
energy flux, $\kappa_{1.4} \sim 10^{-12} - 10^{-11} \> \rm Hz^{-1}$ 
\citep{bicknell2000}, 
the jet power is between $2 \times 10^{45}$ and 
$2 \times 10^{46} \> \mathrm{erg} \> \mathrm{s}^{-1}$. 
For $\beta =0.5$, and
$\eta =1$, the energy flux given by equation~(\ref{e:fe_rel})
is approximately $3 \times 10^{47} \> \mathrm{erg} \> \mathrm{s}^{-1}$,
about an order of magnitude too large.
  
Another problem with
$\eta \sim 1$ is that the implied Mach number is enormous. 
For example, in the non-relativistic
approximation, the jet Mach number,
$M_{\rm jet} \approx 710
\beta_{\rm jet} T_7^{-1/2} \eta^{1/2}$ and the Mach number of 210 
implied by $\beta_{\rm jet} \sim 0.5$, 
and $\eta \sim 1$ is much larger than the value of 6 preferred 
by \citep{meier1991} in their
explanation for the rings. 
Moreover, simulations with such a high 
Mach number are likely to be quite
different from the ones presented by Meier et al. 
We expect that the jet and its cocoon would be much
more needle-like in this case.

Another useful way to constrain the parameters of the jets in 
Hercules~A is to express the
energy flux in terms of $\eta$ and $M_{\rm jet}$. 
For a hypersonic, non-relativistic jet
\begin{eqnarray}
F_E & \approx & \frac {1}{2} \rho_{\rm ism} \,
\left( \frac {p_{\rm jet}}{p_{\rm ism}}\right)^{3/2} \,
  \left( \frac {4kT_{\rm ism}}{3\mu m_p} \right)^{3/2} \,
\eta^{-1/2} \, M^3 \, A_{\rm jet} \nonumber \\
& \approx & 2.9 \times 10^{44} \left( \frac{p_{\rm jet}}{p_{\rm 
ism}}\right)^{3/2} \,
\left( \frac {n_{\rm ism}}{10^{-2} \> \rm cm^{-3}} \right) \,
\left( \frac {\eta}{10^{-4}} \right)^{-1/2} \, T_7^{3/2} \,\left( 
\frac {M}{10} \right)^3 \,
\left( \frac {D}{\rm kpc} \right)^2 \> \mathrm{erg} \> \mathrm{s}^{-1}
\end{eqnarray}
where $T_7 \approx 2.8$ in Hercules~A.  
As we have seen above, and this expression confirms it, 
a value of $\eta \sim 1$ requires a Mach number $\sim 70 - 150$ 
to attain an energy flux 
$\sim 2 \times 10^{45-46} \> \mathrm{erg} \> \mathrm{s}^{-1}$. 
Again, these values
considerably exceed the value of 6 favored by \cite{meier1991}. 
Conventional values of
$\eta \ll 1$ are able to provide the requisite energy flux for more 
modest Mach numbers. 
For example,
$\eta \sim 10^{-4}$ and $M \sim 10-12.5$ gives an energy flux of 
$2\times 10^{45-46} \> \mathrm{erg} \> \mathrm{s}^{-1}$. 
For these reasons, therefore, we have restricted our 
simulations to (fairly wide) ranges in
Mach number of 2 - 50 and density ratio of $10^{-4}$ - $10^{-2}$. 
As we shall see, there is some
indication that the Mach number may be close to transonic 
and that correspondingly, 
the density ratio may be even lower than $10^{-4}$. 
However, we have not carried out simulations for
$\eta < 10^{-4}$ since computationally, 
these are prohibitively expensive as a result of the Courant condition.

An appealing aspect of the
\citet{meier1991} simulations is that, at approximately
$20^{\prime\prime}$ from the core, 
the western jet becomes over-pressured. 
Indeed, inspection of the most recent image \cite{gizani2000} 
shows that the western jet may be behaving like a classical
over-pressured jet over at least part of the region between 
$20^{\prime\prime}$ and $60^{\prime\prime}$. 
The knots of emission in this region could well be 
the result of particle acceleration at reconfinement shocks. 
The characteristic expansion and compression of the
jet is especially manifest in the lower panel of 
Figure~\ref{'fig.radio'} 
where the brightness and contrast of the
\citet{gizani2000} image has been adjusted to highlight faint features.
Reconfinement features are also present in our simulations since the 
production of periodic shocks is relatively easy in a supersonic jet. 
The X-ray data \citep[e.g. ][]{siebert1999} 
provide a natural explanation for the jet becoming over-pressured 
in this region since the core radius is approximately
$35^{\prime\prime}$ and $20^{\prime\prime}$ deprojected is 
approximately $26^{\prime\prime}$. 
At this point the background confining pressure would be starting to decrease 
from the central value.
However, there is nothing in the X-ray data to suggest that the jet 
is becoming extremely overpressured
at this point as required by the \citet{meier1991} models.

Thus, for the above reasons, we have opted for a model for Hercules~A 
in which the jet density ratio is
much less than unity, the Mach number is not excessively large and 
the jet is initially confined. 
In \citet{saxton2002} we simulated the interactions of 
a hypersonic, low density jet with its turbulent
cocoon, covering a parameter-space with two orders of magnitude in 
jet density and Mach numbers from $5$ to $50$. 
We considered the dynamics of point-like and ring-like 
shocks near the head of the jet, in
order to model and interpret the complex shock structures observed 
near the western hot-spot of Pictor~A. 
In the present paper, we use this simulation database, 
augmented by some additional simulations, 
to generate synthetic radio maps resembling the 
morphology and brightness distribution of the features of Hercules~A. 
Our model for some of the observed radio rings is that they are 
open or partially filled annular shocks 
propagating in  the cocoon backflow in the western radio lobe. 
It is feasible that other features
that have been described as rings are in fact either related to jet 
reconfinement or recessed hot-spots in the lobe. 
At this stage we have not attempted 
to model the effects of reconfinement shocks, 
and we have concentrated on the production of 
rings near the head of the lobe.
Therefore, we assume a constant background density in our simulations.

It is useful at this stage to refer to the {\em lower} panel of 
Figure~\ref{'fig.radio'} where we have labeled some of the regions in the
western lobe according to our model and have shown the effects of ellipse 
fitting to the features that could be interpreted as rings. Note that these
fits are not tightly constrained
and depend to some extent upon what
parts of the radio surface brightness are deemed to belong to a particular ring. 
We discuss the 
elliptical rings further in \S~\ref{s:orientation} with respect to the 
orientation of the jets.

\section{Simulations}
\label{s:simulations}

The hydrodynamic simulations were conducted using
the Piecewise Parabolic Method (PPM) \citep{colella84a}
implemented in a code based on the VH-1 code
\citep{blondin93a}.
An advantage of PPM
for this type of simulation is its excellent shock-capturing properties.
We have enhanced the code to achieve greater efficiency, have 
included code to correct for a numerical
shock instability
and have added a scalar tracer variable $\varphi$
that is passively advected with the flow.
By assigning $\varphi=1$ within jet plasma entering the grid,
and initially $\varphi=0$ in the background medium,
we use this variable to distinguish and follow
the evolution of the constituents of the physical system.

Our images of simulated radio surface brightness
were rendered by a ray-tracing program
that projects three-dimensional structures obtained from the PPM output.
Assuming that the magnetic pressure is approximately
proportional to the hydrodynamic pressure, $p$,
the volumetric emissivity,
$j_\nu\propto \varphi p^{(3+\alpha)/2}$. 
This ansatz is only valid at a frequency
that is less than any cooling-induced break in frequency 
so that to be consistent we choose a value of $\alpha =0.6$.
This is flatter than the observed spectral index in, for example,
the rings wherein $\alpha=1.1$
\citep{gizani2002}
indicating that these features are older than the synchrotron cooling time.
Our assumption of $\alpha=0.6$ means that the relative
contrast of features will be affected.
However, the resulting images obtained by
integrating $j_\nu$ along given lines of
sight for each pixel of the simulated intensity map,
should give a good qualitative indication of
the appearance of the flow and also give a good
indication of the relative contrast of features as a function of jet
parameters.
(This is an important point that is discussed below in \S~\ref{s.contrast}.)

Despite the importance of synchrotron and/or inverse Compton cooling
for the radio spectrum,
cooling does not affect the hydrodynamics of the flow since the
lowest energy particles that dominate the internal energy are unaffected.
Therefore, the way the system behaves depends on dimensionless
parameters such as
the density ratio of jet to ambient gas,
$\eta$, the corresponding jet to ambient pressure ratio (assumed 
equal to one),
and the Mach number of the jet,
$M\equiv \eta^{-1/2} v_\mathrm{jet}/ c_\mathrm{s,ism}$ where 
$c_\mathrm{s,ism}$ is the sound speed in the interstellar medium. 
Our simulations cover a parameter space defined by
$\eta=10^{-2}, 10^{-3}, 10^{-4}$ and $M=2, 5, 10, 50$.
Adopting the cooler estimate of the temperature and density in the
ambient cluster medium
as fiducial scales for our simulations,
the thrust and power of the simulated  jets are given in
Table~\ref{table.jet.properties}.
These are classical values,
and can be related to parameters of relativistic jets
with equivalent thrust or power
\citep[e.g. ][]{rosen1999,carvalho2001a}.
The time interval between output frames
of the hydrodynamic data was set so that approximately $600$ frames
were generated before the jet reached
the right side (large $z$) boundary of the grid.

We assume that the outer boundaries of the computational grid
are open to outflow,
except at the base of the jet at the left (low-$z$) boundary.
The initial properties of the jet are copied to those cells
at the start of each time step,
in order to maintain a jet with constant mass flux, thrust and power.
In \citet{saxton2002} we concentrated on cases in which
a reflecting condition applies at the left boundary,
representing effects near the plane of symmetry near the nucleus
producing equally strong, opposite jets.
The closed boundary condition was found to lead to
a conical cocoon which is typically much wider than
the cylindrical cocoons that develop about jets
in simulations with an open left boundary.
The latter represent
the effects of systems where the head of the jet
is causally distant from the nucleus
and the plane of symmetry about the galaxy.
We will initially focus our discussion on simulations
with this type of boundary condition,
as this seems physically appropriate for Hercules~A,
since the radio lobes extend hundreds of kiloparsecs from the galaxy.

\section{Results}

\subsection{Morphology}

The morphology of our simulated brightness maps is determined
by the distribution of bright shock features.
The termination shock of the jet is often,
but not always, the brightest of such features.
Biconical (``diamond'') shocks also occur within the jet,
initiated by disturbances from turbulence in the surrounding cocoon.
These shocks are similar to reconfinement shocks
and indeed involve regions of overpressure
and underpressure in the jet.
Transient annular shocks also occur 
in the inner parts of the jet backflow; 
these features are particularly frequent near the head of the jet.

In order to show the effect of orientation on the appearance of the 
lobe, we show, in
Figure~\ref{f:orientation}, a logarithmically scaled density image, 
together with synthetic radio images
at orientations of $0^\circ-90^\circ$ at intervals of $15^\circ$. 
As  shown in \citet{saxton2002} 
jets rendered at orientations almost in the plane of the sky,
$70^\circ \la \theta \la 90^\circ$,
may show round hot-spots and knots on the jet axis, 
as well as more spatially extended and transient shock features
that appear almost linear in projection.
These do not resemble the arcs of Hercules~A.
Neither do we find morphologies resembling Hercules~A
when jets are rendered at a small angle to the line of sight,
$\theta \la 30^\circ$:
in such cases the extended shock features appear
in projection as a clutter of tightly overlapping circles and knots.
We find the best resemblance in frames rendered at
$\theta\approx 45^\circ$.

Examples from simulations with a range of jet parameters
$(\eta,M)$
are shown\footnote{
Animations of the ray-traced frames from 
all our choices of the system parameters 
are presented at {\tt http://maker.anu.edu.au/radiojets/herculesa/}.} 
in Figures~\ref{f:pageant.pxit-4ml}-\ref{f:pageant.pxit-4m2}. 
These are discussed further below.

\subsection{Entrainment}

As detailed in \citet{saxton2002},
the hot-spot vanishes and reforms in an irregular cycle
in which the jet is pinched off
and temporarily obstructed by turbulence
and/or the entrainment of dense gas in the cocoon,
followed by the jet burrowing forward again.
The effect of turbulence was documented in \citet{saxton2002}.
This surging is most extensive and frequent for jets with low $\eta$.
In the present paper, we also show in
Figure~\ref{f:fingers}, the effect of entrainment.
Dense gas is swept into the cocoon and directly impedes the jet.
This effect is particularly prominent at low Mach numbers,
as one would expect from the Mach number dependence of 
the classical Kelvin-Helmholtz instability.
In the density images in the figure we show a time series of snapshots
that show dense (white) fingers of gas entrained into the lobe
and partially, then totally, obstructing the passage of
an $(\eta,M) = (10^{-4},2)$ jet.
For a time,
the extreme end of the lobe is being starved of jet plasma
and consequently does not show a hot-spot.

In a significant fraction of instants in
each of our simulations of light jets, the hot-spot is temporarily absent
or dimmer than one or more brighter shock structures in the backflow,
such as the annular rings.
This occurs approximately $20\% - 50\%$ of the time
at all Mach numbers.
To quantify the location of the hot-spot,
we show in Figure~\ref{f:surging}
the trends in the cumulative distributions of the ratios
of the hot-spot to bow shock locations.
That is, the plots show the probability
that the ratio of hot-spot distance to bow-shock distance 
is less than a certain value. 
Note that the probability of a embedded hot-spot
is greater at $M=2$ and also, for a given Mach number, at lower $\eta$.

\subsection{Dependence on the Mach number and density ratio}

Let us discuss some of the features of the simulations
and in particular their dependence on Mach number and density ratio.
The jet Mach number has a substantial
effect on the typical brightness contrast
between the background radio lobe
and the shock features near the head of the jet.
As expected on general theoretical grounds, and as shown in our 
previous studies
\citep{saxton2002},
for a given value of $\eta$
the cases with a larger jet Mach number
show a greater brightness contrast.
As $M$ increases,
the ray-traced intensity maps are increasingly dominated by
the transient frontal shock features near the head of the jet:
a bright hot-spot comprising the jet termination shock,
one of the nearer biconical shocks within the jet,
or else a singularly bright annular shock
emerging from the division of an older hot-spot.
The background emission of the cocoon
and secondary rings at lower $z$ positions,
are relatively faint.

\subsection{Extracts from the simulations}

It is not feasible to show all of the frames from all of the 
simulations that we have carried out.
In this subsection we show a small sample that is designed to 
illustrate the trends with Mach number
and density ratio that we have noticed.

Snapshots from an $(\eta,M)=(10^{-4},50)$ simulation that exhibit 
ring structure  are shown
in Figure~\ref{f:pageant.pxit-4ml}.
Although approximately $3\% -10\%$ of the frames from the entire simulation
are qualitatively similar to the multi-ringed morphology of Hercules~A,
the brightness contrast is too great.
Where there appears a prominent ring,
it is usually a lone product of a hot-spot's temporary disruption,
and there are no other features of comparable brightness.

Simulations with lower Mach number produce a multiple ring
morphology typically with a lower brightness contrast
than for higher Mach numbers. 
In Figures~\ref{f:pageant.pxit-4m5} and \ref{f:pageant.pxit-2m5}
we show snapshots of the $(\eta,M)=(10^{-4},5)$ and 
$(\eta,M)=(10^{-2},5)$ simulations that again
illustrate the production of multiple rings of comparable brightness,
separated appreciably along the length of the jet.
The rings occurring furthest behind the head of the jet
often appear in proximity to diamond shocks in the jet.
Approximately $7\% - 15\%$ of the $M=5$ images
are qualitatively similar to Hercules~A. 
The contrast between the bright features in the lobe compared
to the rest of the lobe is less than for $M=50$. 
Nevertheless, note that the $\eta=10^{-2}$ images
display a larger contrast than $\eta = 10^{-4}$. 
This is a definite trend that is also evident in the
intermediate $\eta=10^{-3}$ images (not shown).

In the $(\eta,M)=(10^{-4},2)$ simulation (Figure~\ref{f:pageant.pxit-4m2})
the rings have a brightness contrast relative to the cocoon
that is less than or equal to the observed factor ($\sim 3$).
Compared to the hot-spots in the simulations with higher $M$,
the brightest point on the jet axis
is typically quite far to the left of the brightest rings.
In $M=2$ simulations,
the cocoon and hot-spot together are dim enough
that the jet itself appears prominent in comparison. 
This reduced contrast is obvious in all $M=2$ simulations. 
The dependence on $\eta$ is not as marked as it is 
for higher Mach numbers.


\section{Discussion and Conclusions}

\subsection{Morphology}

Our simulations show that a jet with constant fluxes
of mass, momentum and energy
may nonetheless vary dramatically in its morphology and extent
within a jet dynamical time scale,
$t_\mathrm{dyn}\equiv 2r_\mathrm{j}/c_\mathrm{s,j}$.
In extremely light, low Mach number simulations,
e.g. with $(\eta,M)=(10^{-4},2)$,
turbulence and dense obstructions in the backflow
are sufficiently important
that they can pinch and break the jet
essentially anywhere along its length.
The disconnected forward parts of the jet are then rapidly mixed
into the rest of the cocoon.
A new hot-spot occurs at the new, recessed, jet terminus and this
hot-spot gradually burrows
forward  through the cocoon until it reaches
the forward working surface of the lobe once more.
In the transitions of these extreme surging events,
arrays of ring-like shocks,
derived from the disconnected jet plasma,
often exceed in brightness the hot-spots and diamond shocks within the jet.
This is our explanation for
the rings of enhanced radio intensity in the western lobe of Hercules~A.

The surging jet dynamics revealed in simulations
puts the western radio emitting structures into a new perspective.
If we overlook the distraction of
the rings and outer parts of the western radio lobe,
the innermost ``shell'' identified in
\citet{mason1988} and \cite{morrison1996}
actually resembles a hot-spot in a traditional FR~2 radio source.
This  is located at approximately $46\%$
of the distance from the nucleus to the outer edge of the lobe.
This may be the present western hot-spot,
and  the larger, outer western radio lobe may be the product of an
earlier stage when the jet was not disrupted and reached out to a 
larger distance.

The apparent absence of hot-spots
near the outer edges of the radio lobes of Hercules~A
is a distinctive feature
reproduced in many instants in all of our simulations.
However, this phenomenon is most dramatic at $M=2$.
Moreover, the occurrence of recessed hot-spots in the
$M=2$ simulations is greatest for $\eta = 10^{-4}$.
The probability of an
$(\eta,M) = (10^{-4},2)$ jet showing a jet terminus located at less
than $50\%$ of
the distance to the outer edge of the lobe, is approximately $0.4$.

The differences between the $M=2$ and $M \ge 5$ simulations probably
lies in the
effective entrainment of matter at lower Mach numbers
that is reflected in the higher rate of growth
of unstable modes for low $M$.
The minor dependence on $M \ge 5$ of the cumulative distribution functions
shown in Figure~\ref{f:surging}
may be due to the increasingly vigorous nature of
turbulent eddies in the cocoon.
For these Mach numbers, as well, the lower $\eta$
jets are more prone to disruption.

The notion of a low Mach number in Hercules~A is also consistent with
the morphology of the {\em eastern} jet,
since both jets presumably have similar
properties when  they emerge from the nucleus.
However, inherent instabilities may  cause differences of appearance
at any particular moment on arcminute scales.
The long ``helical'' portion of the eastern jet
(upper panel of Figure~\ref{'fig.radio'})
has the appearance of a jet that has become unstable,
that is possibly decelerating
to a transonic velocity and, as a result, becoming turbulent.
The transition of a relativistic jet 
to turbulent transonic flow occurs when
$\rho_\mathrm{j} c^2 \sim 4 p$
\citep{bicknell1995},
which implies
$\eta = \eta_{\rm crit} \approx 4\mathrm{k}T/\mu m_\mathrm{p} c^2\sim10^{-4}$.
For jets that have not made the
transition to turbulent flow, $\eta < \eta_{\rm crit}$.
If the eastern jet is indeed transonic at the base of the eastern lobe,
then both jets may emerge from
the galaxy  with a low Mach number, perhaps $2\la M\la 5$.
Since the western jet
does not appear to have become fully turbulent, $\eta < 10^{-4}$
confirming our preference for parameters for the Western jets
that involve both a low Mach number and density ratio.

The brightness asymmetry of the prominent outer ring in the western lobe
is an important characteristic that is reproduced
in many bright shocked rings appearing in our simulations.
This effect occurs because
some shock-rings are ribbon-like,
i.e. flat in cross-section, with the rings tilted in the $r-z$ plane.
If the greater length of the cross-section is not exactly parallel
to the $z$ axis then one side of the ring
presents the observer with a wider projected area than the opposite side,
at a particular orientation.
The side with the narrower projected area
is more extended along the line of sight,
and therefore  appears brighter than
the side that is viewed at a less grazing angle.
Therefore simple orientation effects are sufficient to explain
the ring morphology of Hercules~A,
without resort to radical departures from axisymmetry.
Nor is it necessary to consider the possibilities of
directed Doppler boosting of emission from
shocked plasma moving at highly relativistic velocities
within the cocoon
\citep[c.f. discussion of the hot-spot of Pictor~A in ][]{saxton2002}.

The relative sizes of the rings also naturally occurs as a result of the flow
dynamics described here.
As Figures~\ref{f:pageant.pxit-4ml}-\ref{f:pageant.pxit-4m2} show,
ring shocks appear with different relative sizes in different instances.
In particular, the smallest ``ring'' in the observations
corresponds to the structure surrounding a recessed hot-spot.

\subsection{Radio brightness contrast}
\label{s.contrast}

The high-resolution radio intensity maps of Hercules~A
\citep{dreher1984,gizani2002}
indicate that the brightest parts of the rings are
on the order of $\sim 2 - 5$
times the intensity of the radio lobe background.
The flatter radio spectra of the rings indicates that they
are indeed shock features
like those that we have modeled in simulations.
Note however, that the ring spectra are not as flat as what we expect
from shocked plasma.
This may indicate that they are relatively long-lived structures
(in terms of synchrotron time-scales)
and this is also consistent with the rings lasting longer
in the lower Mach number simulations.
However, assessment of this possibility must await
the publication of more detailed surface brightness and spectral index
observations.

If the bright part of the ring is the thinnest projected area
of the underlying ribbon-like annular shock,
then by inspection, the outer ring has a thickness
$\ga 0.08$ times the diameter of the radio lobe in that vicinity.
The surface brightness enhancement of the ring over the lobe $\sim 2.5$;
hence the volumetric emissivities of material
in average conditions within the ring and within the lobe generally
has a ratio
$j_{\nu,\mathrm{ring}}/j_{\nu,\mathrm{lobe}} \la 31$.
This contrast suggests shocks of a certain strength,
since $j_\nu \propto \varphi p^{(3+\alpha)/2}$
where $\alpha$ is the radio spectral index.
The shock pressure increases by a factor $\sim M^2$.
This implies that the Mach number of the ring shocks is
$\sim 3$.
Our simulations with $M=2, 5$
are the best matches for
this strength of shocks in the cocoon.
This conclusion is constrained by the
assumptions of the emissivity model (see \S~\ref{s:simulations});
however it is probably indicative.

Inspection of the synthetic radio images
shows that in cases with $M=10$ or $M=50$,
the contrast between the rings and the cocoon
is typically much greater than
the observationally required value $\sim 3$.
Many of the morphologically selected
instances from the $M=2$ and $M=5$ simulations
show the appropriate contrast.
For each choice of $M$,
the densest jets, $\eta=10^{-2}$
produce more frames showing  excessive contrast
in the shock features near the terminus of the jet,
relative to the secondary rings.
Cases with $\eta=10^{-4}$ give the more frames with
the desired contrast relationship between rings,
and we expect this to be so for a real jet
that is lighter than the parameter regime our calculations could attain,
$\eta\ll 10^{-4}$.
Thus, on this basis, the simulations reinforce the suggestion
that the jets of Hercules~A are very light
and have Mach number in the range $2 \la M\la 5$.

\subsection{Orientation of Hercules~A}
\label{s:orientation}

The spherical shell models of
\citet{mason1988} and \citet{morrison1996}
imply geometric constraints on the orientation of the trail of shells.
Assuming that such a model is correct,
the shells probably are not in physical contact,
otherwise their spherical form and brightness distribution
would be affected greatly.
If consecutive shells of radius $r_n$ and $r_{n+1}$
do not intersect in three dimensions,
and yet are partially superimposed in projection on the sky,
then they are aligned on a line at an inclination $\theta$,
which is constrained by the inequality
\begin{equation}
\sin\theta < {{\Delta z_{n,n+1}'}\over{r_n+r_{n+1} }}
\end{equation}
where $\Delta z'_{n,n+1} = z_{n+1}'-z_n'$ is
the separation of the shell origins projected onto the plane of the sky.
For Hercules~A, the constraints are
$\theta<46^\circ, 47^\circ, 23^\circ$
for the pairs of visually overlapping rings 
(assumed to be circular)
as shown in the Figures~2(a) and 3(a) of \citet{morrison1996}.

However in our model, the luminous radio-emitting rings
may in principle be separated by distances much less than their radii,
and larger values of $\theta$ are feasible.
On the other hand,
our model explains the non-circularity of the rings,
implying that $\cos\theta\approx b'/a'$
where $a'$ and $b'$ are the semi-major and semi-minor axes
of the elliptical projection of a circular ring onto the plane of the sky.
Apart from the innermost bright feature
which we re-identify as a buried hot-spot
or reconfinement shock,
we distinguish up to five rings clearly in the radio images,
(lower panel of Figure~\ref{'fig.radio'})
from east to west:
a partially filled ellipse,
one dimmer partial elliptical arc,
a bright nearly filled ellipse,
a relatively dim ring
and then the  bright half-ring.
In the lower panel of Figure~\ref{'fig.radio'}
we exhibit the elliptical fits to some of the 
observed features,
and we infer inclinations of $\theta \approx 45^\circ \pm 10^\circ$.
These fits are demonstrative but not unique;
the thickness of the radio structures allows some freedom
in fitting ellipses as does the inclusion or exclusion of other faint filamentary
features near each ring.
However the fitted ellipses of projected ring-shocks,
with their projected major axes 
approximately locally perpendicular to the jet are consistent with our model.

\subsection{Implications for the cluster environment}

The two temperatures of diffuse X-ray emission
from the cluster medium surrounding Hercules~A
may reflect the different states of gas
upstream and downstream of the bow shock
driven by the outer radio lobes.
The observed temperature ratio is $1.7 \la T/T_\mathrm{ism} \la 2.1$,
which is comparable to the temperature ratios
of the shocked thermal gas in some of our simulations.
In Table~\ref{table.temperature.ratios}
we show temperature ratios found at the point of the bow shock,
and on the sides of the bow shock parallel to the jet axis.
The tabulated temperature ratios indicate a lower Mach number $\sim 5$,
if this explanation for the two-temperature structure of the local cluster
medium is correct.
The temperature ratios are too small ($\approx 1$)
for the $M=2$ simulations.

\subsection{Closing remarks}

We have presented a model for the western lobe of Hercules~A 
in which the ring structures are explained as shocks in the jet cocoon, 
with a ribbon-like, approximately annular geometry.
The  absence of a strong hot-spot at the
extremity of the radio lobe may be attributed to the surging of the
head of the jet on length scales of
$\sim 10^2 \> \mathrm{kpc}$,
as a result of the dynamics of the turbulent and
entraining backflows in the radio lobe.
The lack of a hot-spot near the edge of the lobe, the low contrast of
the radio rings and the appearance
of the Eastern jet all suggest a low Mach number
$M \sim 2-5$, low density ratio $\eta < 10^{-4}$ jet.
In order that the jet power be consistent with the
luminosity of the source, $\eta$ may be as low as $10^{-6}$
and this is consistent with our view of radio galaxy jets
as being composed of relativistic plasma.
We have suggested that the recessed hot-spot may be associated with 
one of the bright knots some
$35^{\prime\prime}$ west of the core. 
Another possibility is that this knot is associated with a
reconfinement shock and the hot-spot could well be absent. 
However, the fraction of the time when a
hot-spot is {\em entirely} absent in these simulations is small.

Another appealing aspect of our modelling is that we have estimated 
an orientation such that the eastern
jet makes an acute angle to the line of sight,
$30^\circ \la \theta < 70^\circ$,
consistent with the value  $\theta\approx 50^\circ$
determined from the radio depolarization measurements of
\citet{gizani1999}.

With the simulations that we have carried only three out of the 
twelve have an energy flux that is close
to the range of 
$\sim 2 \times 10^{45-46} \> \mathrm{erg} \> \mathrm{s}^{-1}$ 
identified in \S~\ref{s:pars}. 
However,
the other nine simulations are useful in that their morphologies and 
contrasts serve to identify a
trend towards the low Mach numbers and density ratios that is 
consistent with the region of parameter
space defined by the energy budget.

The notion of a reasonably low Mach number for the western jet in 
this source is compatible with the
appearance of {\em both} sides. 
Both jets seems to be on the verge of a transition to full turbulence. 
It is noteworthy that this may be occurring in a source 
which is so much more powerful than borderline FR~1/FR~2 sources. 
The production of turbulent jets in Hercules~A is
most likely the result of the large core-radius of the hot X-ray 
emitting gas that gradually decelerates
the initially relativistic jets to a low Mach number. 
Normally one expects the X-ray core-radius to be of
the order of the optical core radius since at small radii 
the matter distribution is dominated by the optical potential. 
At present we can only speculate that the large 
X-ray core radius in Hercules~A is
related to the merger apparent in the  optical observations.
\citep{minkowski1957,sadun1993}.

Finally we should comment on the relevance of two-dimensional simulations.
These are justified {\em a priori}
on the basis of observed ring-like structure
and the simulations seem to
capture a lot of the physics that is relevant to the observations.
Nevertheless, it would be desirable
to carry out three-dimensional simulations
and test some of the features of the present work.
For example, how much of a perturbation from axi-symmetry is allowable
before the ring structure is
destroyed and how sensitive is the Mach number dependence of the
entrainment effects to the
dimensionality of the simulations?

\section*{Acknowledgments}

\begin{figure}
\begin{center}
$
\begin{array}{c}
\includegraphics[width=12cm]{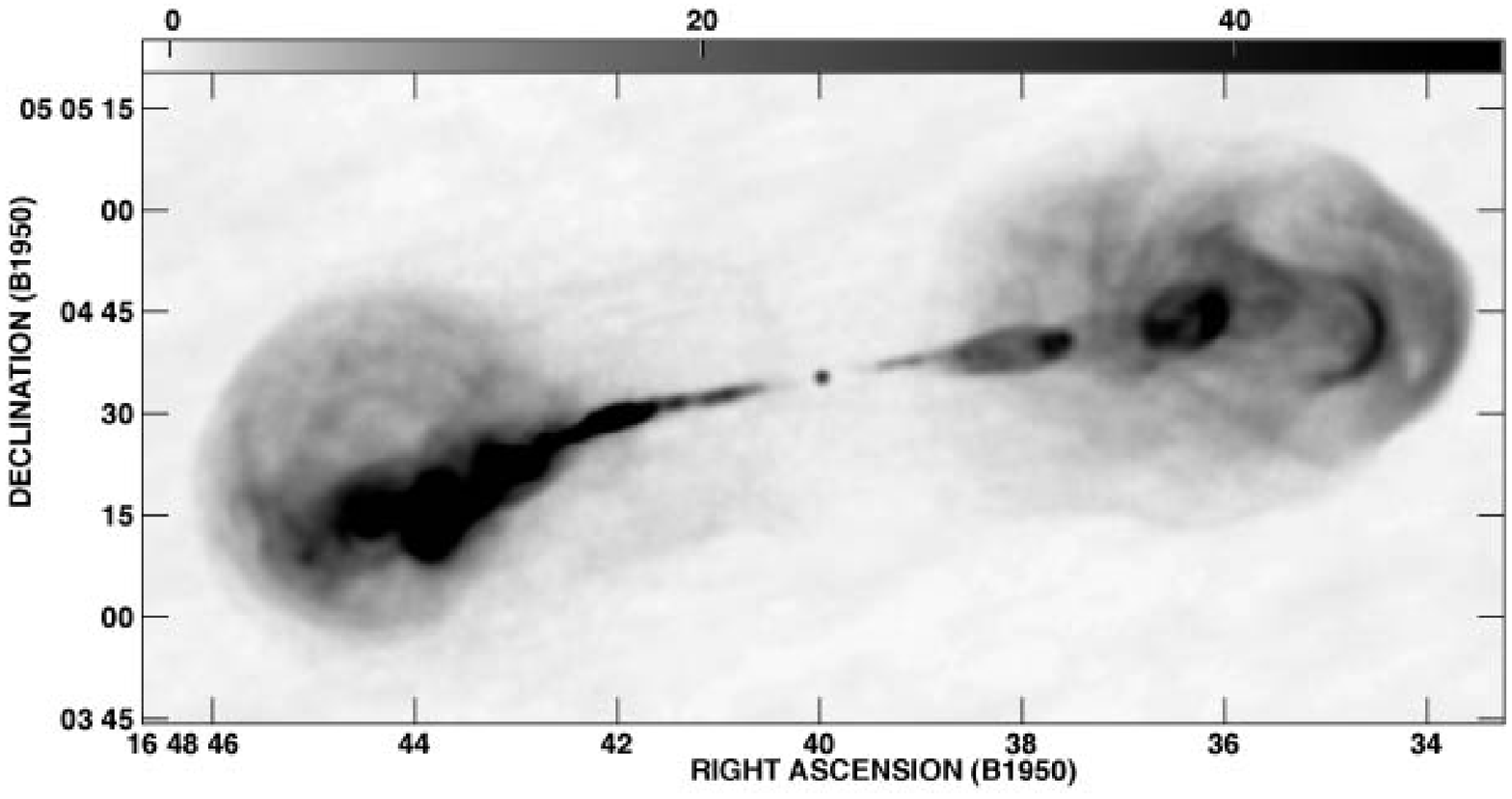}
\\
\includegraphics[width=10cm]{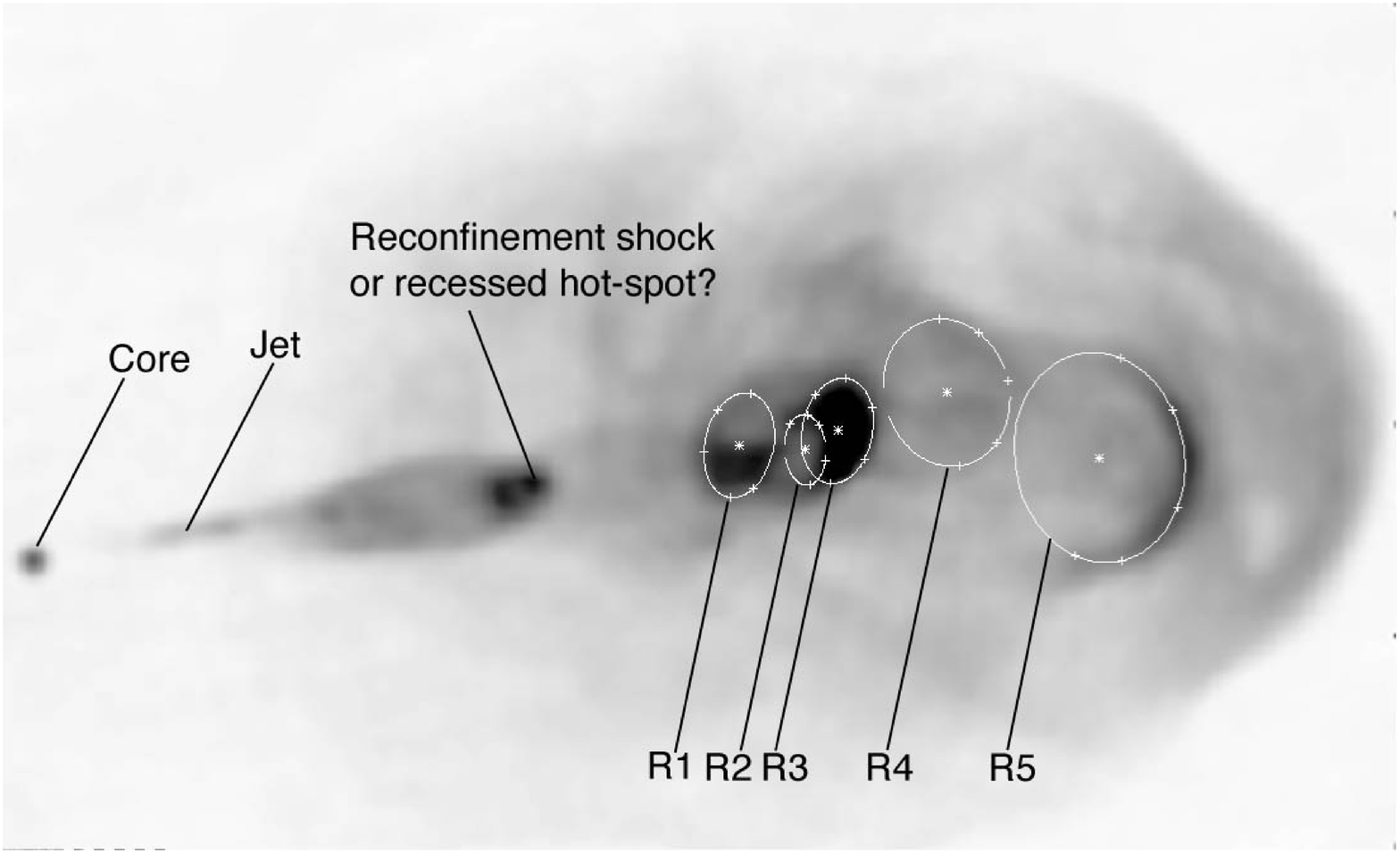}
\end{array}
$
\caption{
Radio image (at 18 cm) of Hercules~A,
its lobes, jets and associated rings,
from \citet{gizani2002}
and published with the permission of those authors.
The lower panel is an enhanced-contrast sub-region
showing the core, jet and shock features.
Some apparent ring-like structures are marked $R1 - R5$.
The illustrative fitted ellipses are at inclinations
$\theta=36^\circ, 32^\circ, 38^\circ, 50^\circ, 47^\circ$
respectively.
}
\label{'fig.radio'}
\end{center}
\end{figure}

\begin{figure}
\begin{center}
$
\begin{array}{cc}
  \includegraphics[width=7cm]{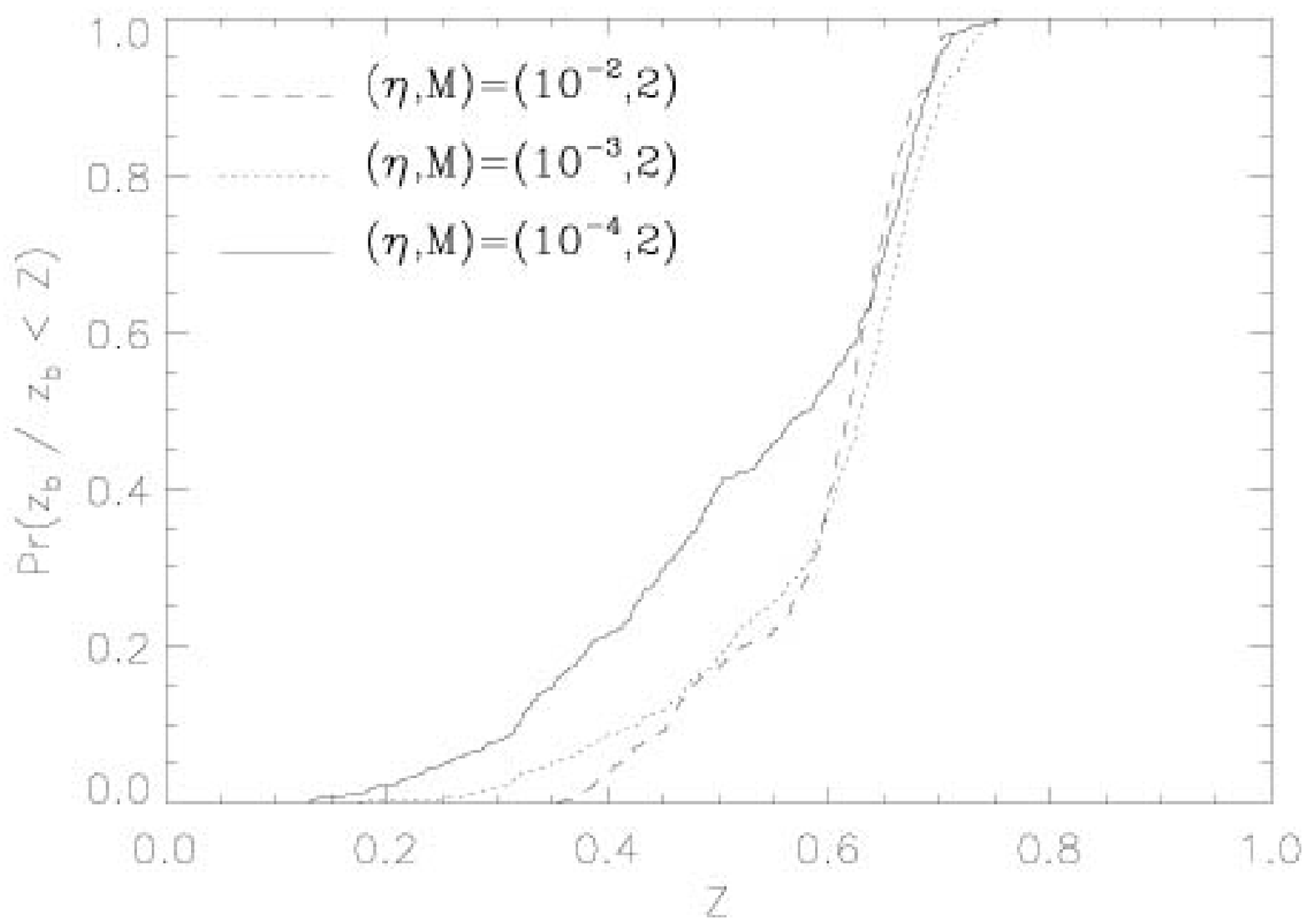}
&\includegraphics[width=7cm]{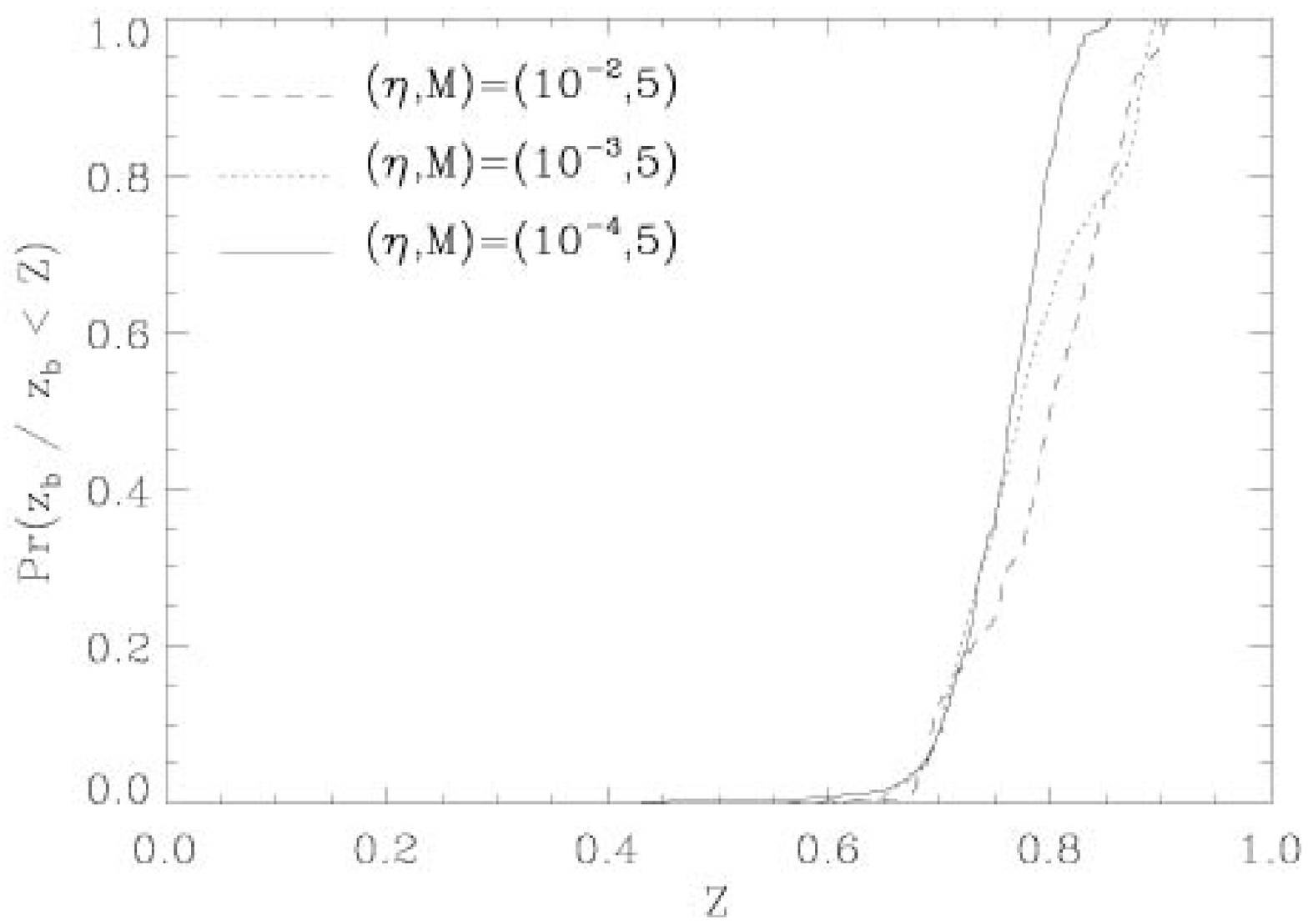}
\\
  \includegraphics[width=7cm]{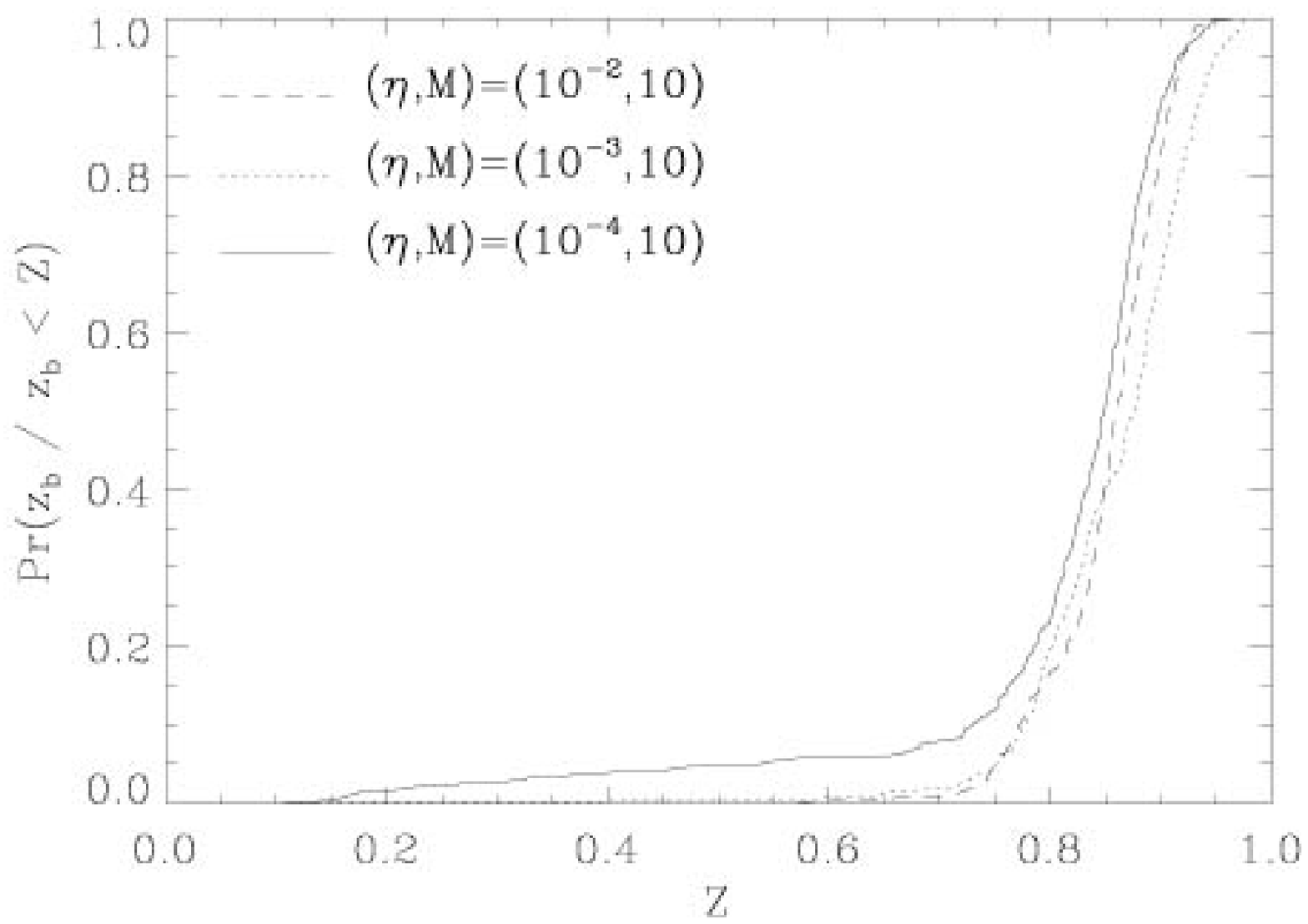}
&\includegraphics[width=7cm]{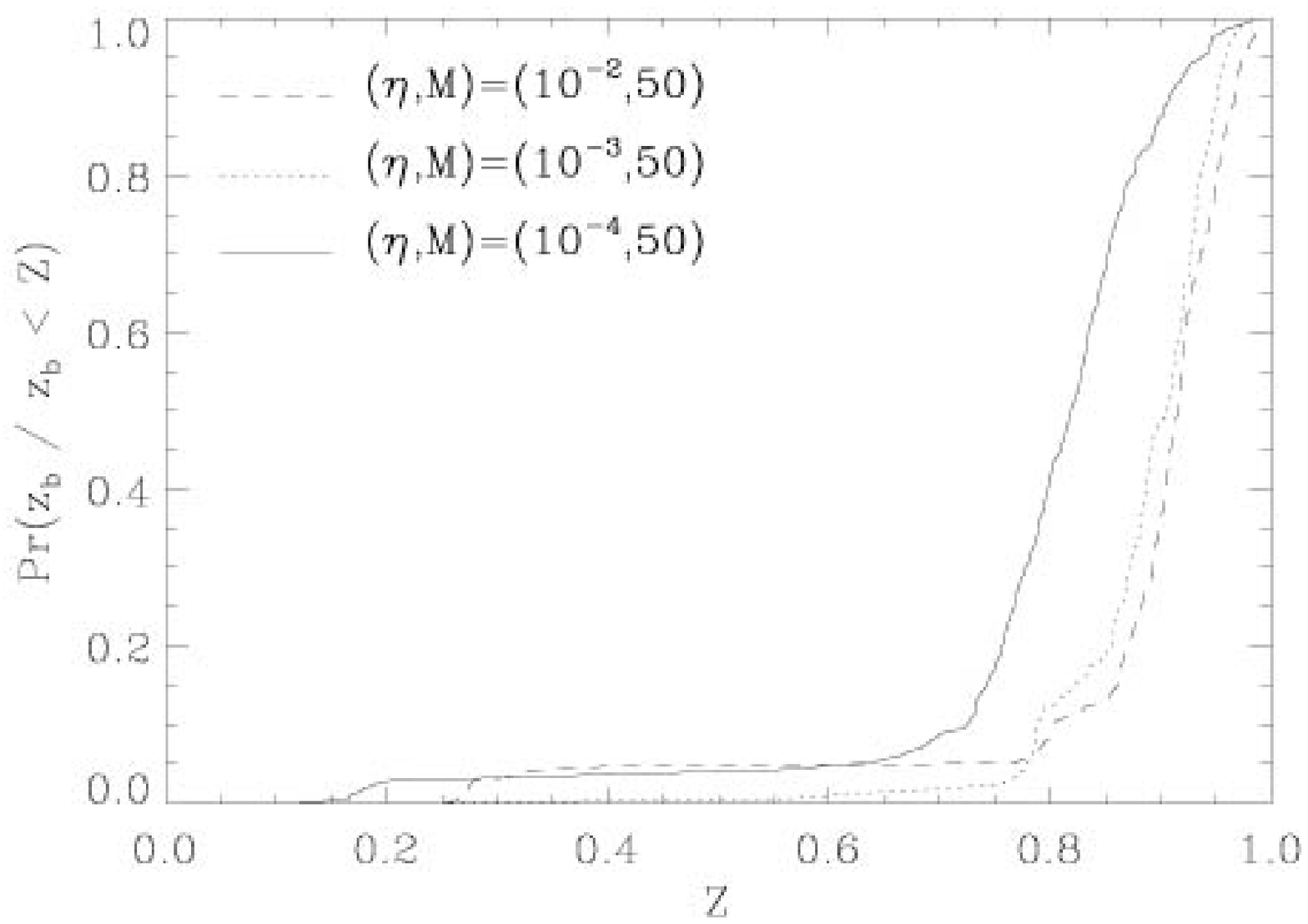}
\\
\end{array}
$
\caption{
Cumulative distribution functions expressing
the frequency and extent of surging
in terms of ratios of the $z$ positions of
the bow shock and jet head.
The jet's head position $z_\mathrm{h}$
is defined as the forwardmost cell
where the jet plasma remains unmixed with external gas,
i.e. $\varphi=1$.
We show cases with $M=2, 5, 10, 50$.
These are simulations with an open left boundary,
and $\eta=10^{-2}, 10^{-3}, 10^{-4}$
for dashed, dotted and solid lines respectively.
}
\label{f:surging}
\end{center}
\end{figure}

\begin{figure}
\begin{center}
$
\begin{array}{ccc}
  \includegraphics[width=5cm]{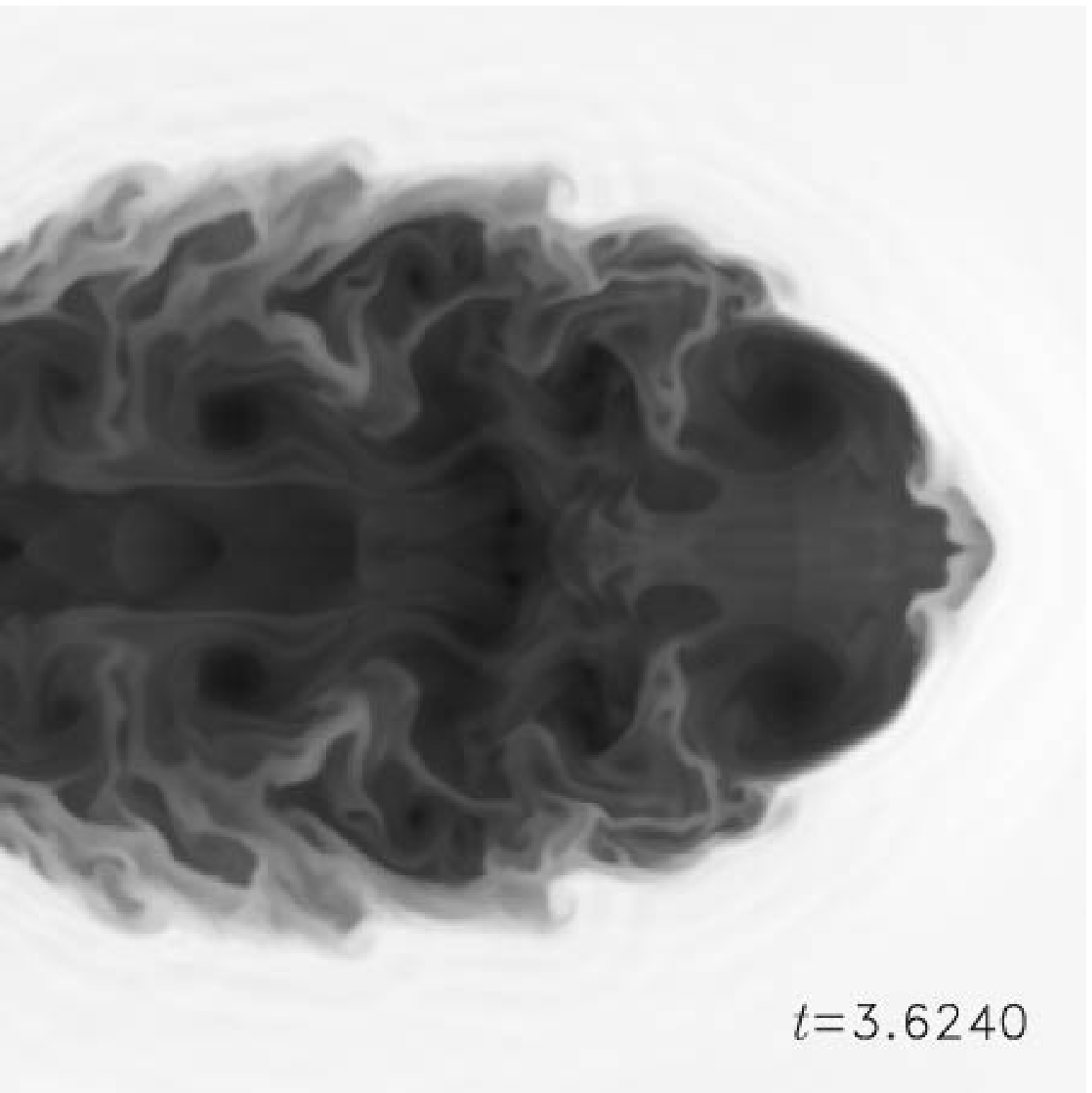}
&\includegraphics[width=5cm]{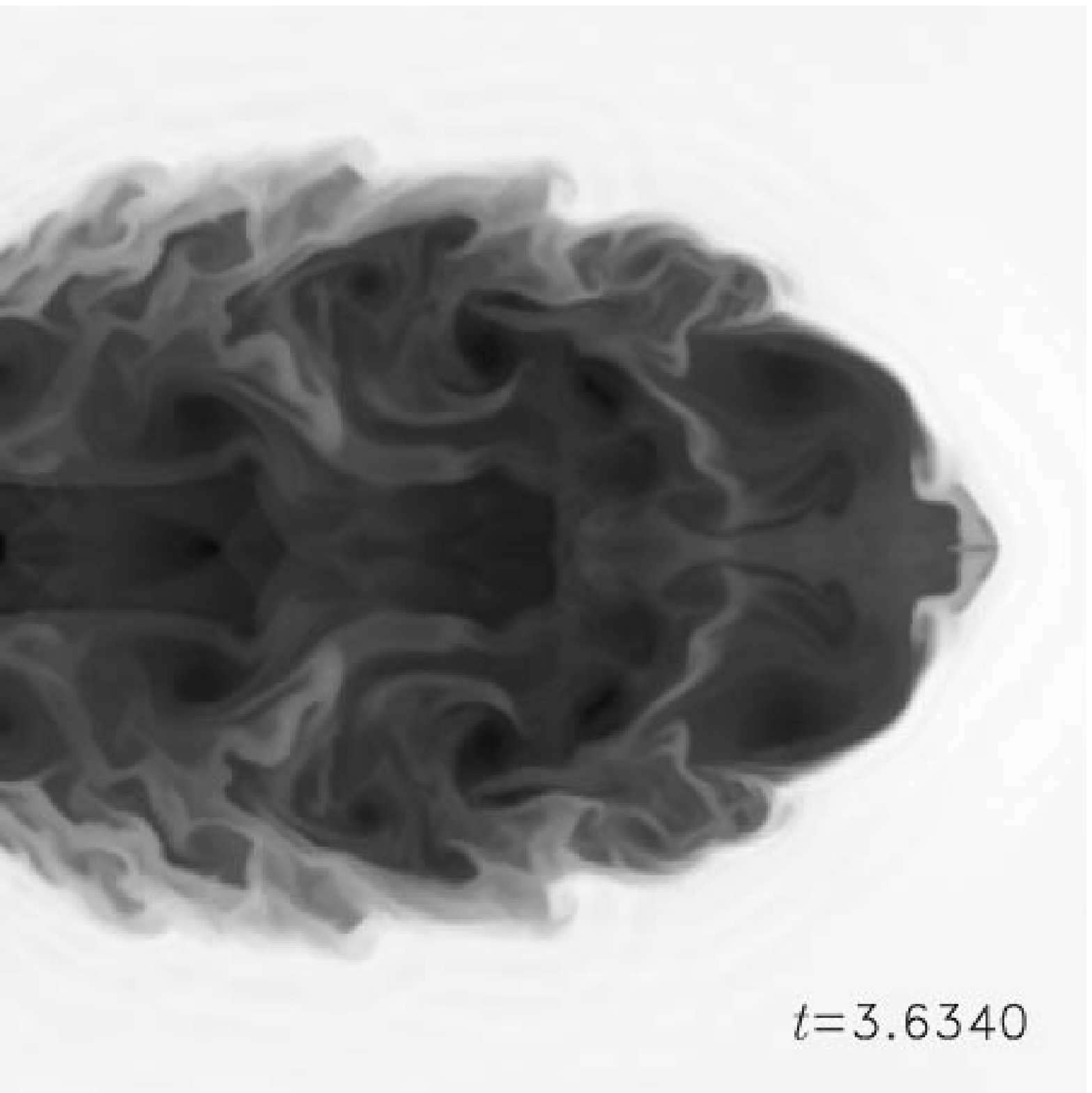}
&\includegraphics[width=5cm]{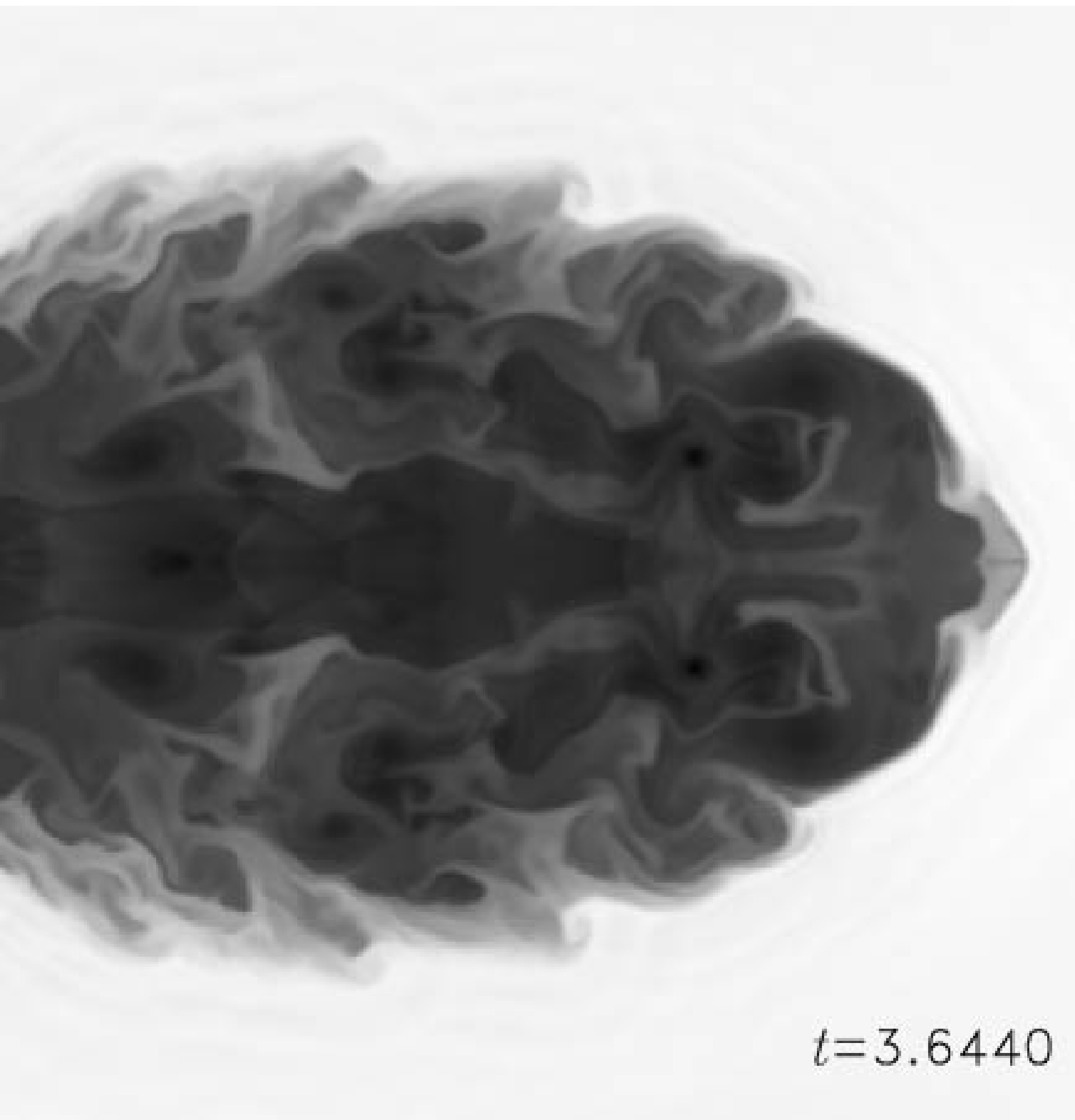}
\\
  \includegraphics[width=5cm]{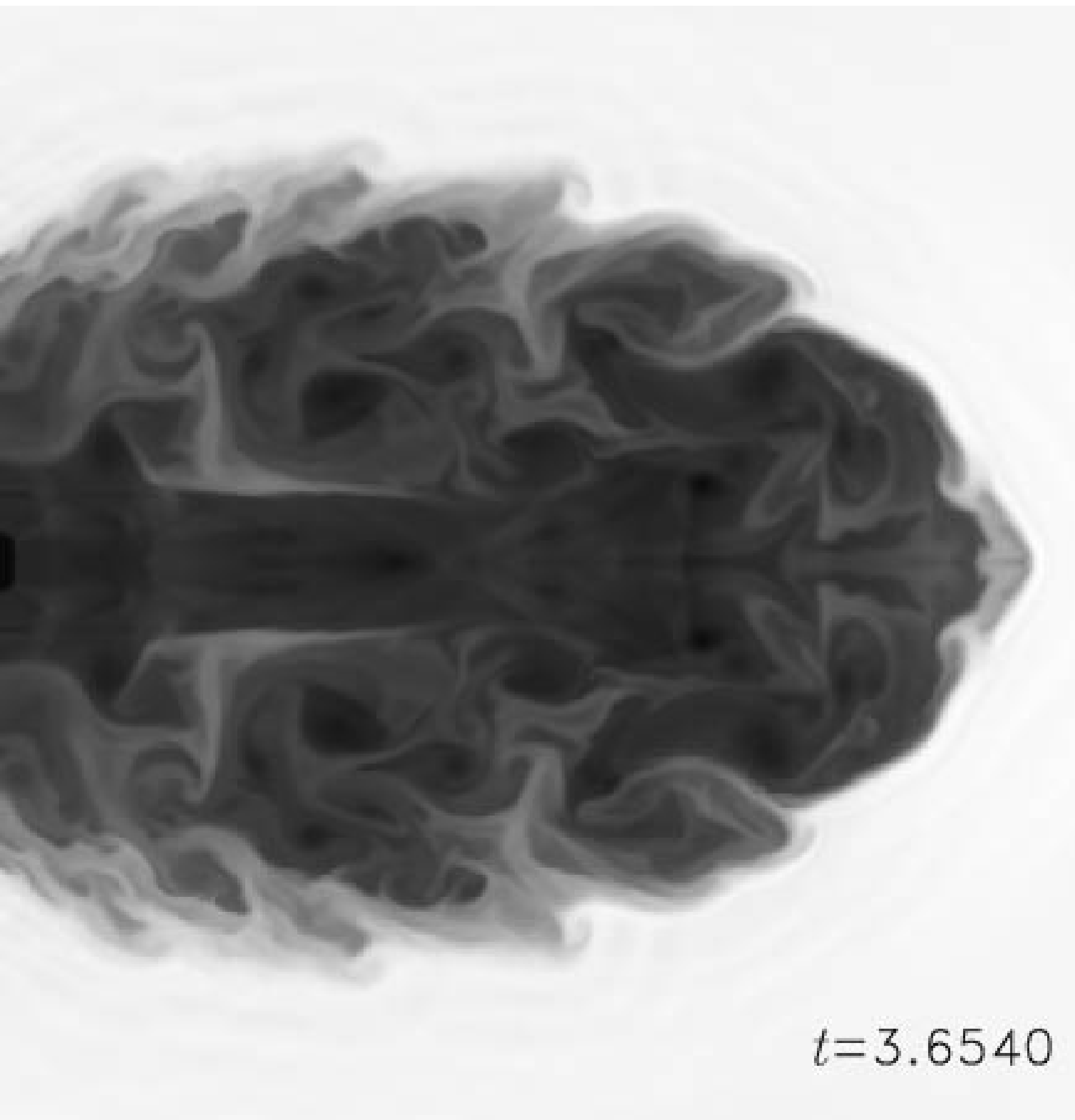}
&\includegraphics[width=5cm]{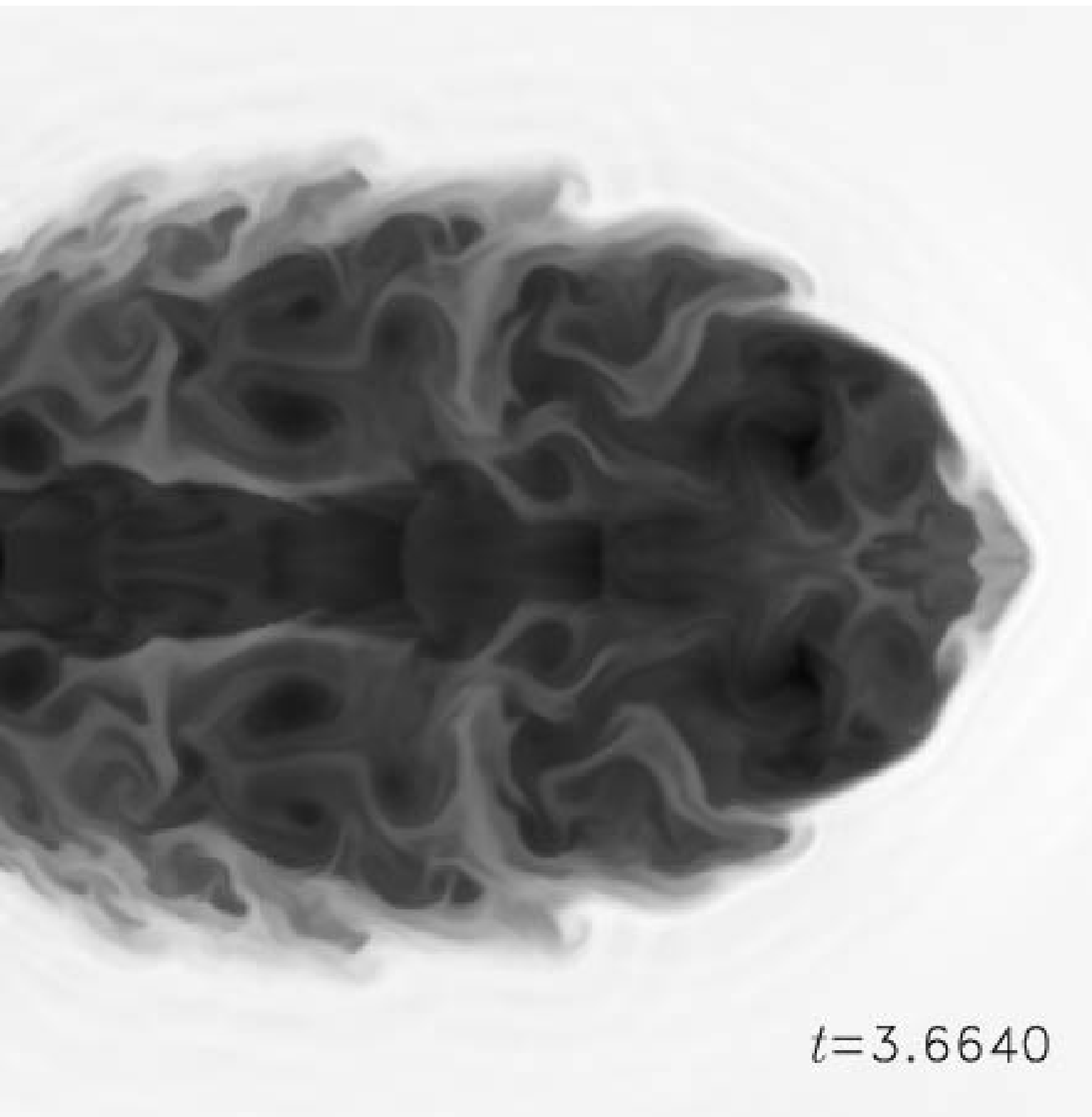}
&\includegraphics[width=5cm]{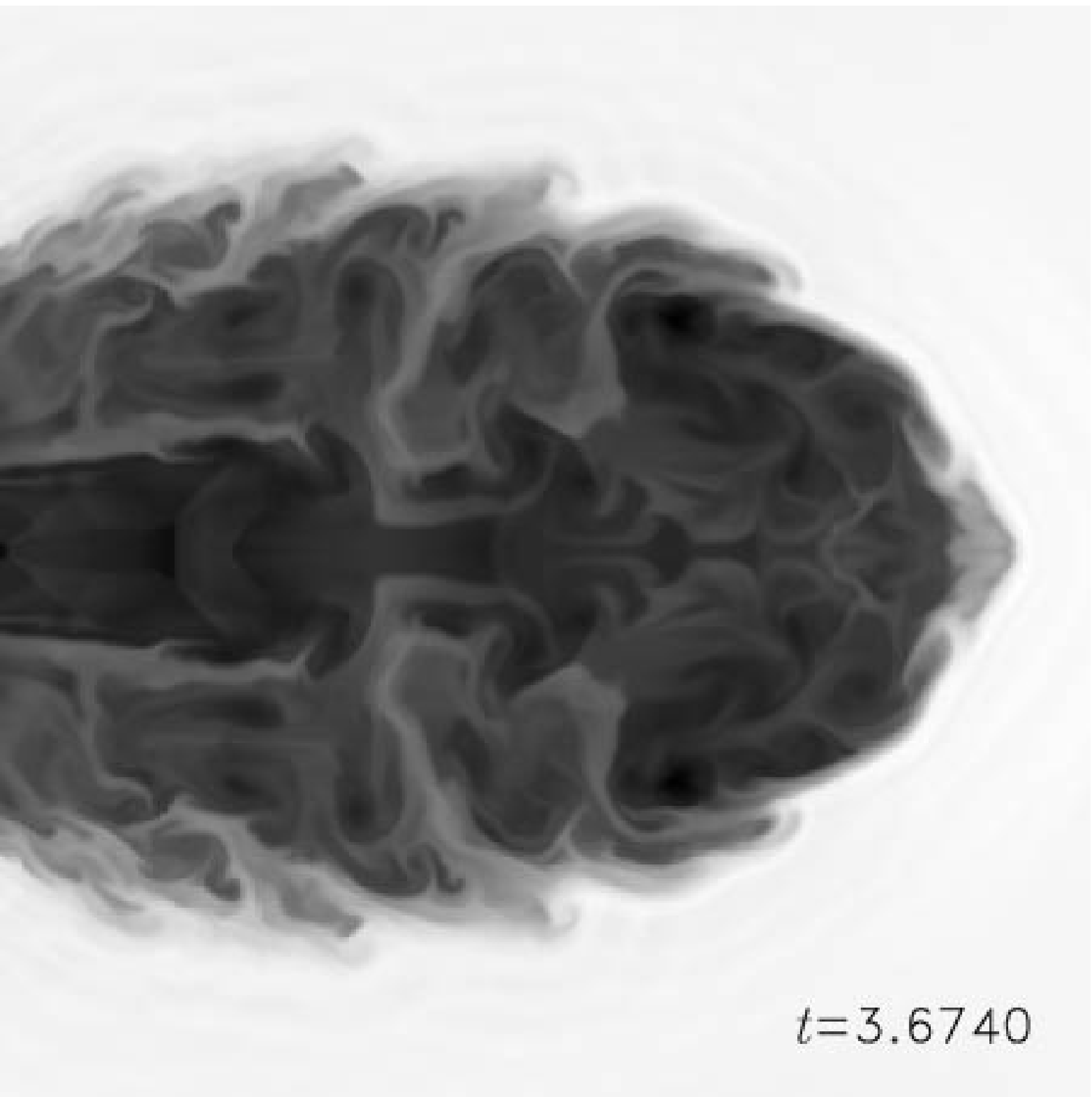}
\\
  \includegraphics[width=5cm]{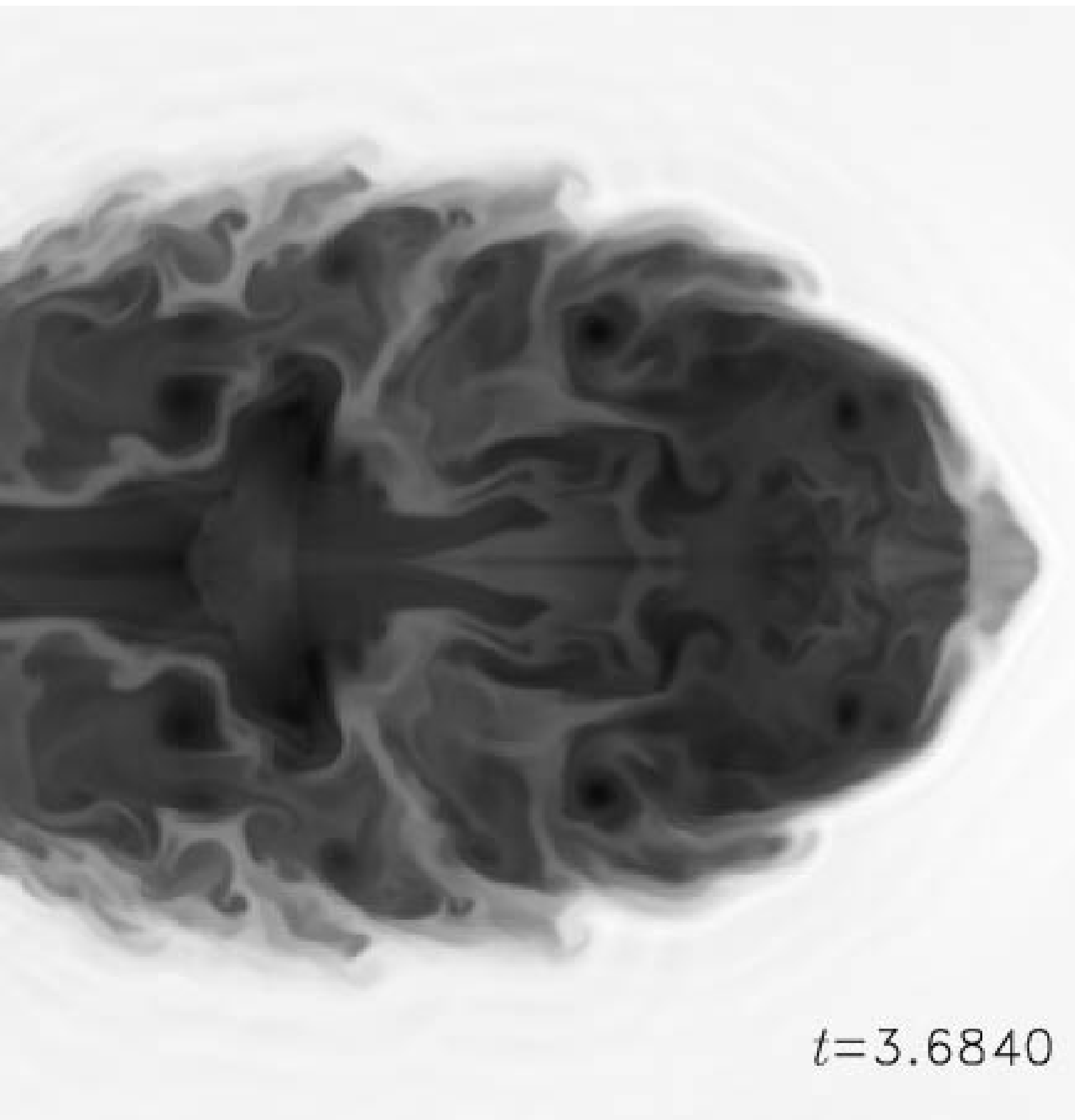}
&\includegraphics[width=5cm]{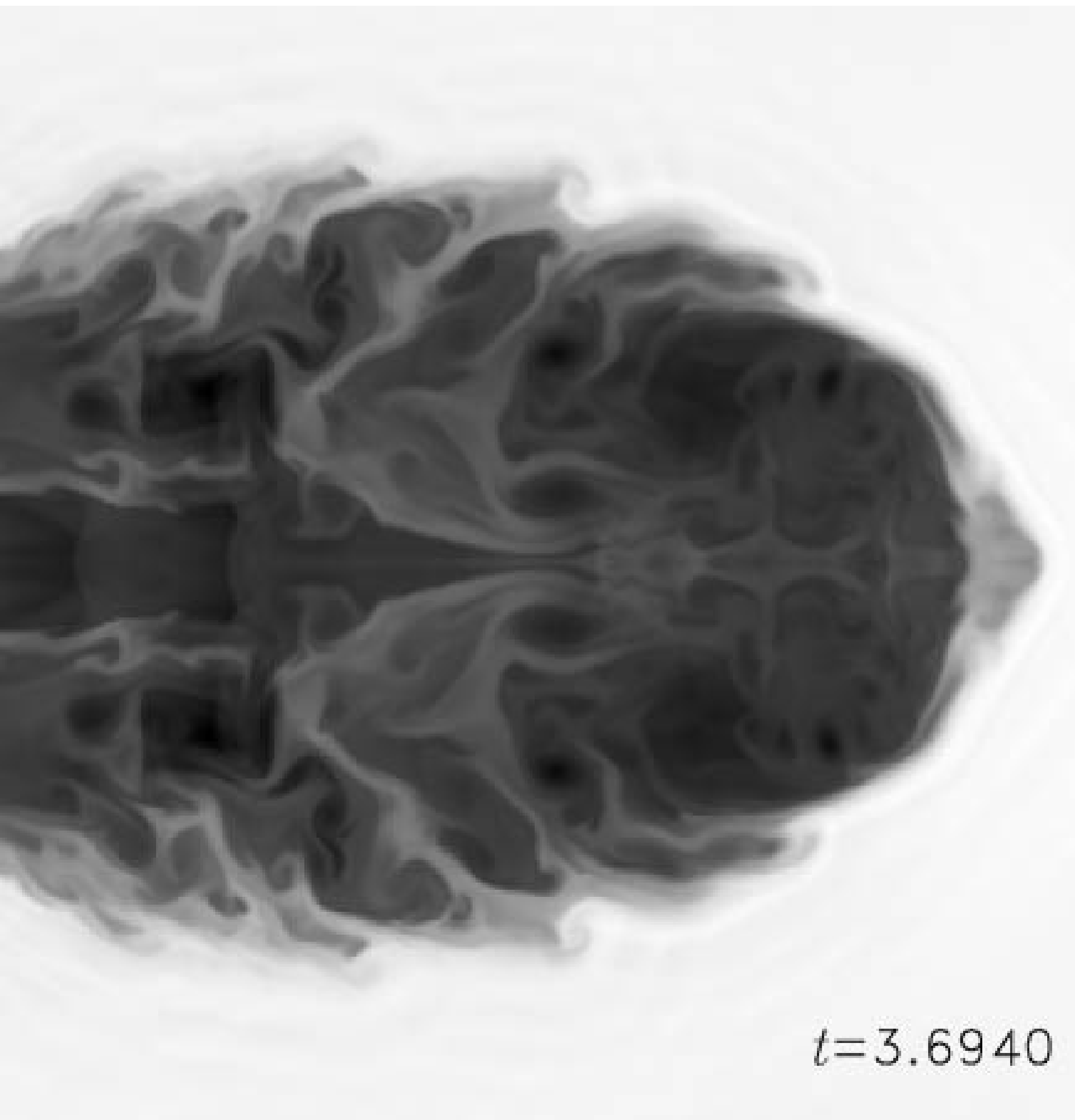}
&\includegraphics[width=5cm]{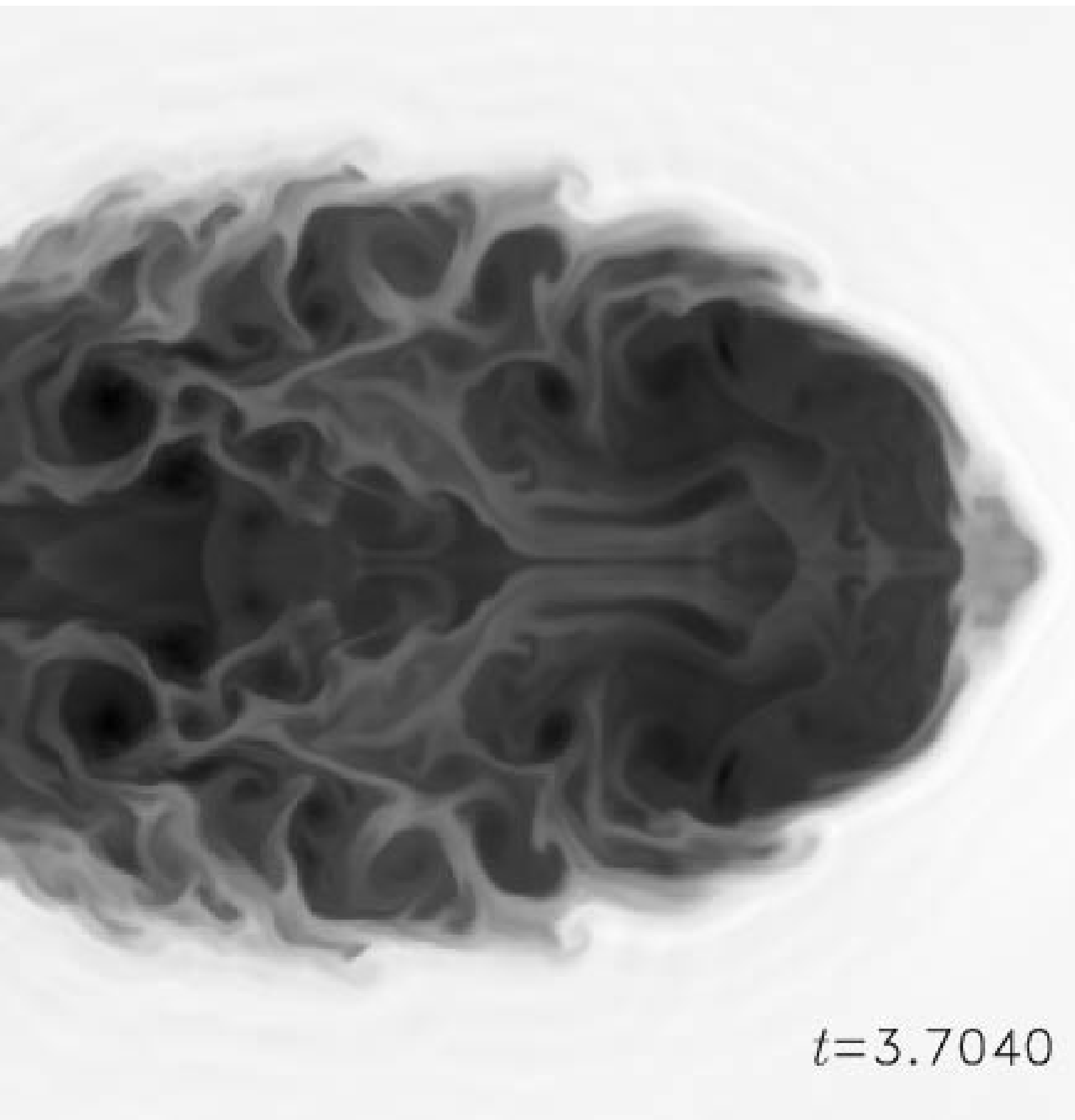}
\\
\end{array}
$
\caption{
A sequences of $\log\rho$ cross-sections of the
jet showing as it is affected by dense fingers
of external gas entrained into the cocoon. 
In this sequence, $(\eta,M) = (10^{-4},2)$, 
the dark color represents the low density jet material; 
the lightest areas represent the dense ambient gas.
}
\label{f:fingers}
\end{center}
\end{figure}

\begin{figure}
\begin{center}
$
\begin{array}{ccc}
  \includegraphics[width=5cm]{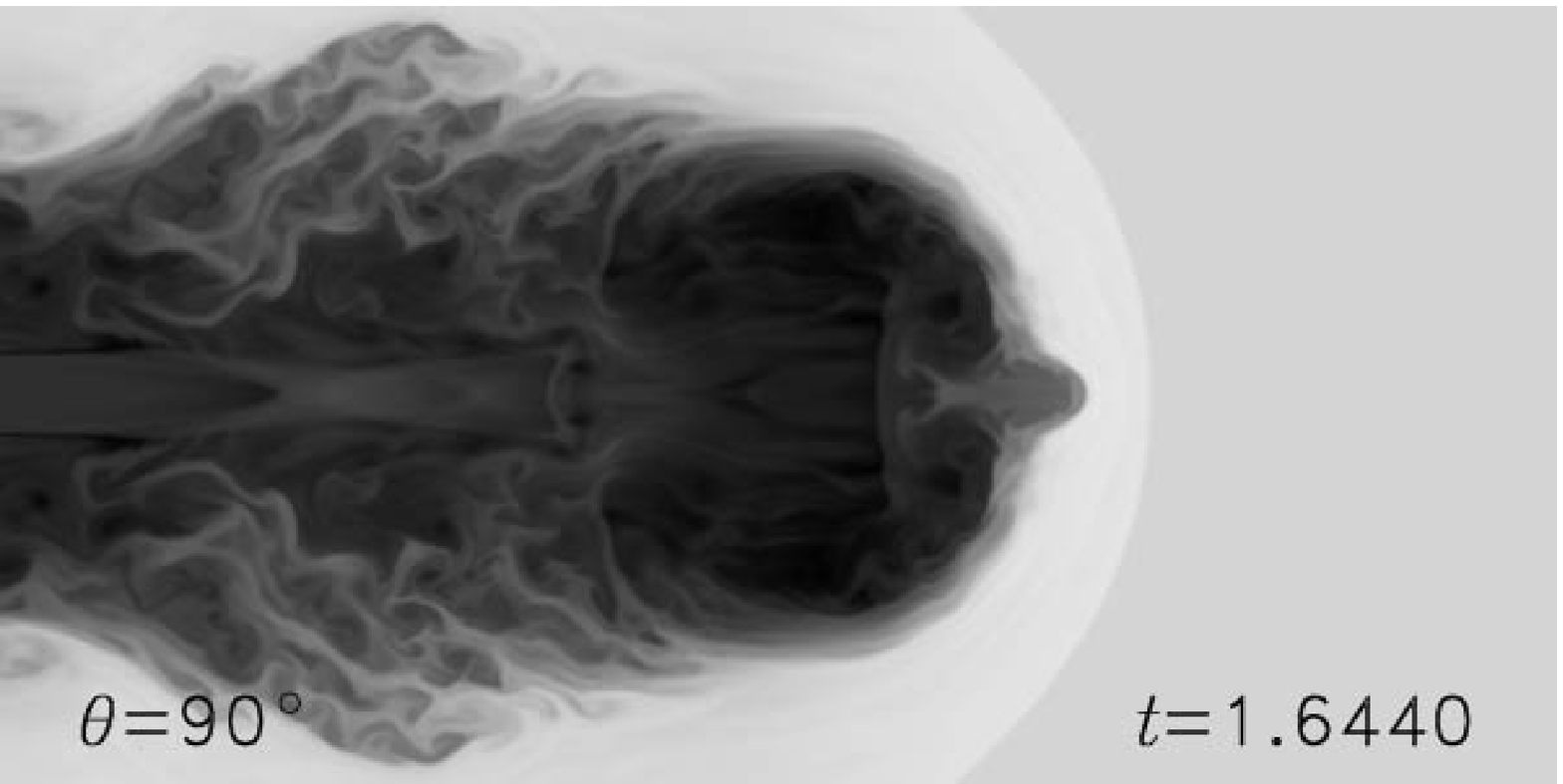}
&\includegraphics[width=5cm]{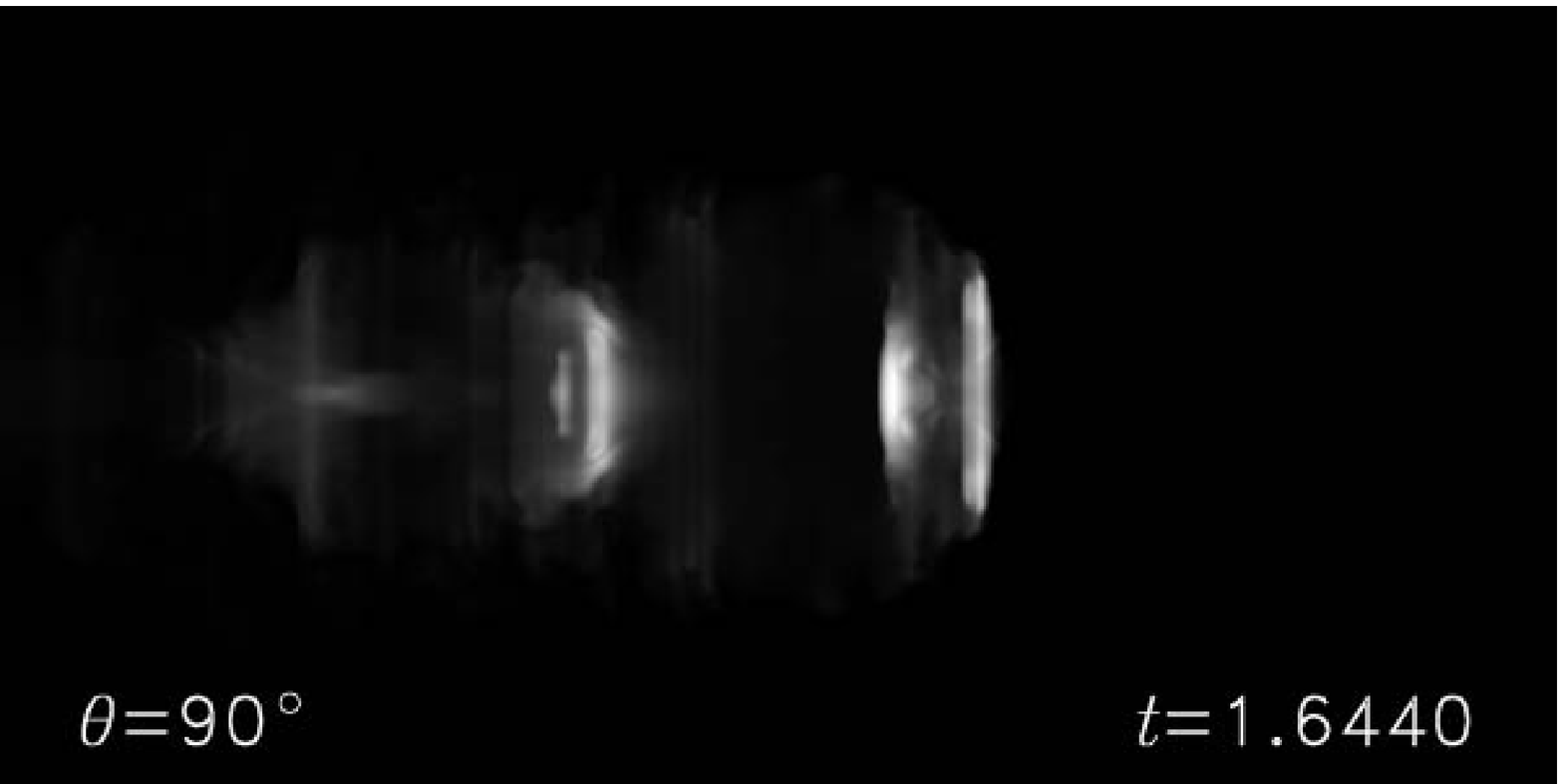}
&\includegraphics[width=5cm]{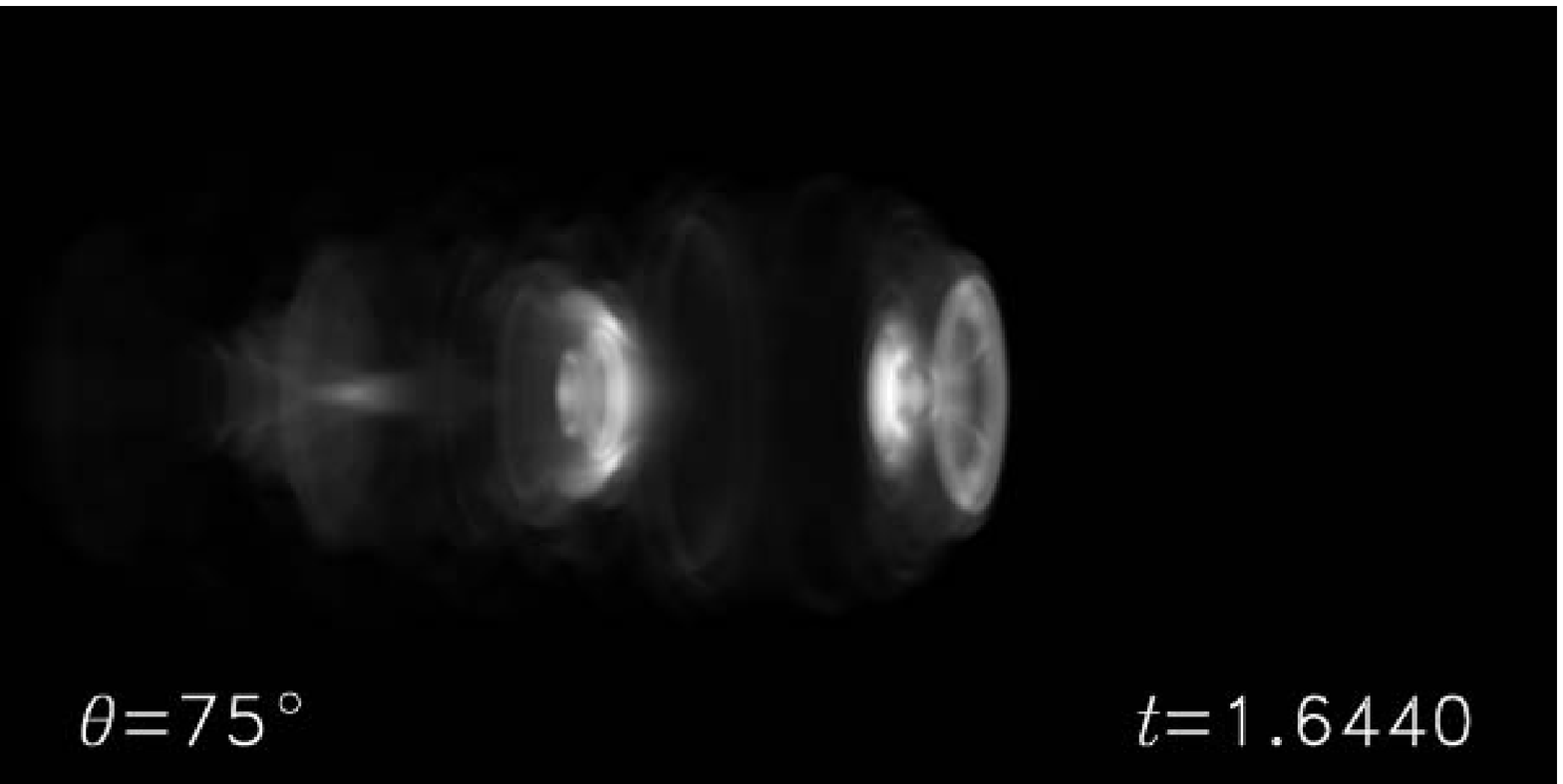}
\\
  \includegraphics[width=5cm]{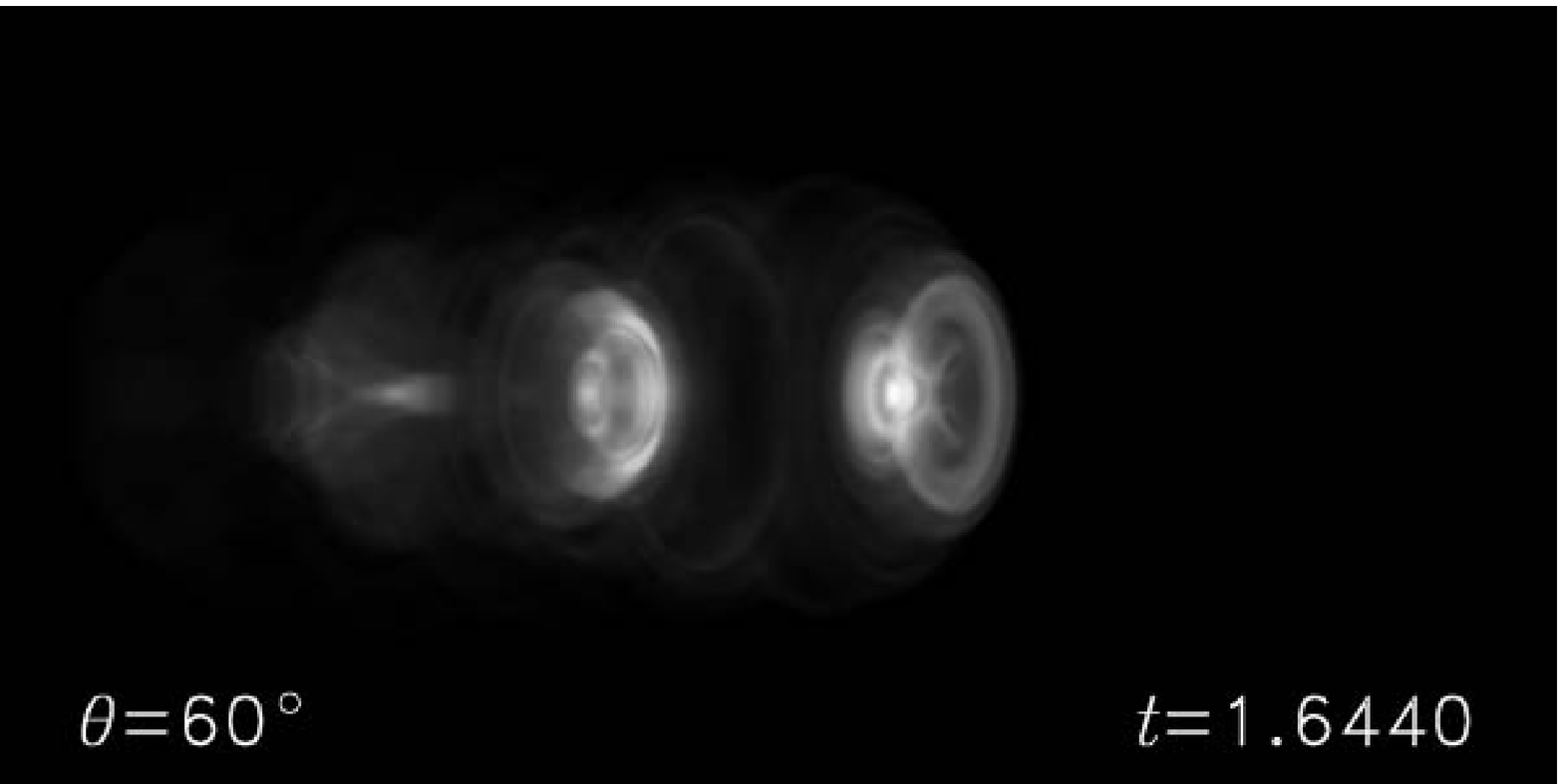}
&\includegraphics[width=5cm]{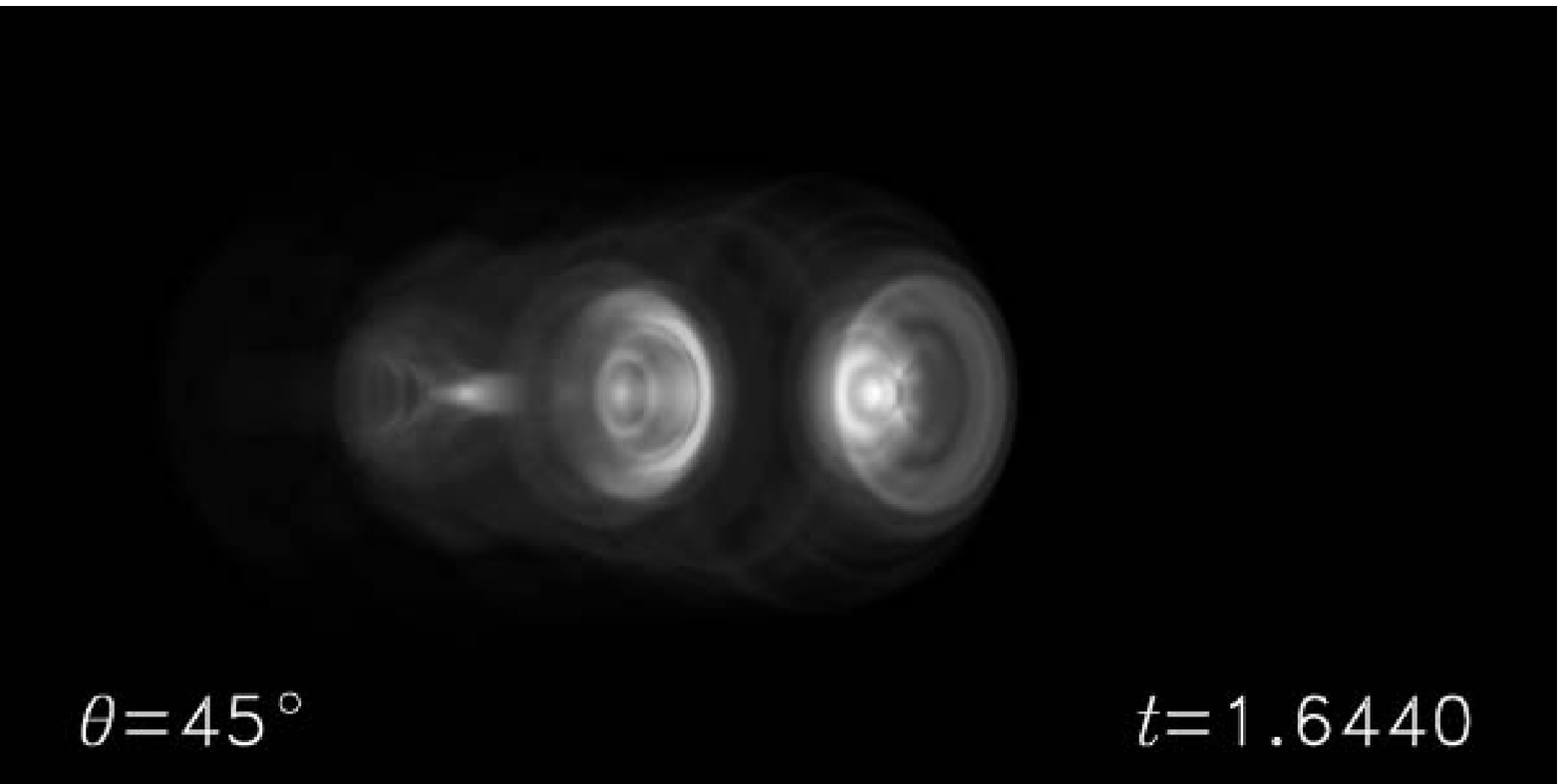}
&\includegraphics[width=5cm]{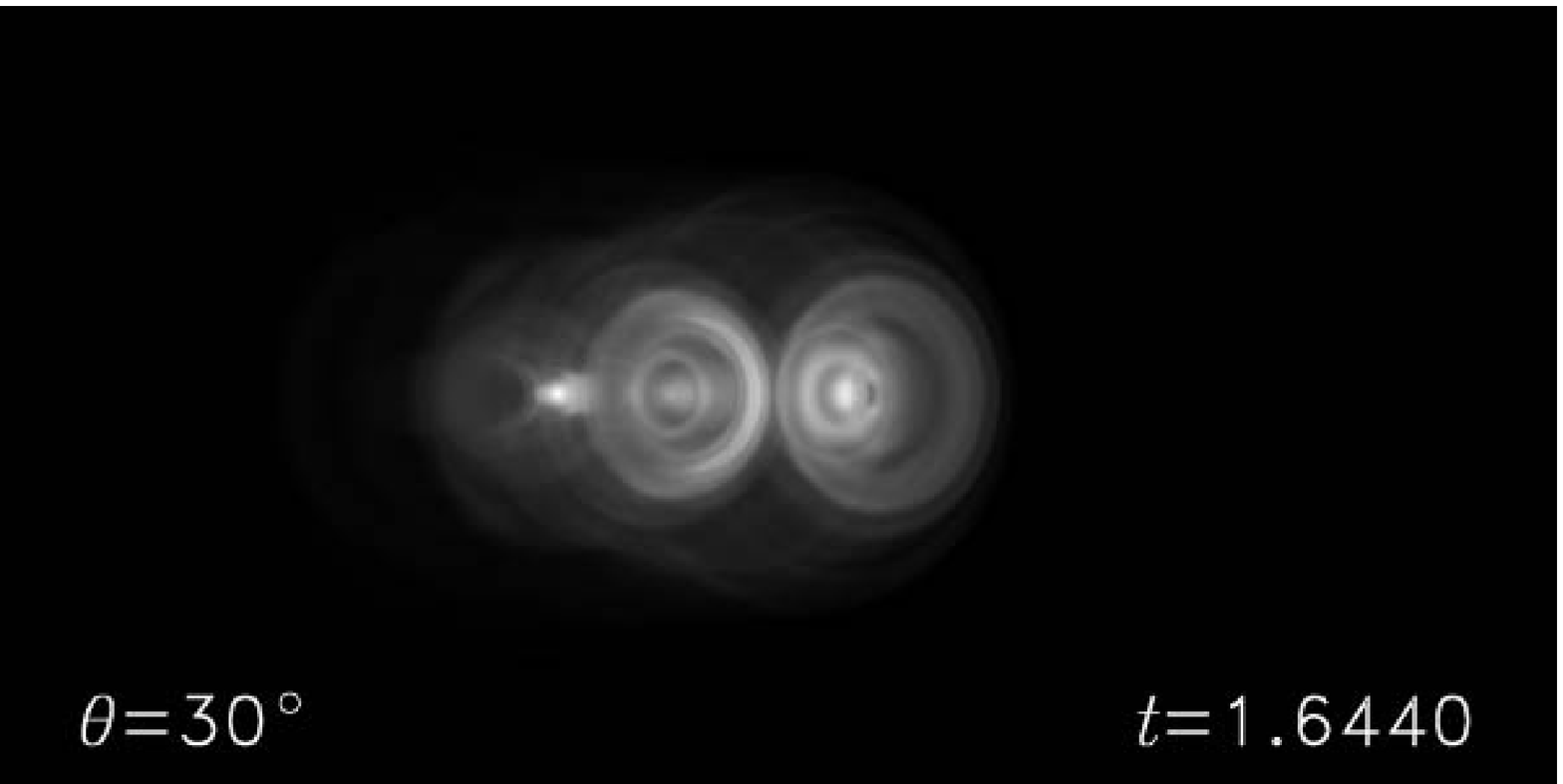}
\\
  \includegraphics[width=5cm]{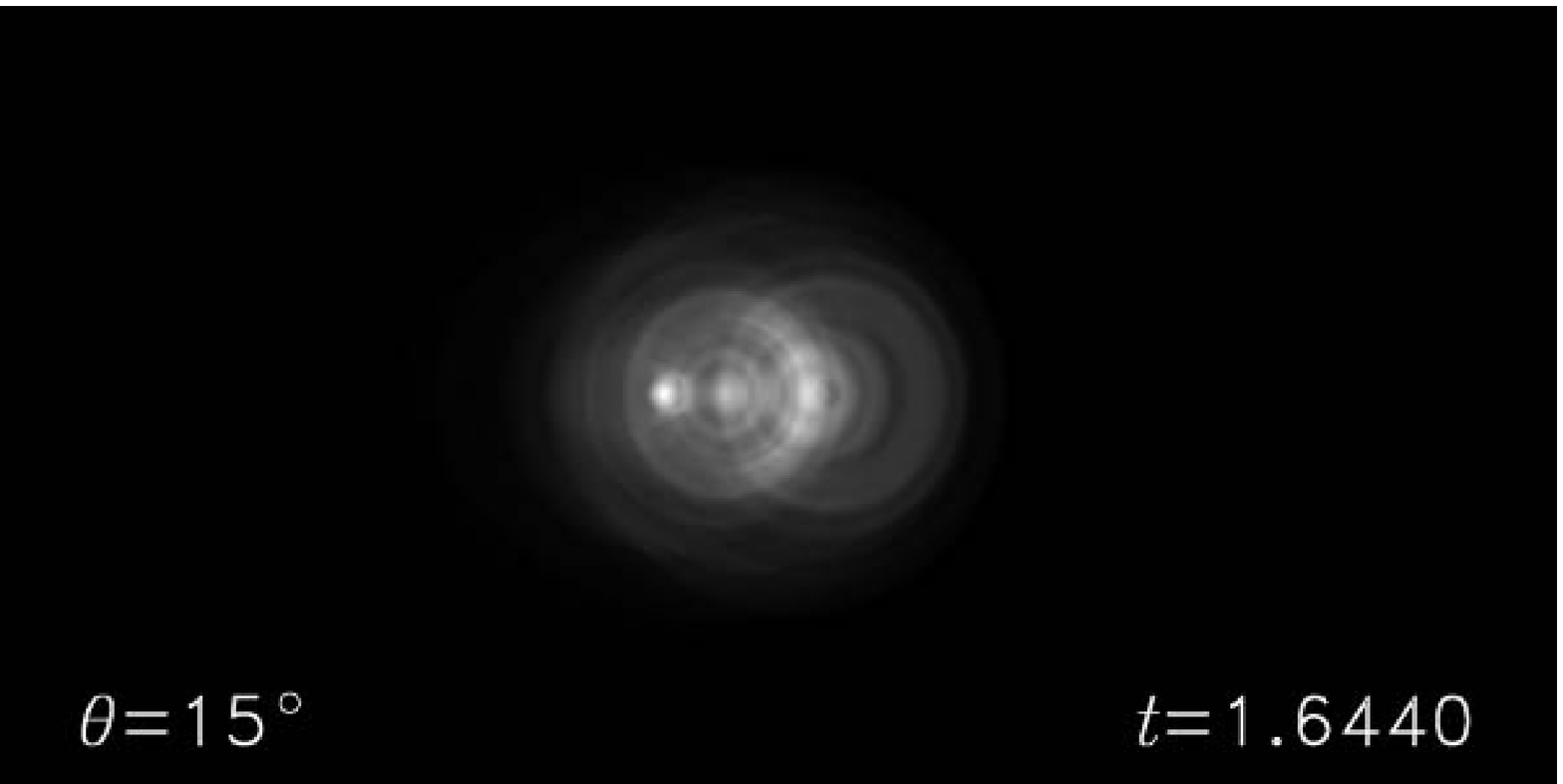}
&\includegraphics[width=5cm]{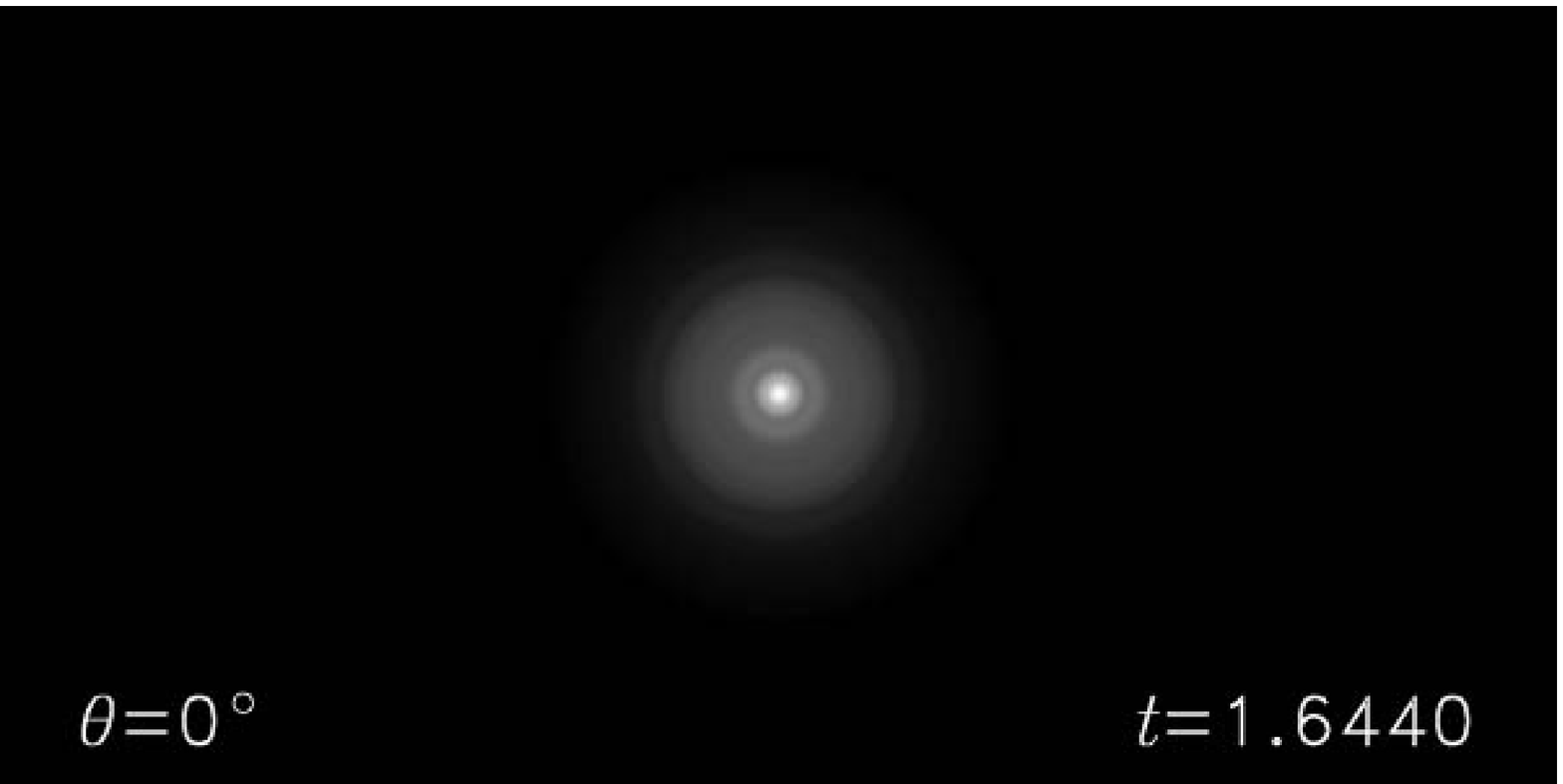}
\\
\end{array}
$
\caption{
A sub-region of the $(\eta,M)=(10^{-4},5)$ simulation,
showing $\log\rho$ (top left panel),
and a corresponding sequence of simulated intensity maps
for different orientations,
with an angle $\theta$ between the jet axis and the line of sight.
}
\label{f:orientation}
\end{center}
\end{figure}

\begin{figure}
\begin{center}
$
\begin{array}{ccc}
\includegraphics[width=4.5cm]{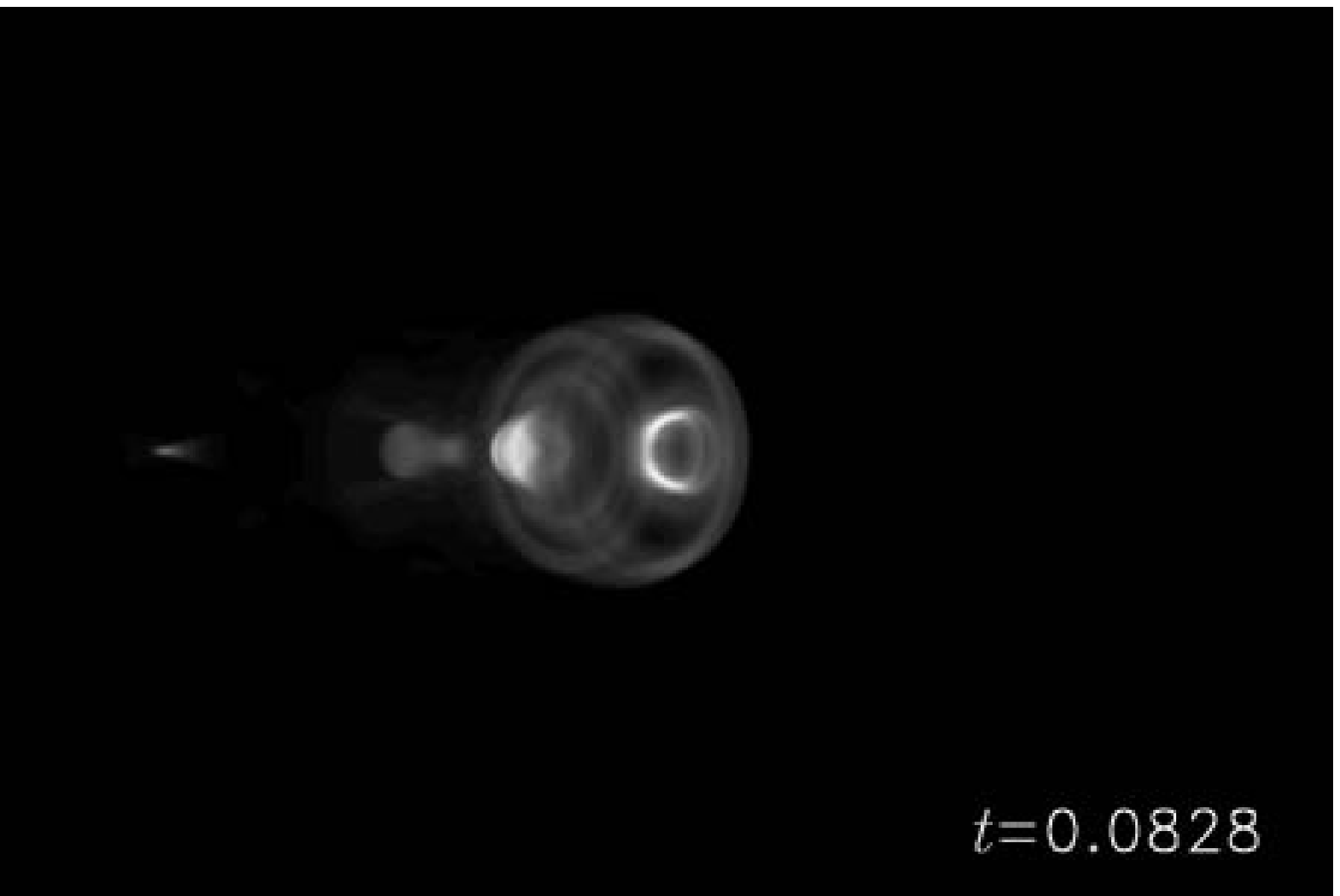}
&\includegraphics[width=4.5cm]{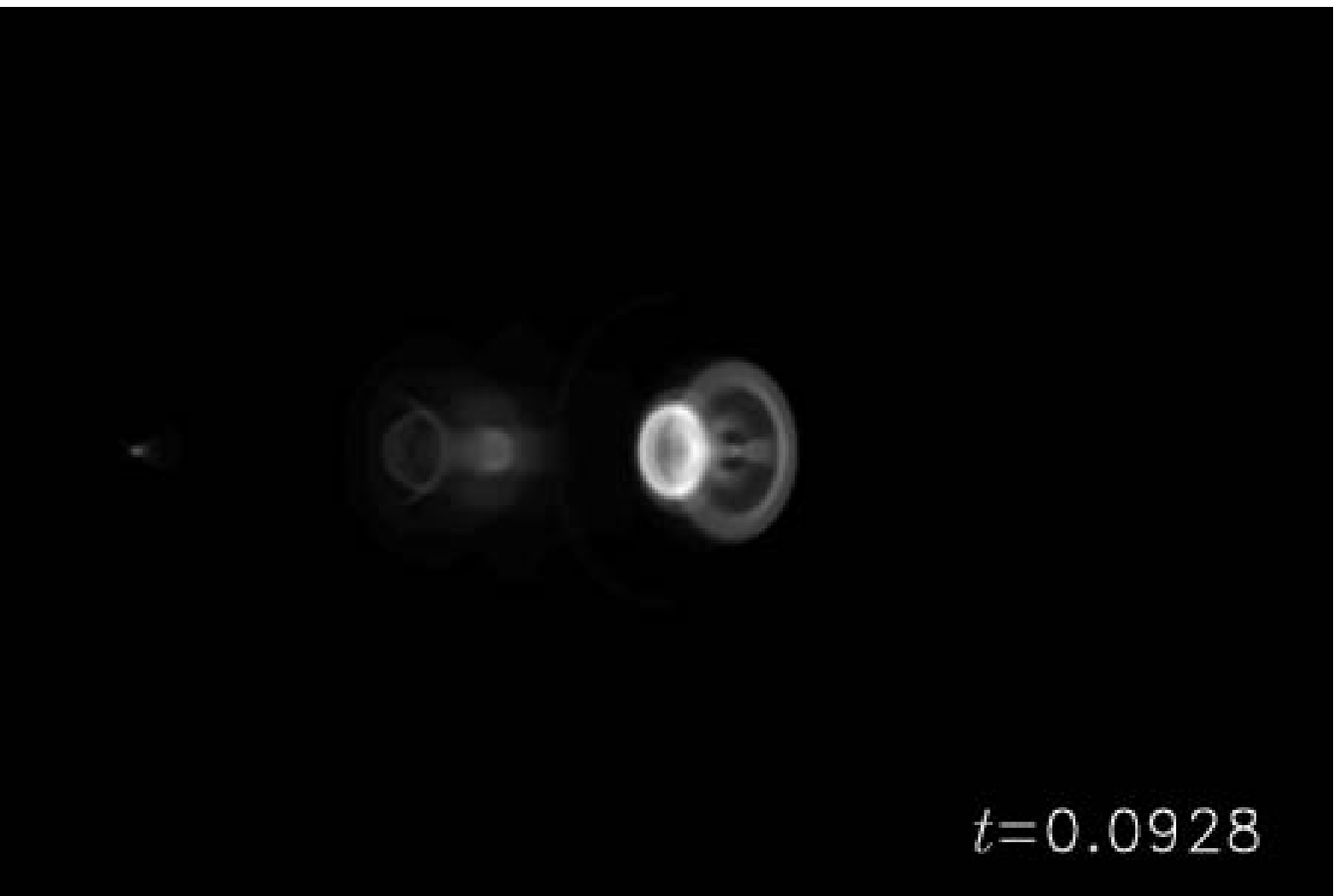}
&\includegraphics[width=4.5cm]{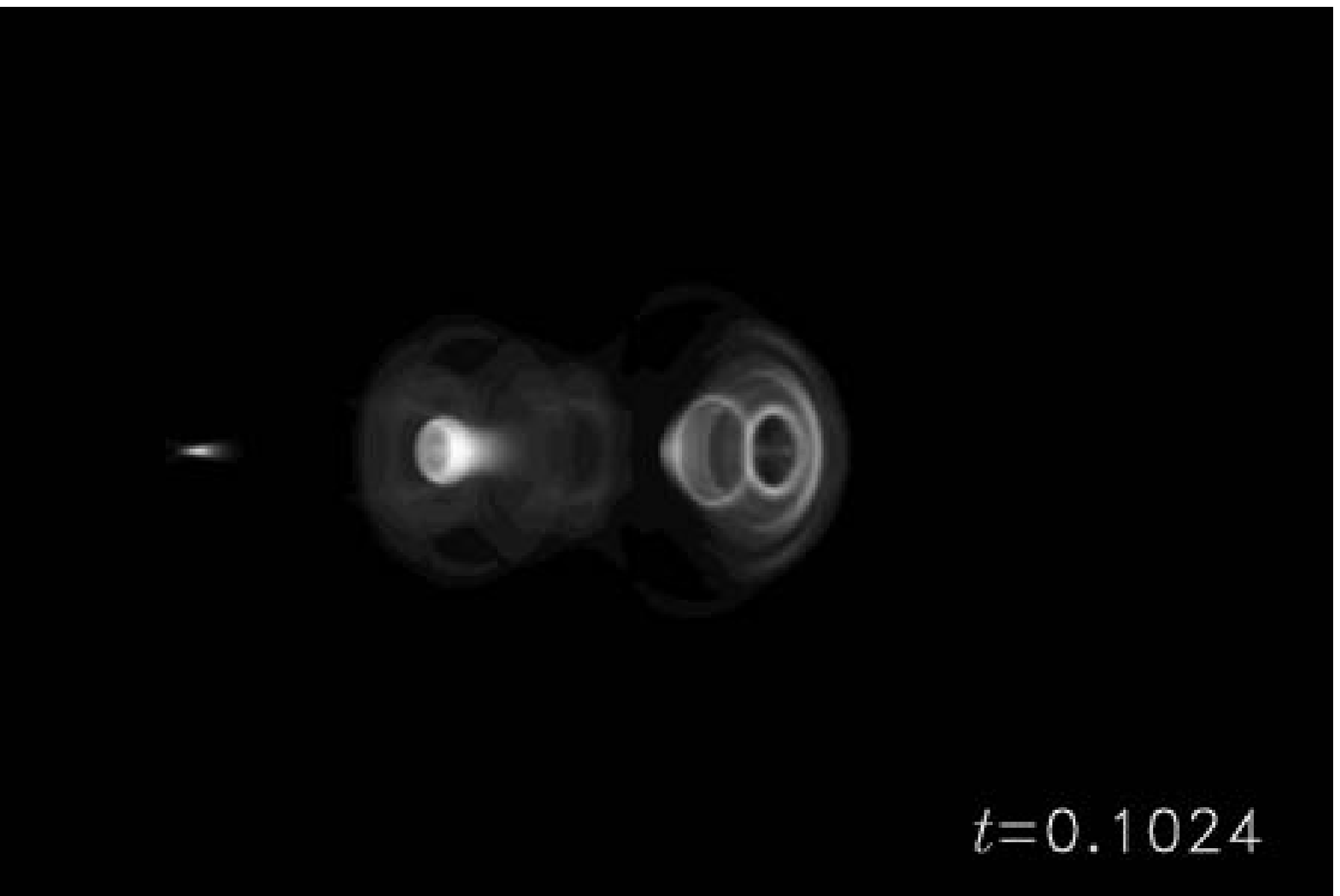}
\\
\includegraphics[width=4.5cm]{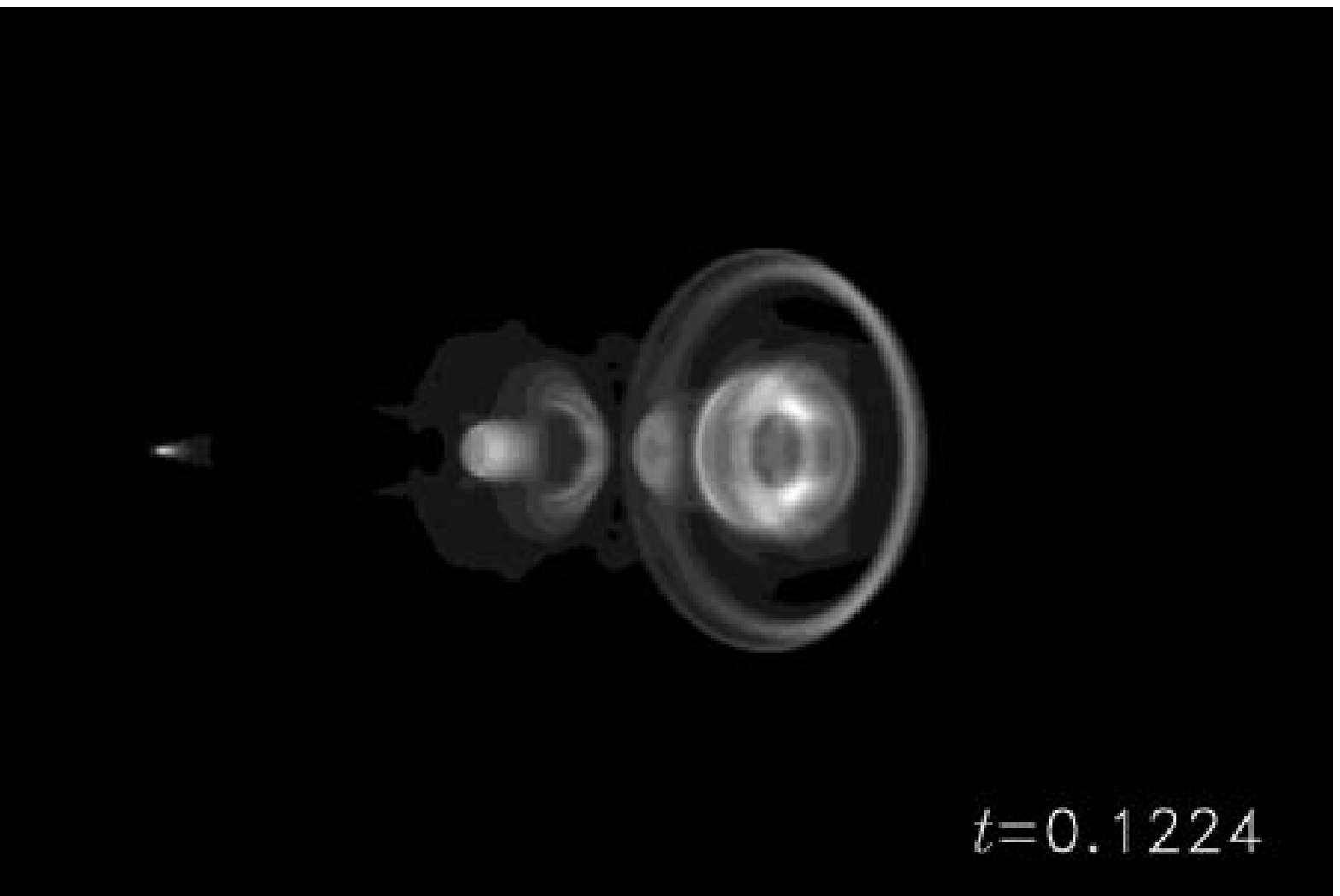}
&\includegraphics[width=4.5cm]{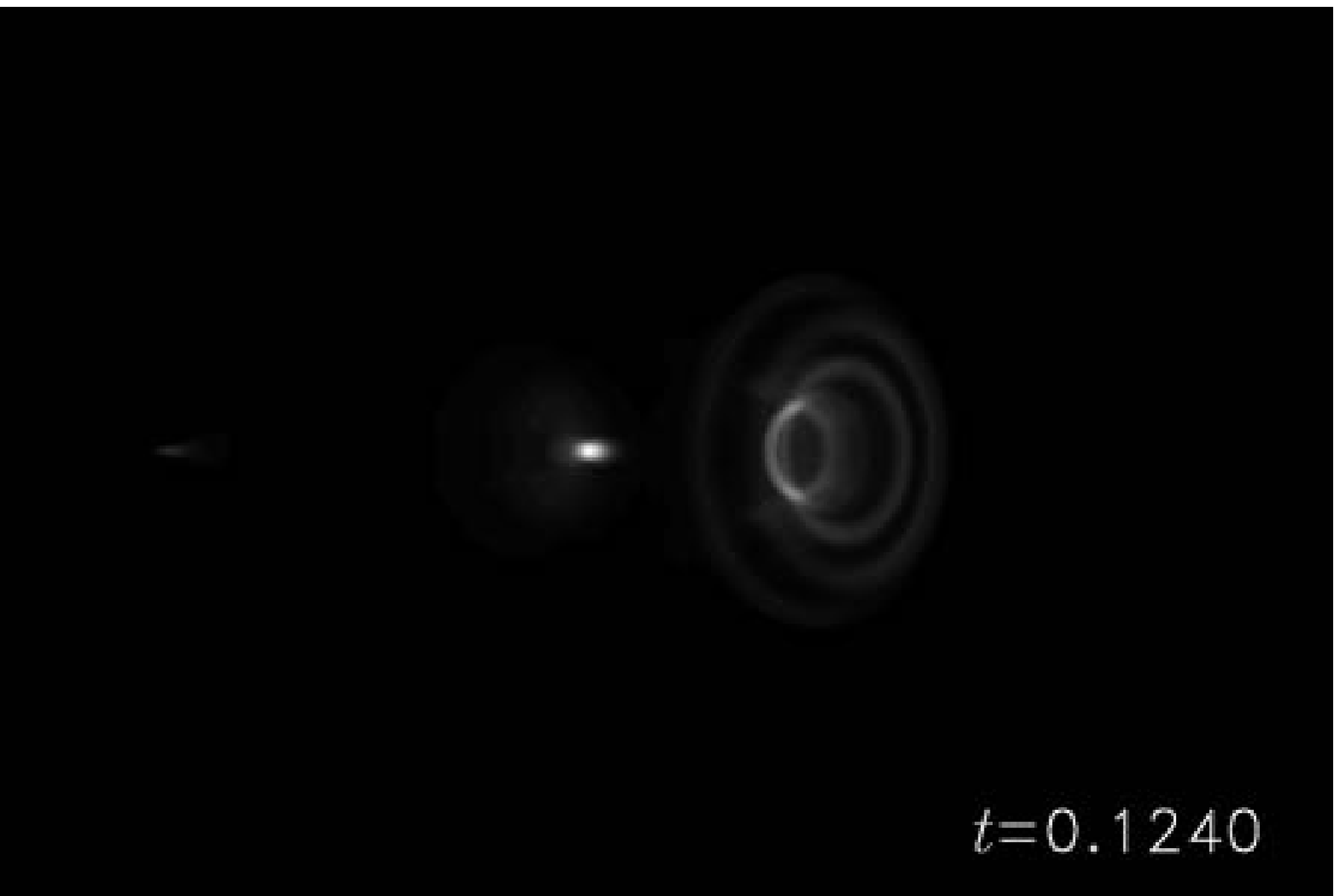}
&\includegraphics[width=4.5cm]{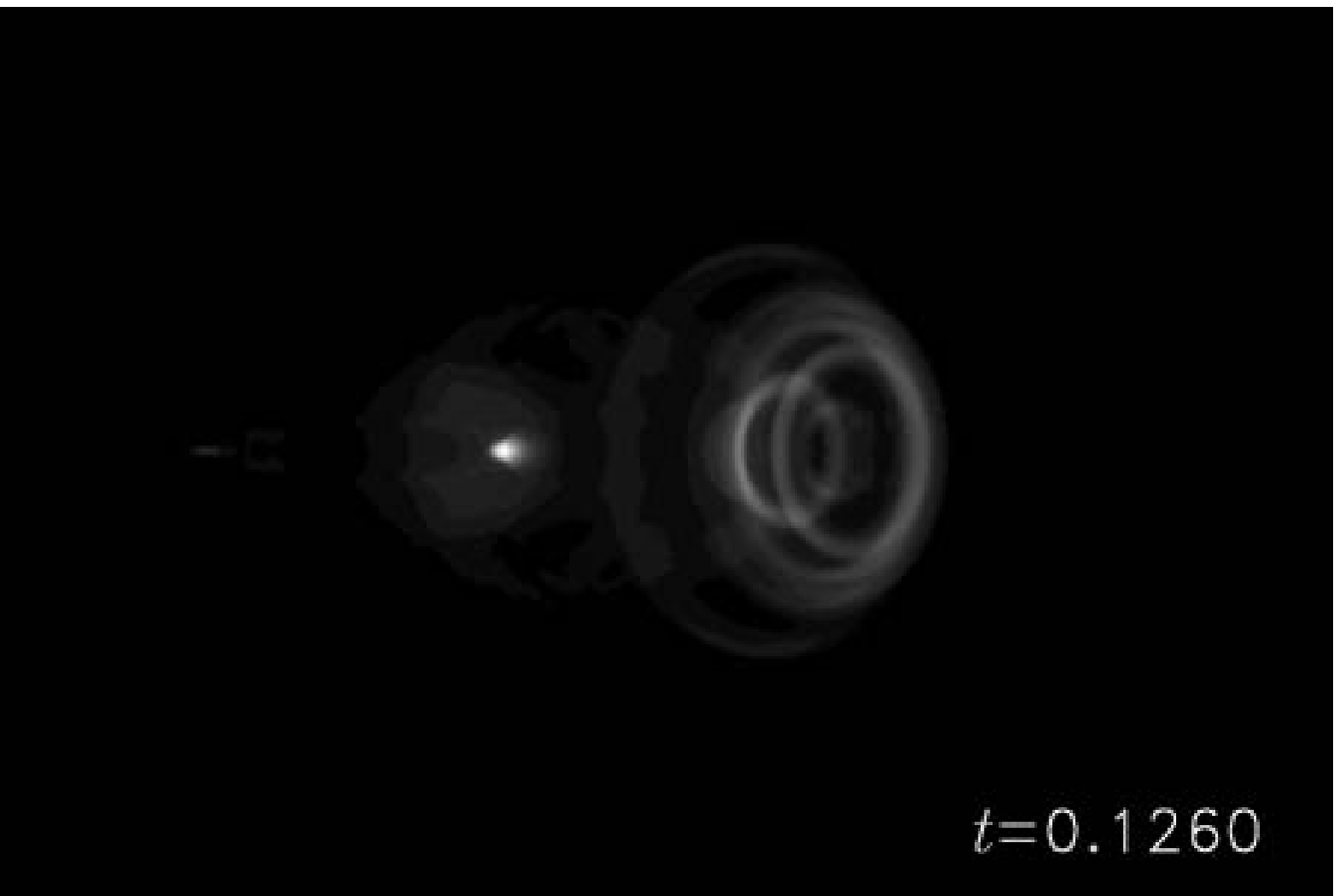}
\\
\includegraphics[width=4.5cm]{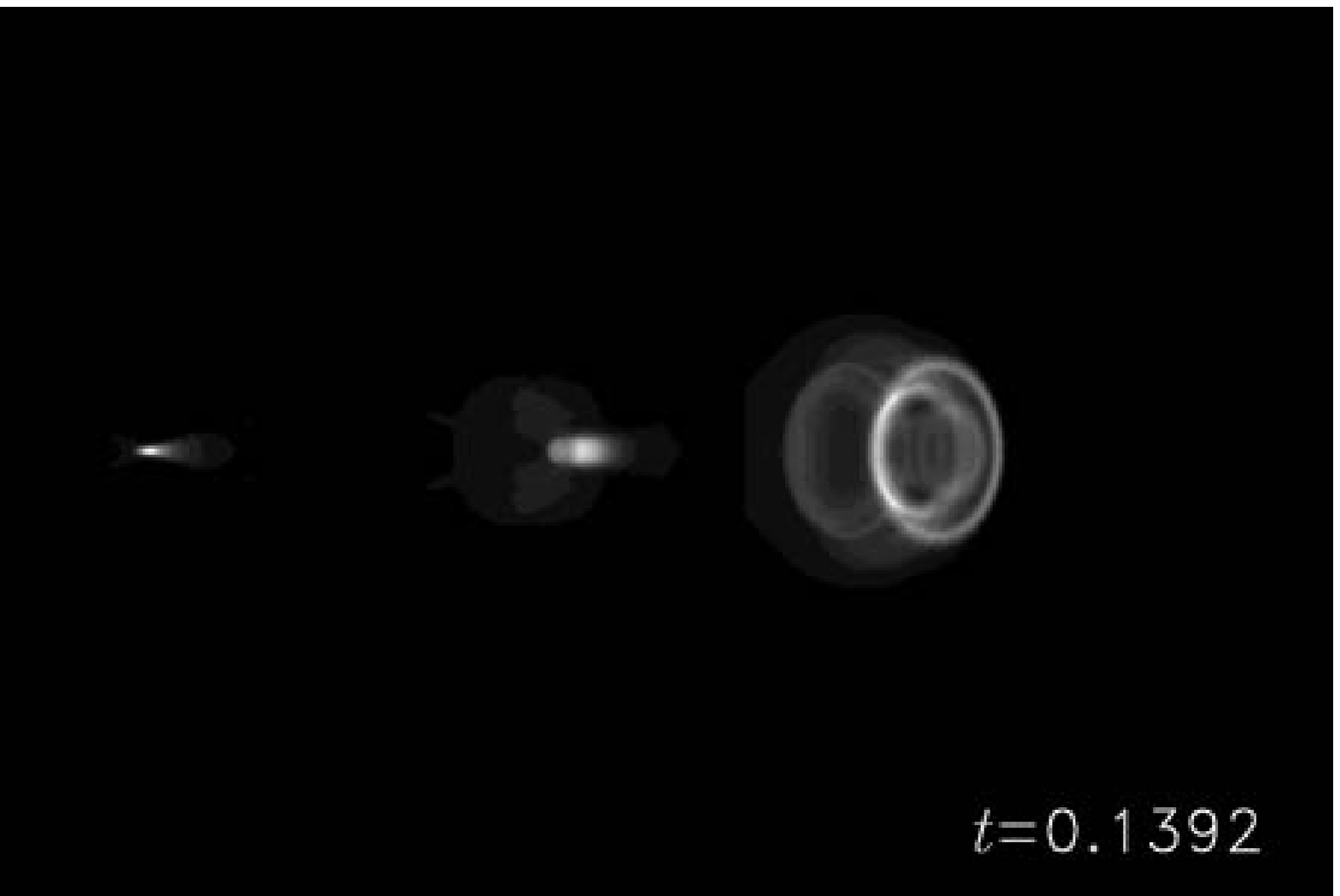}
&\includegraphics[width=4.5cm]{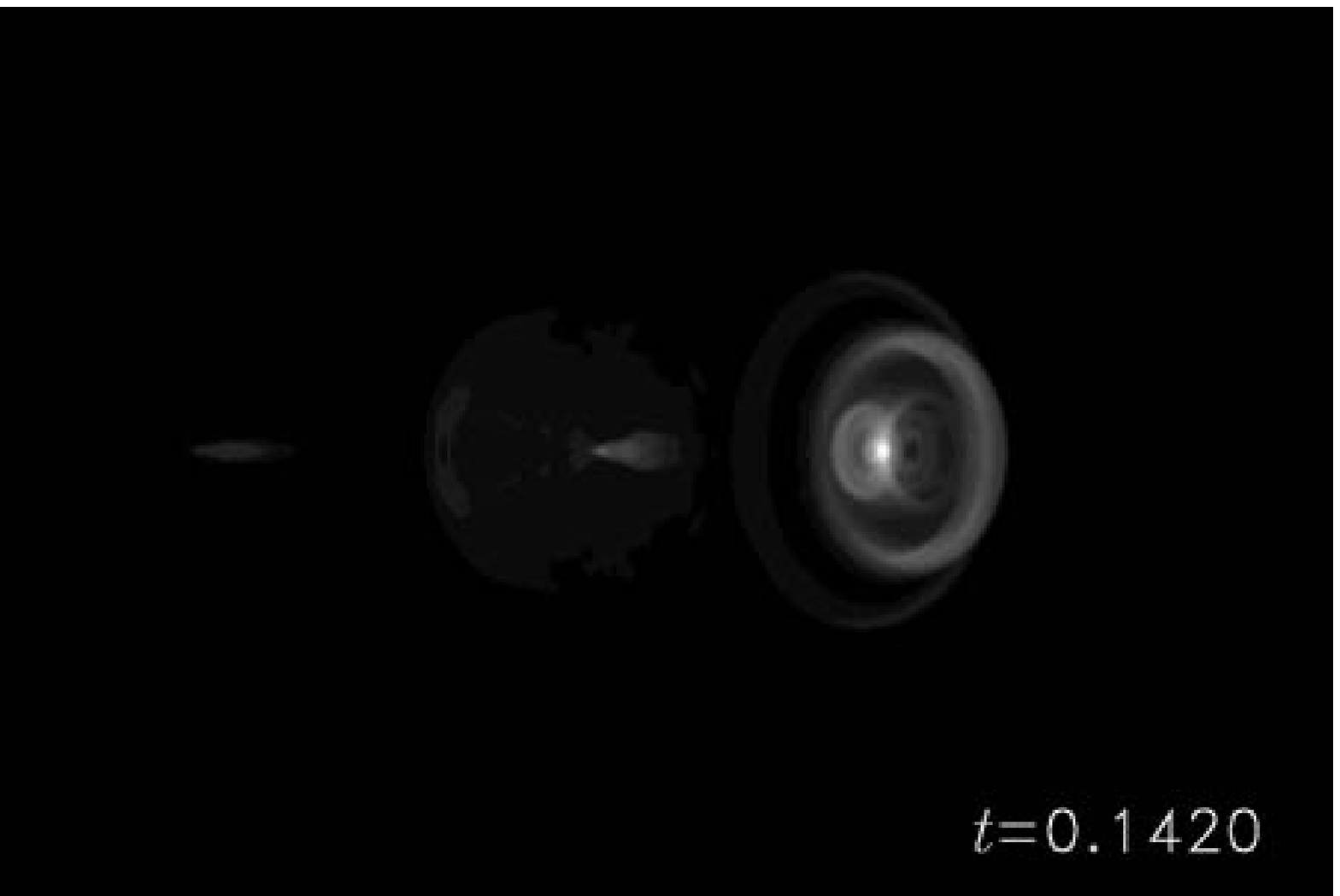}
&\includegraphics[width=4.5cm]{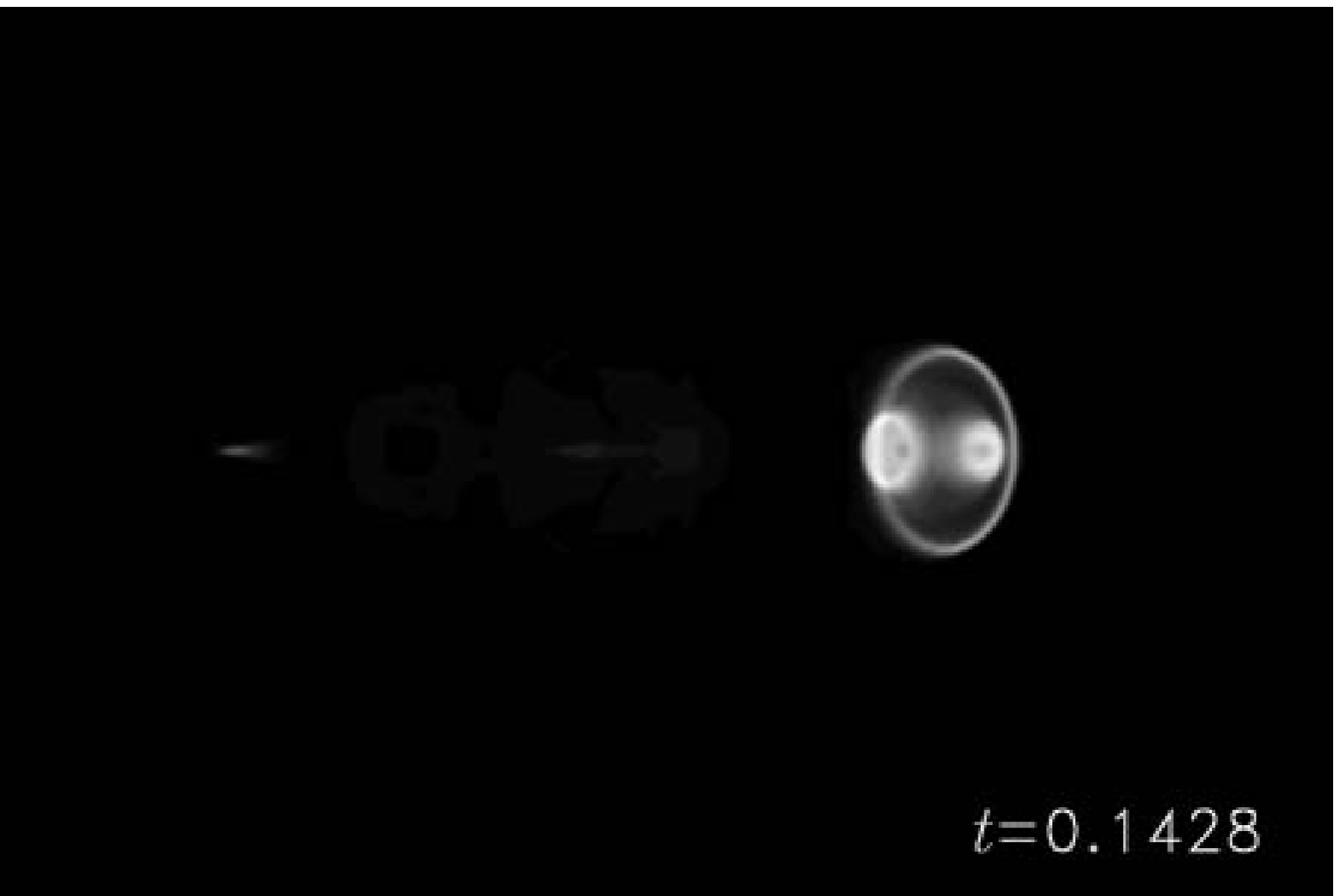}
\\
\end{array}
$
\caption{
Selected images rendered with the jet
at an orientation of $\theta=45^\circ$
to the line of sight,
from the simulation with jet parameters $(\eta,M)=(10^{-4},50)$
and an open left boundary. 
This figure shows the high brightness contrast of the rings and hot-spot 
with respect to the rest of the lobe when the Mach number is high.
}
\label{f:pageant.pxit-4ml}
\end{center}
\end{figure}

\begin{figure}
\begin{center}
$
\begin{array}{ccc}
 \includegraphics[width=4.5cm]{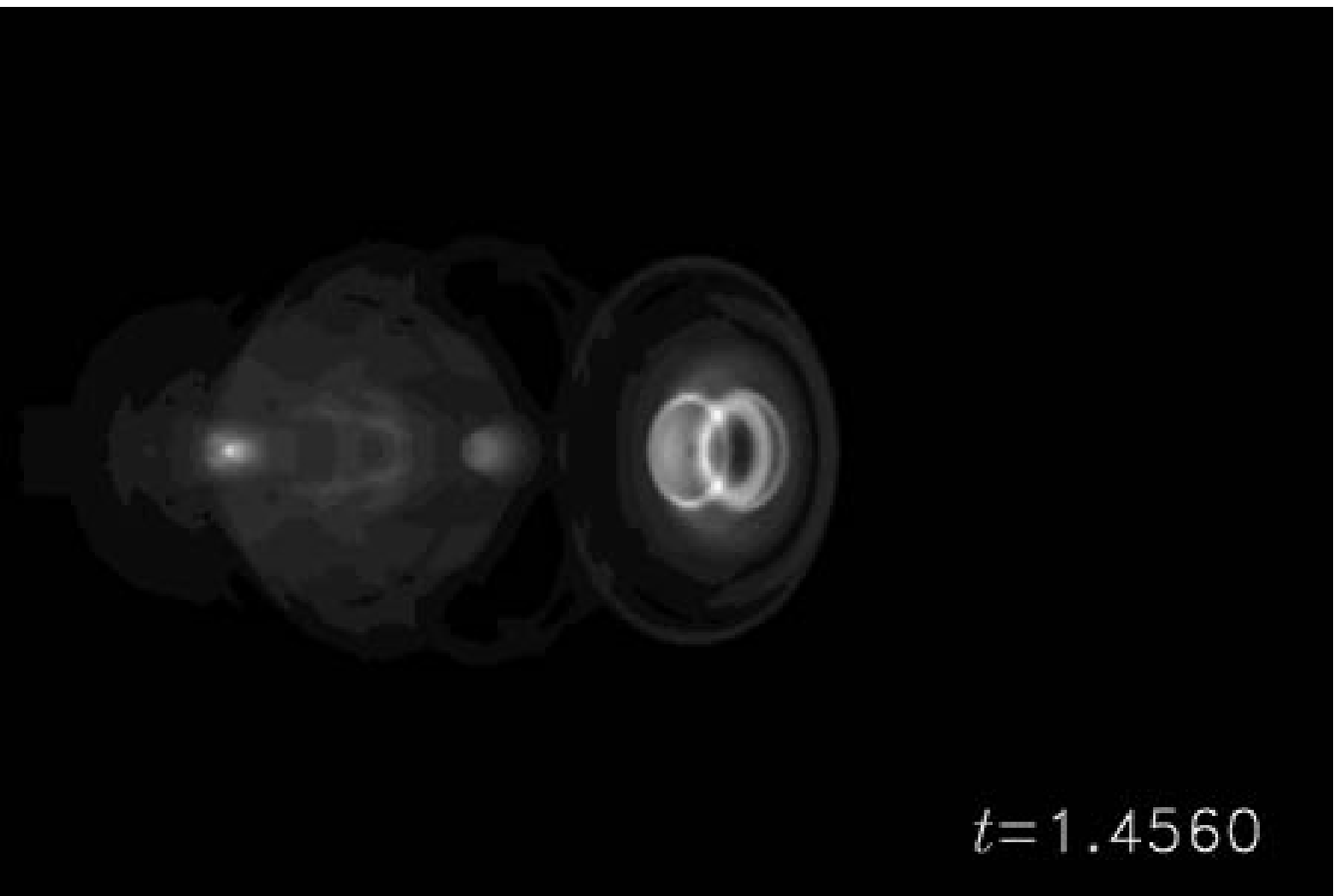}
&\includegraphics[width=4.5cm]{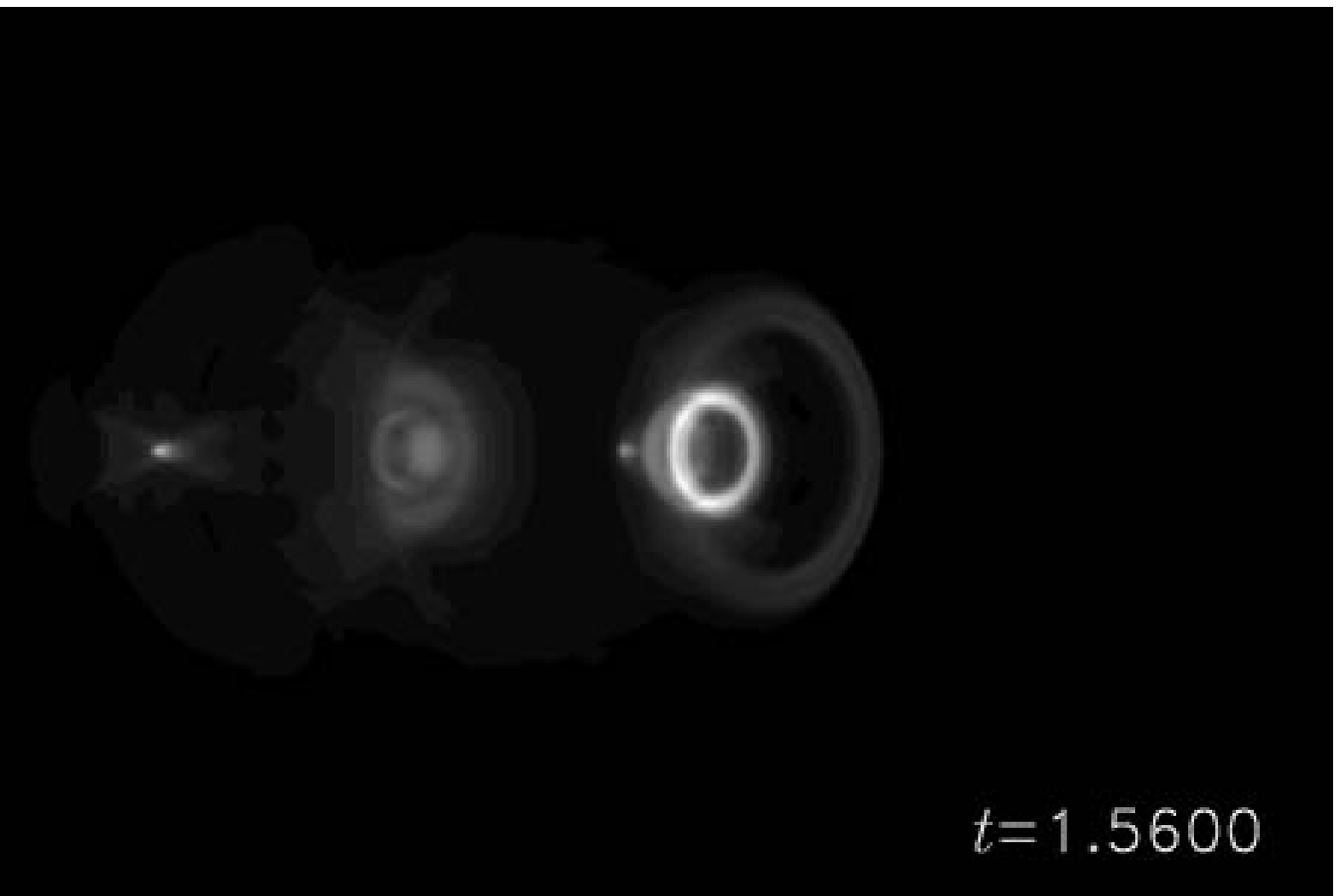}
&\includegraphics[width=4.5cm]{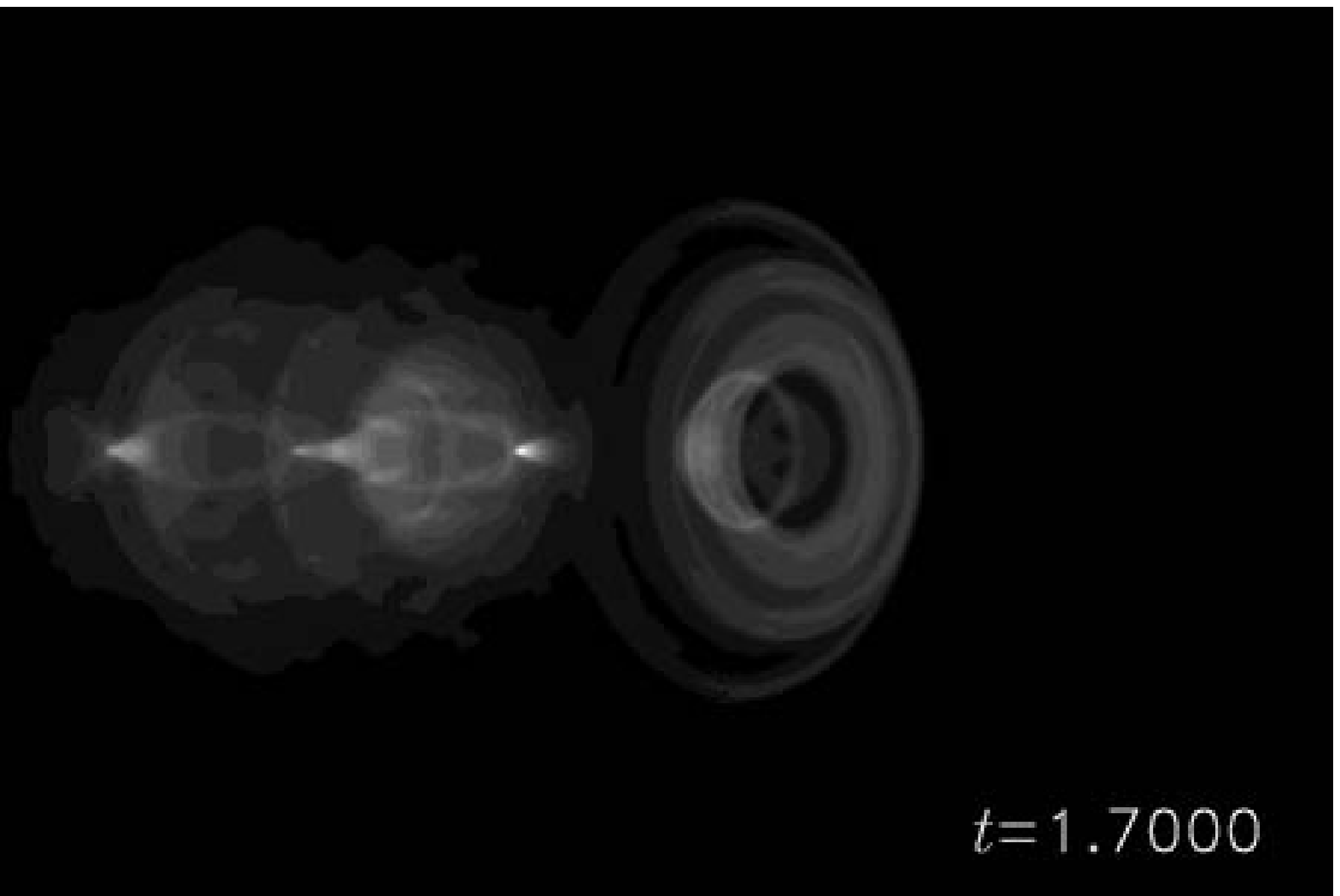}
\\
 \includegraphics[width=4.5cm]{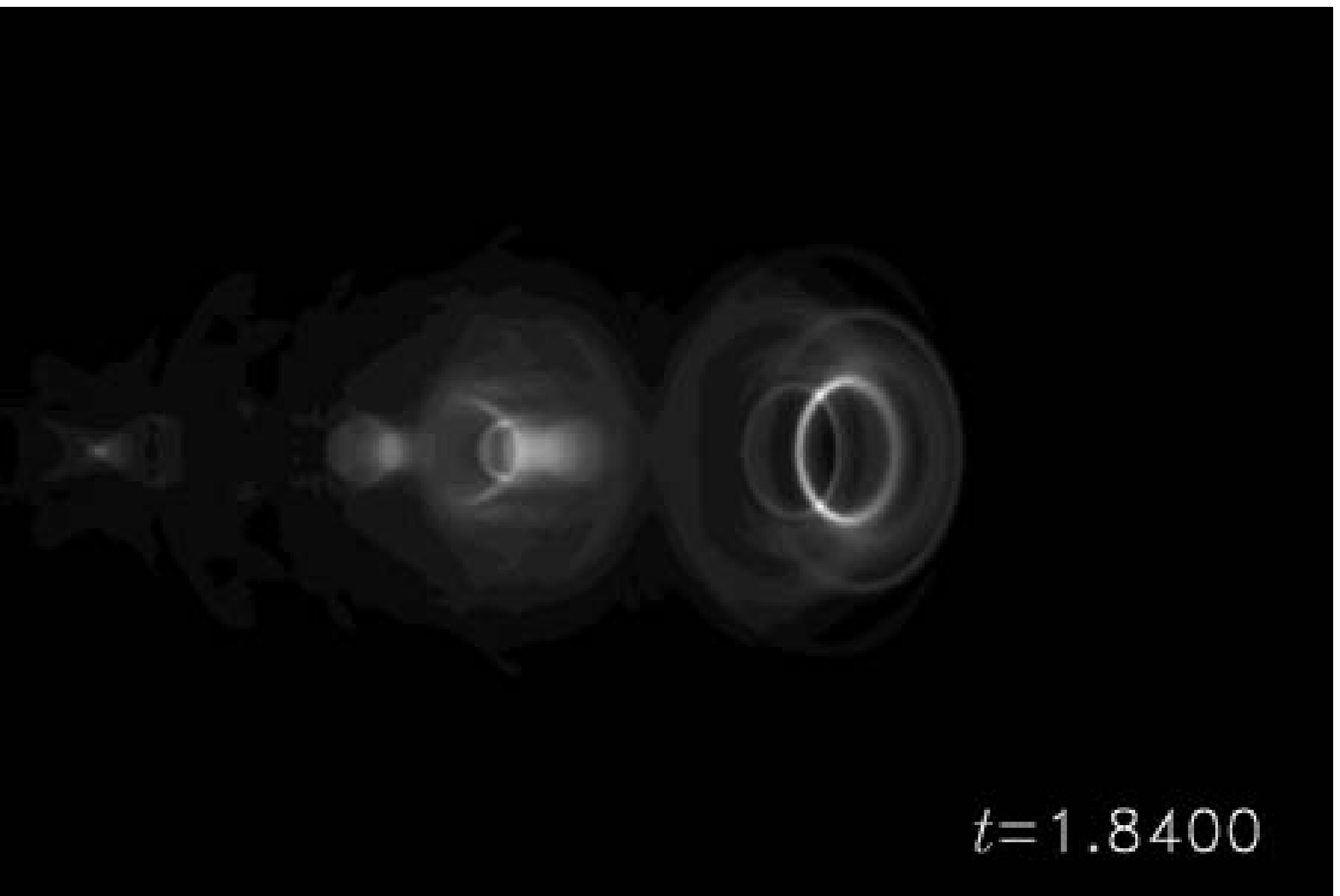}
&\includegraphics[width=4.5cm]{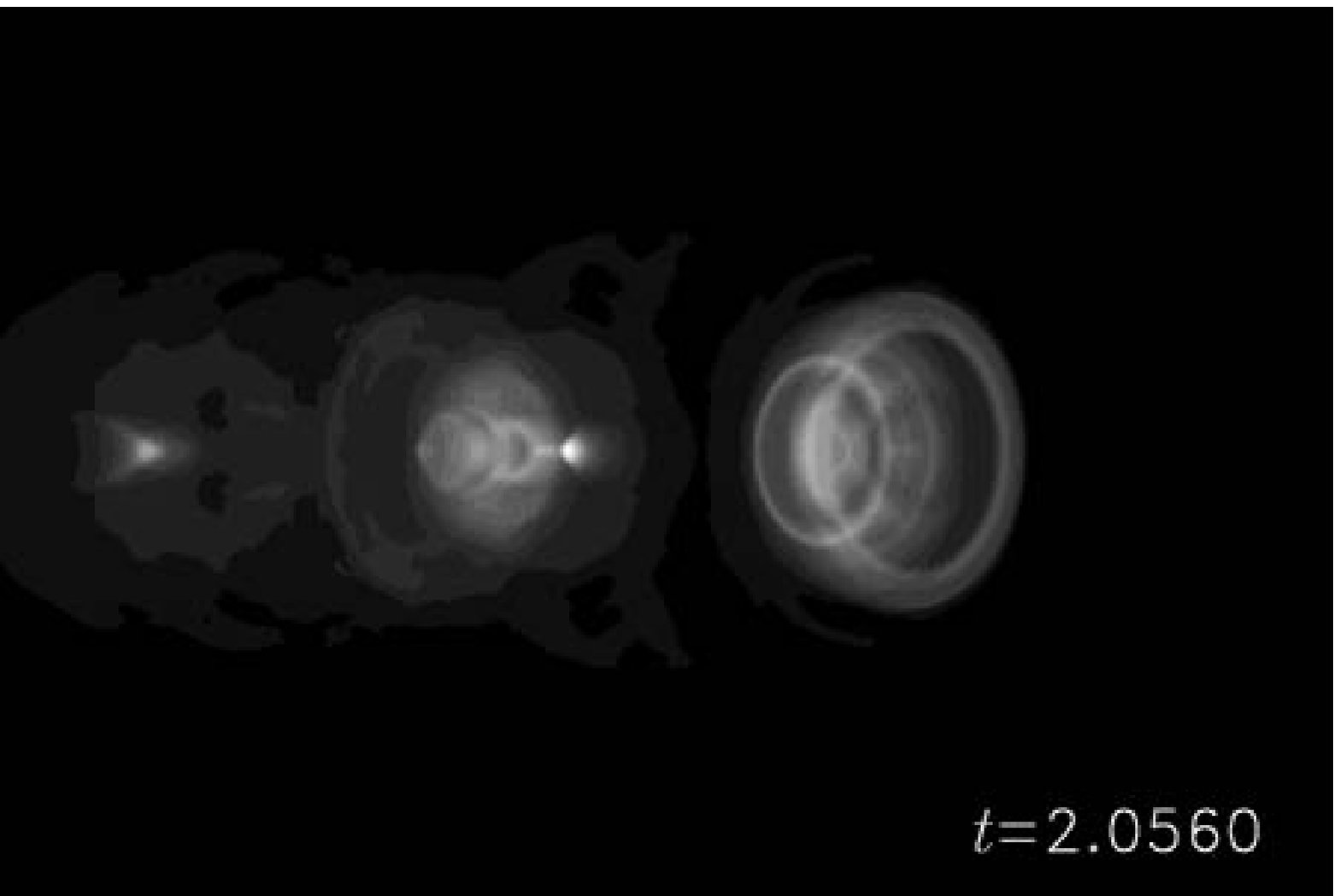}
&\includegraphics[width=4.5cm]{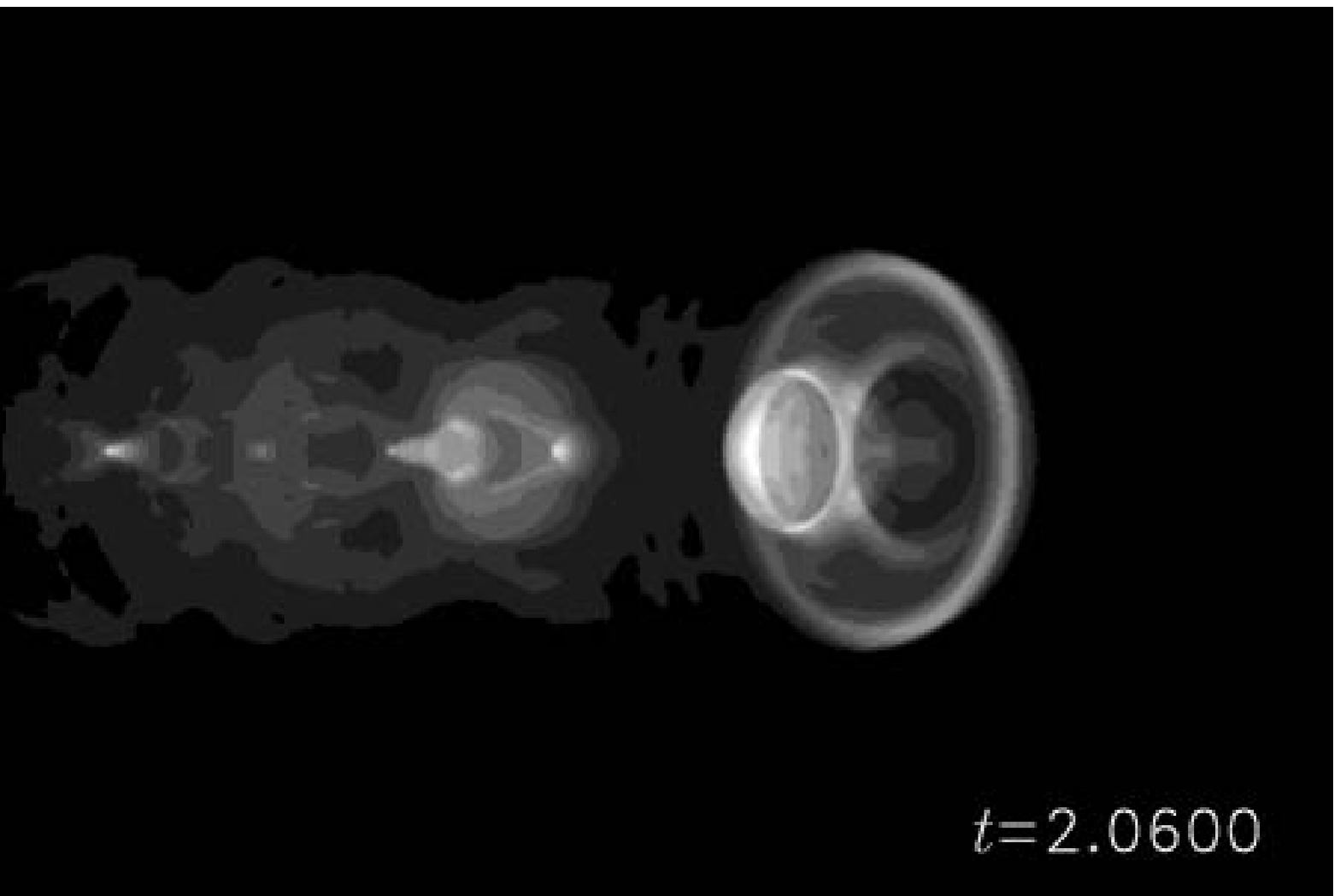}
\\
 \includegraphics[width=4.5cm]{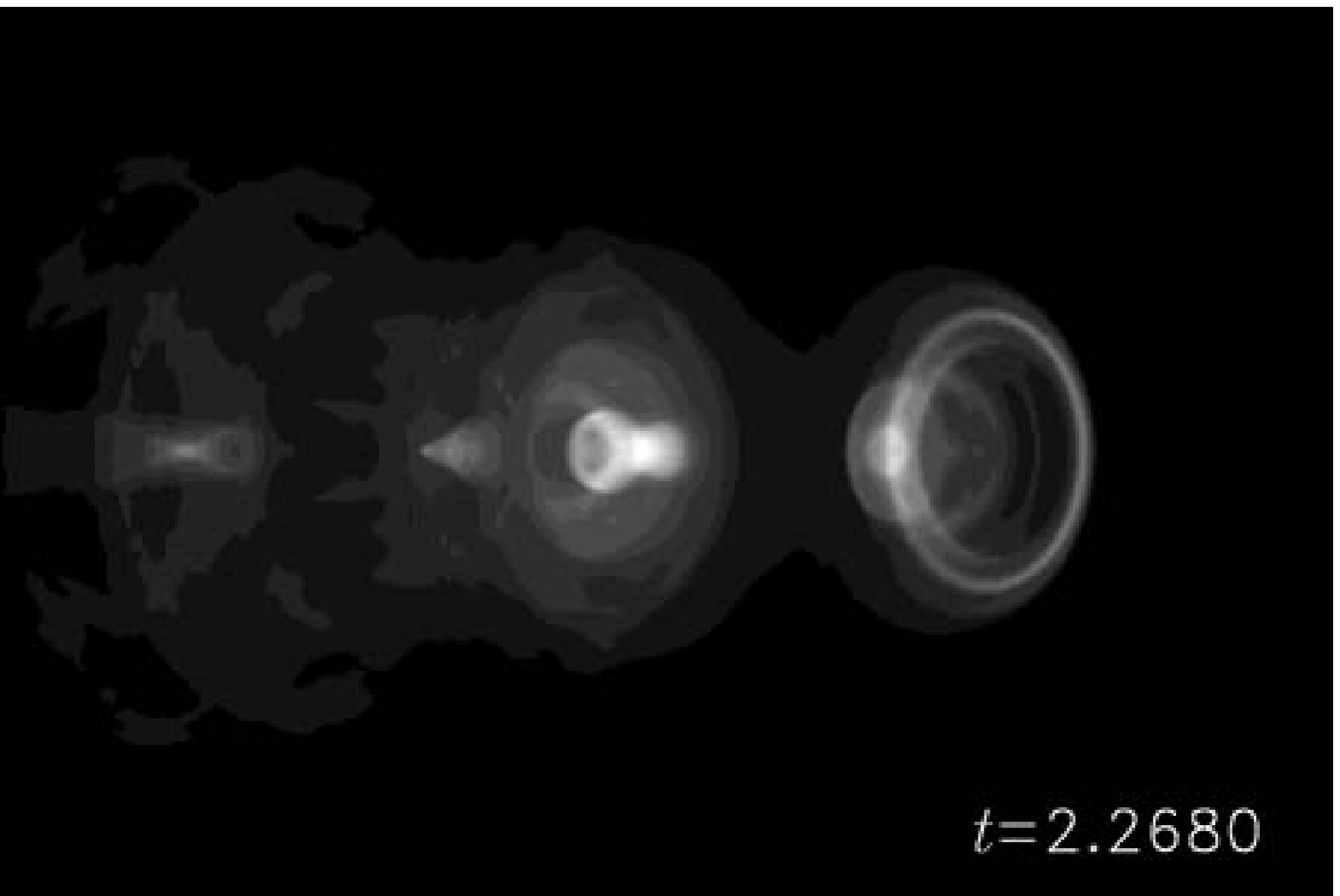}
&\includegraphics[width=4.5cm]{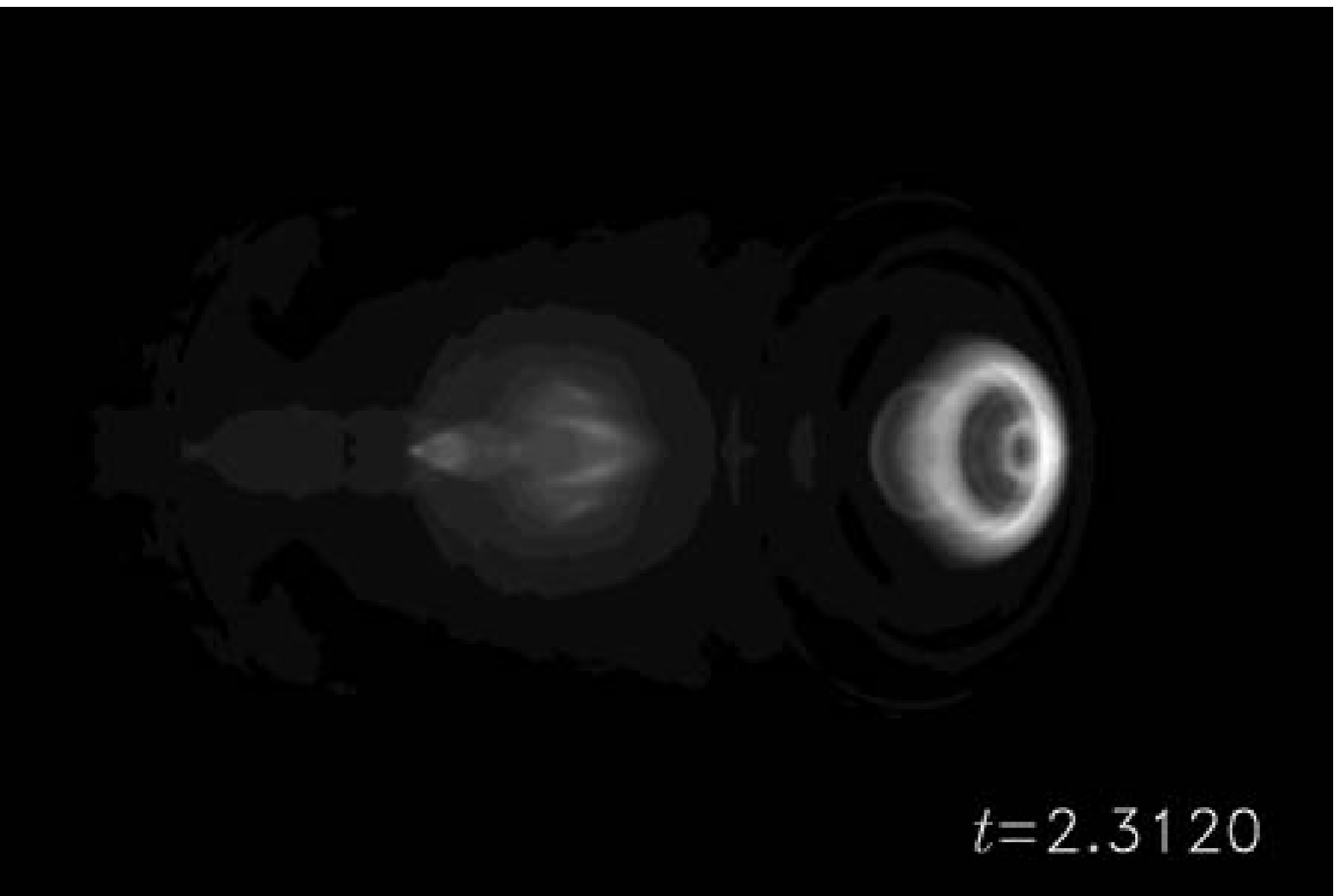}
&\includegraphics[width=4.5cm]{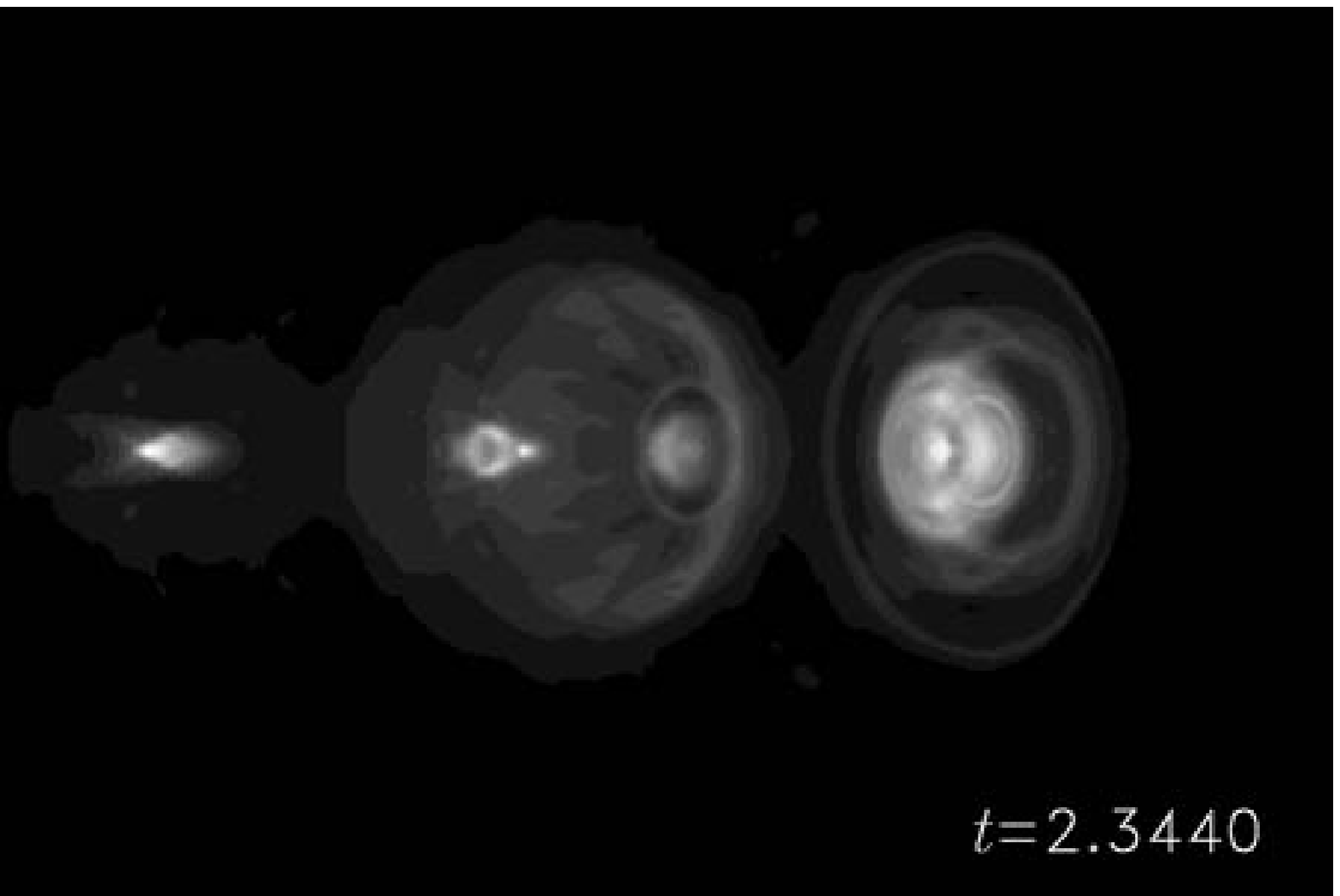}
\\
&\\
\end{array}
$
\caption{
Selected images rendered with the jet
at an orientation of $\theta=45^\circ$
to the line of sight,
from the simulation with jet parameters $(\eta,M)=(10^{-4},5)$
and an open left boundary. 
The contrast of the rings and hot-spots against the radio lobe
is less marked than for $(\eta,M)=(10^{-4},50)$.
}
\label{f:pageant.pxit-4m5}
\end{center}
\end{figure}

\begin{figure}
\begin{center}
$
\begin{array}{ccc}
 \includegraphics[width=4.5cm]{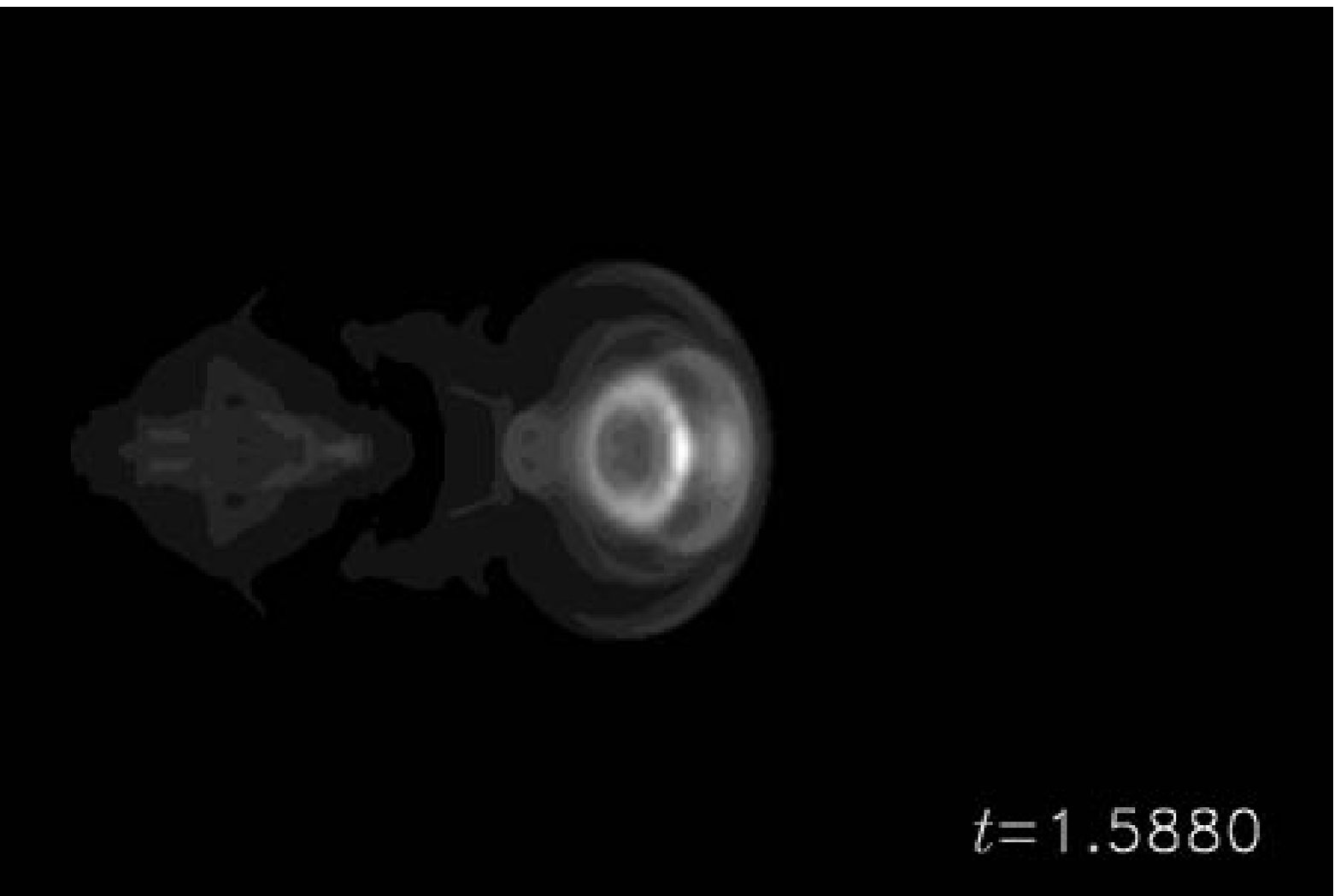}
&\includegraphics[width=4.5cm]{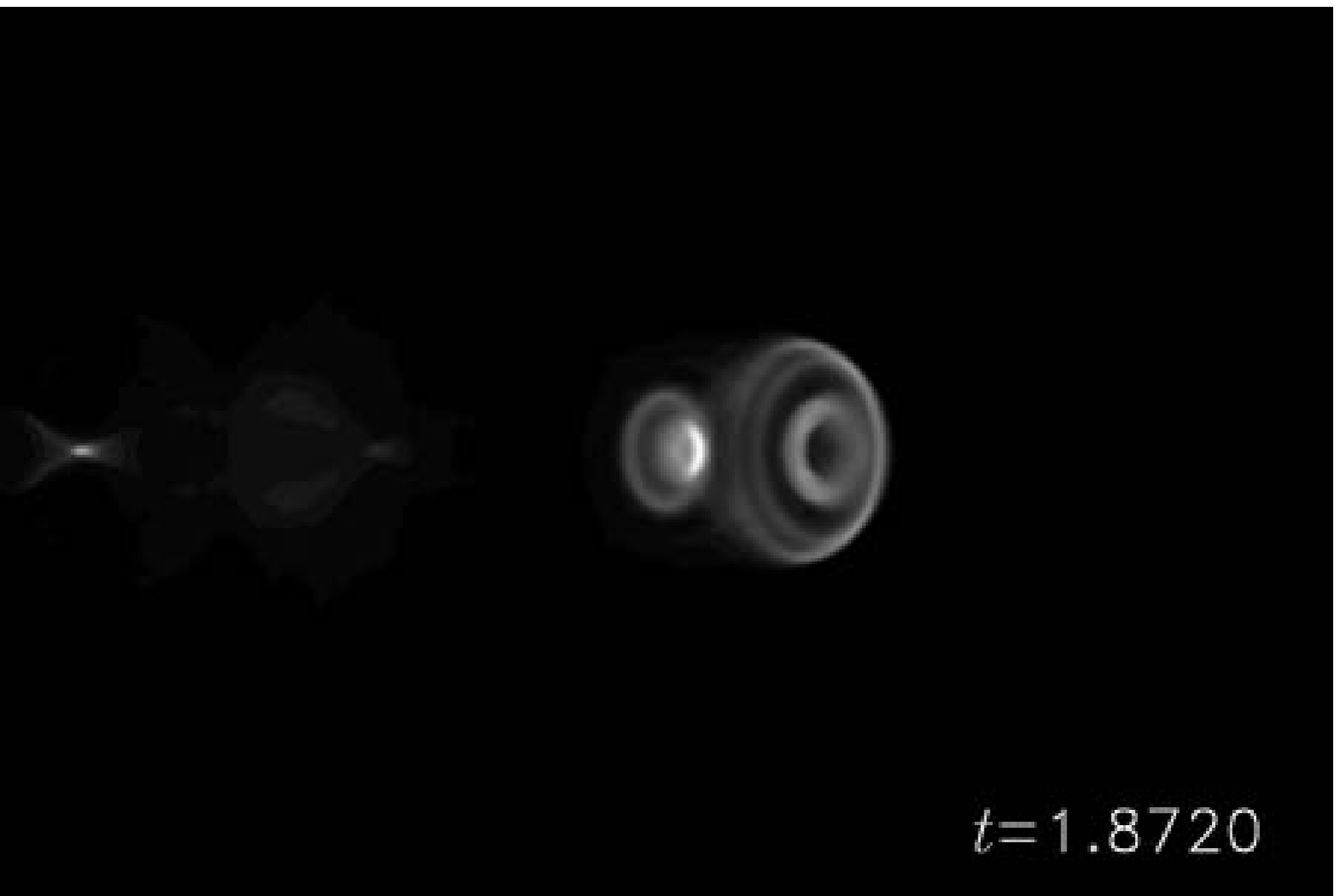}
&\includegraphics[width=4.5cm]{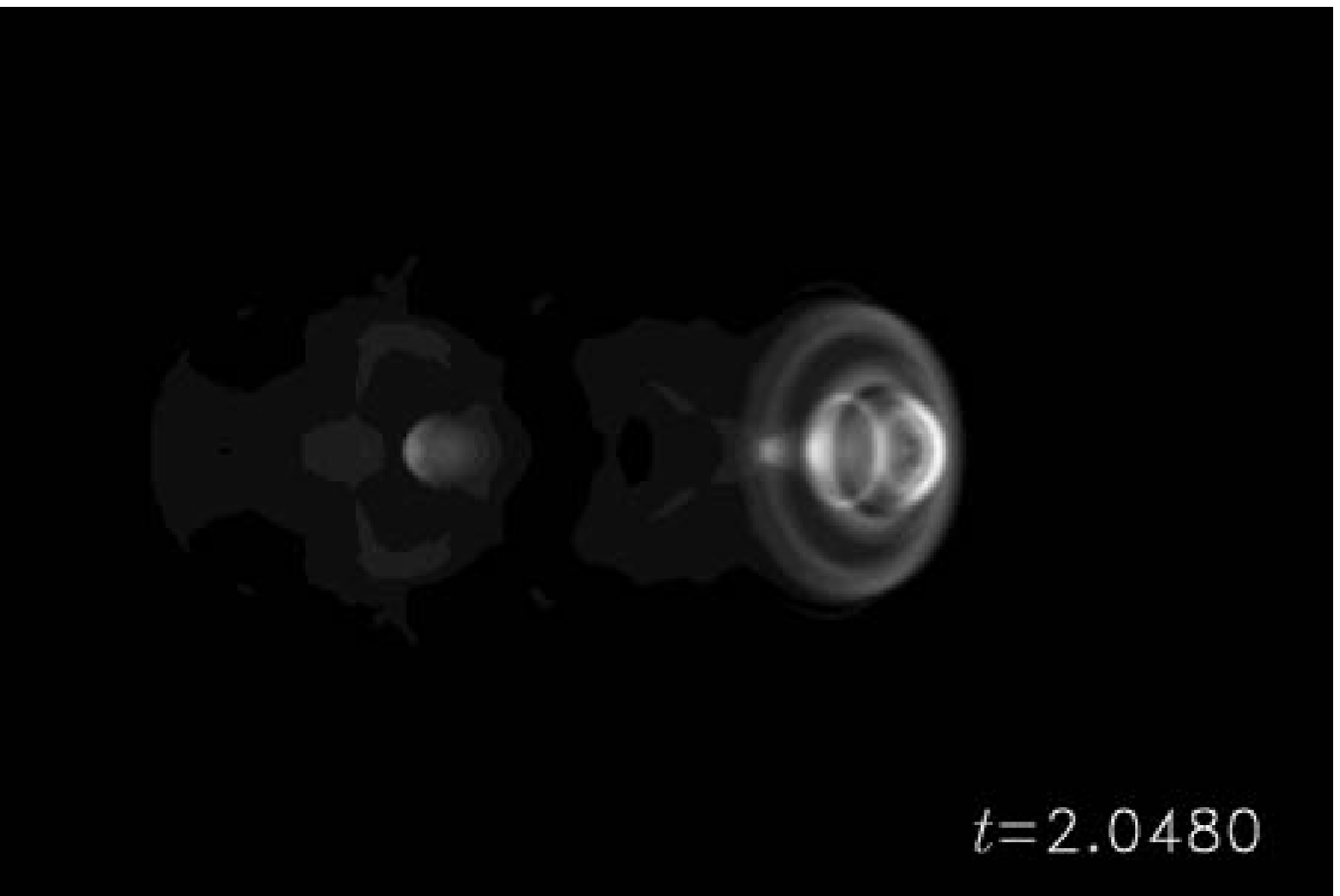}
\\
 \includegraphics[width=4.5cm]{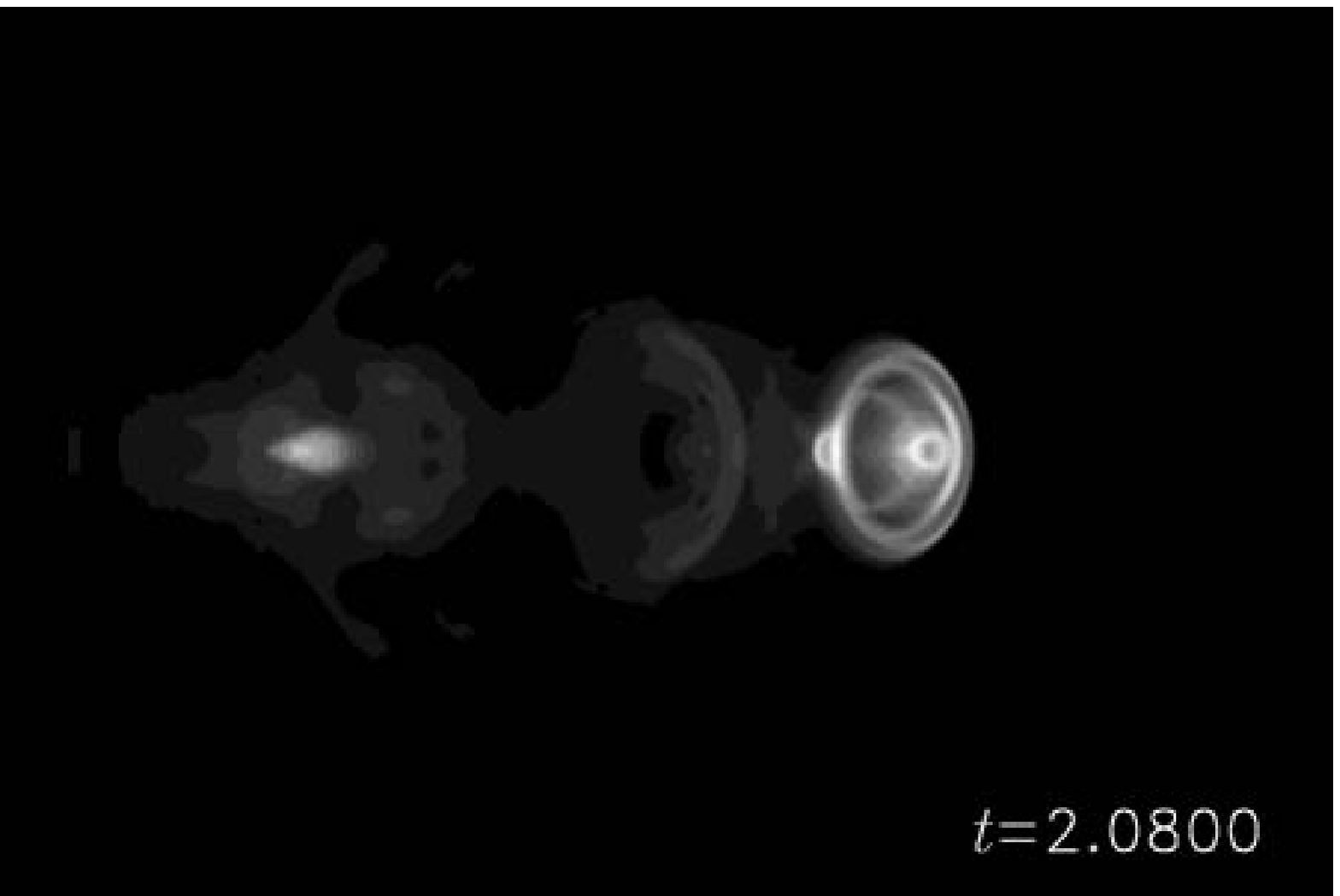}
&\includegraphics[width=4.5cm]{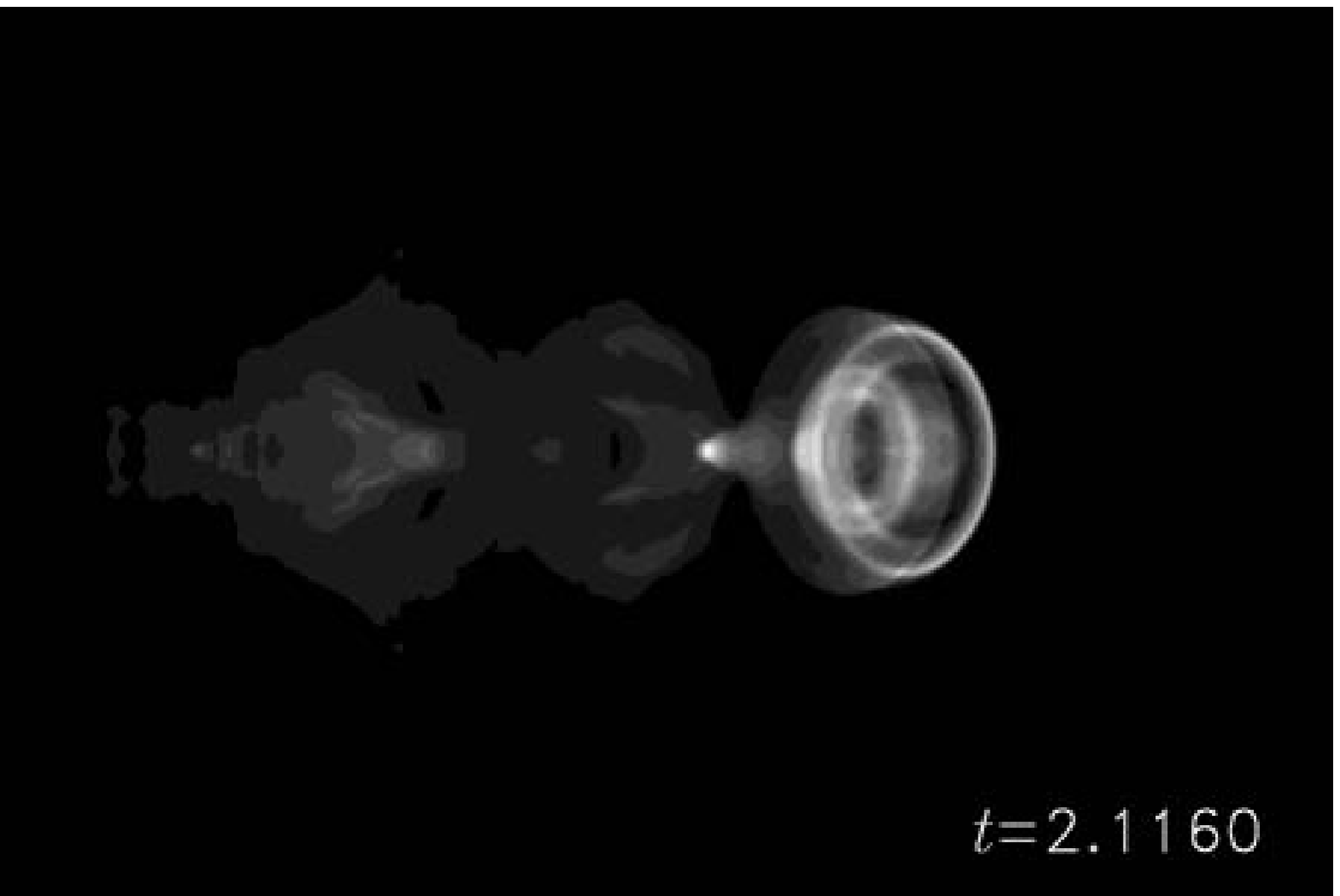}
&\includegraphics[width=4.5cm]{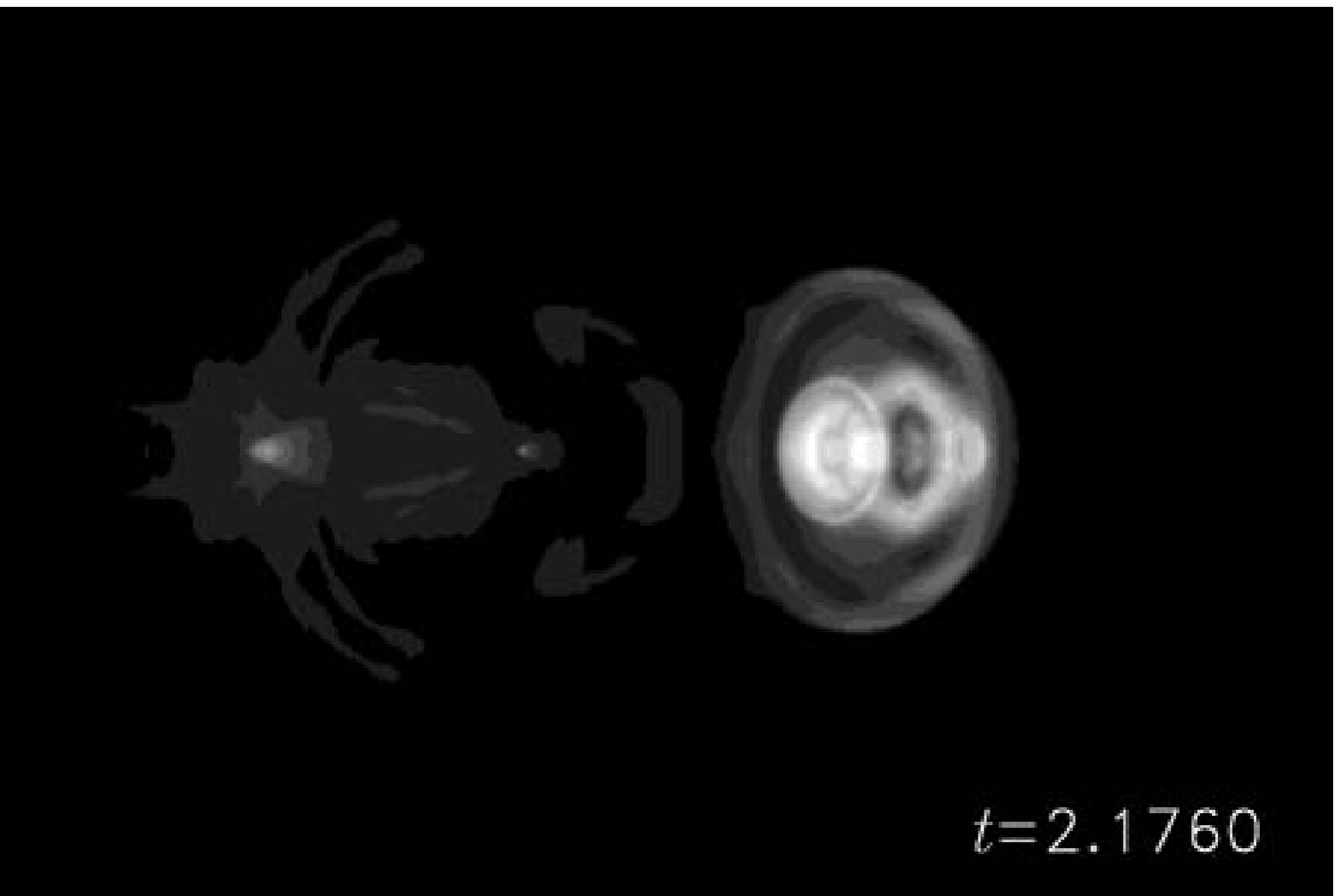}
\\
 \includegraphics[width=4.5cm]{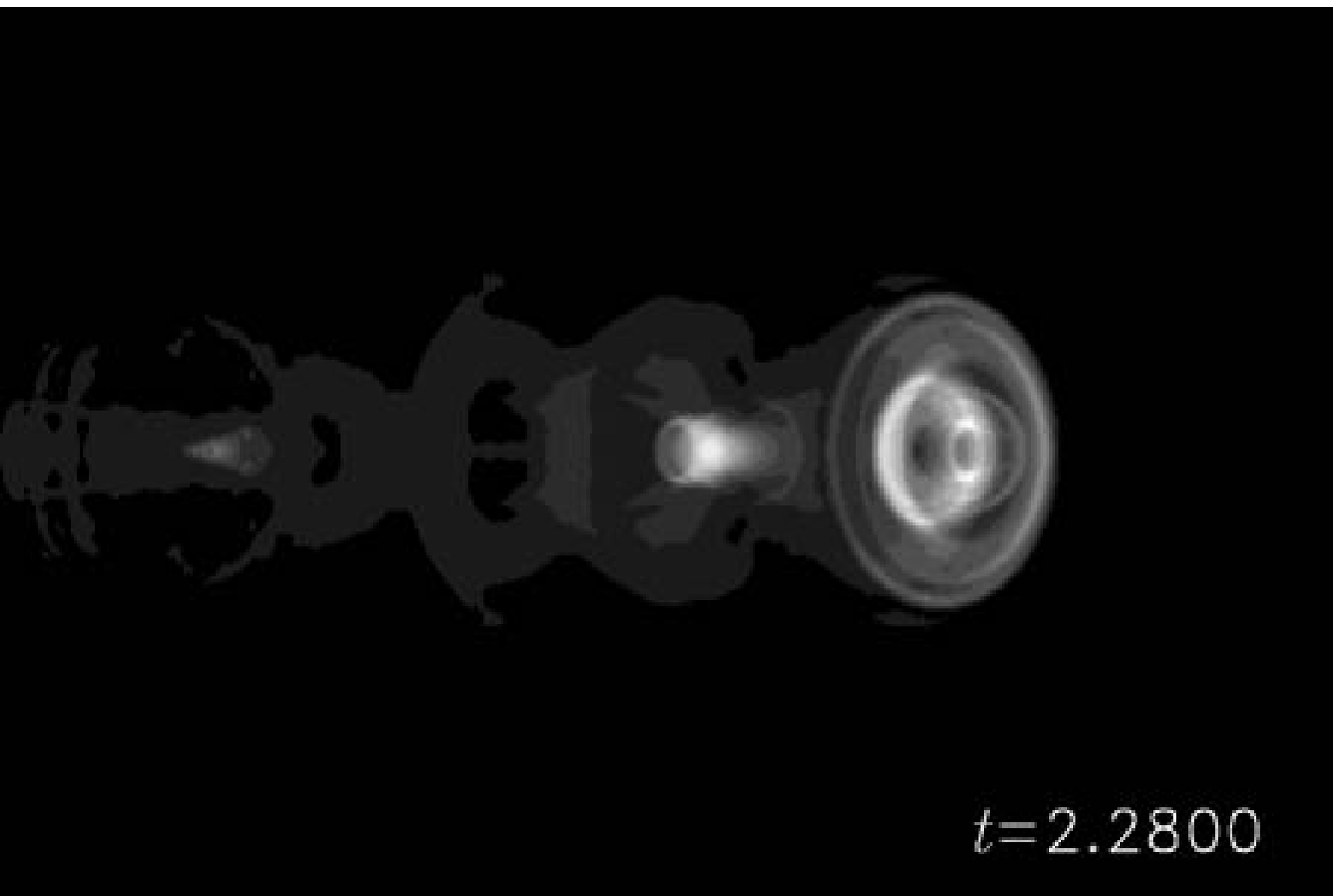}
&\includegraphics[width=4.5cm]{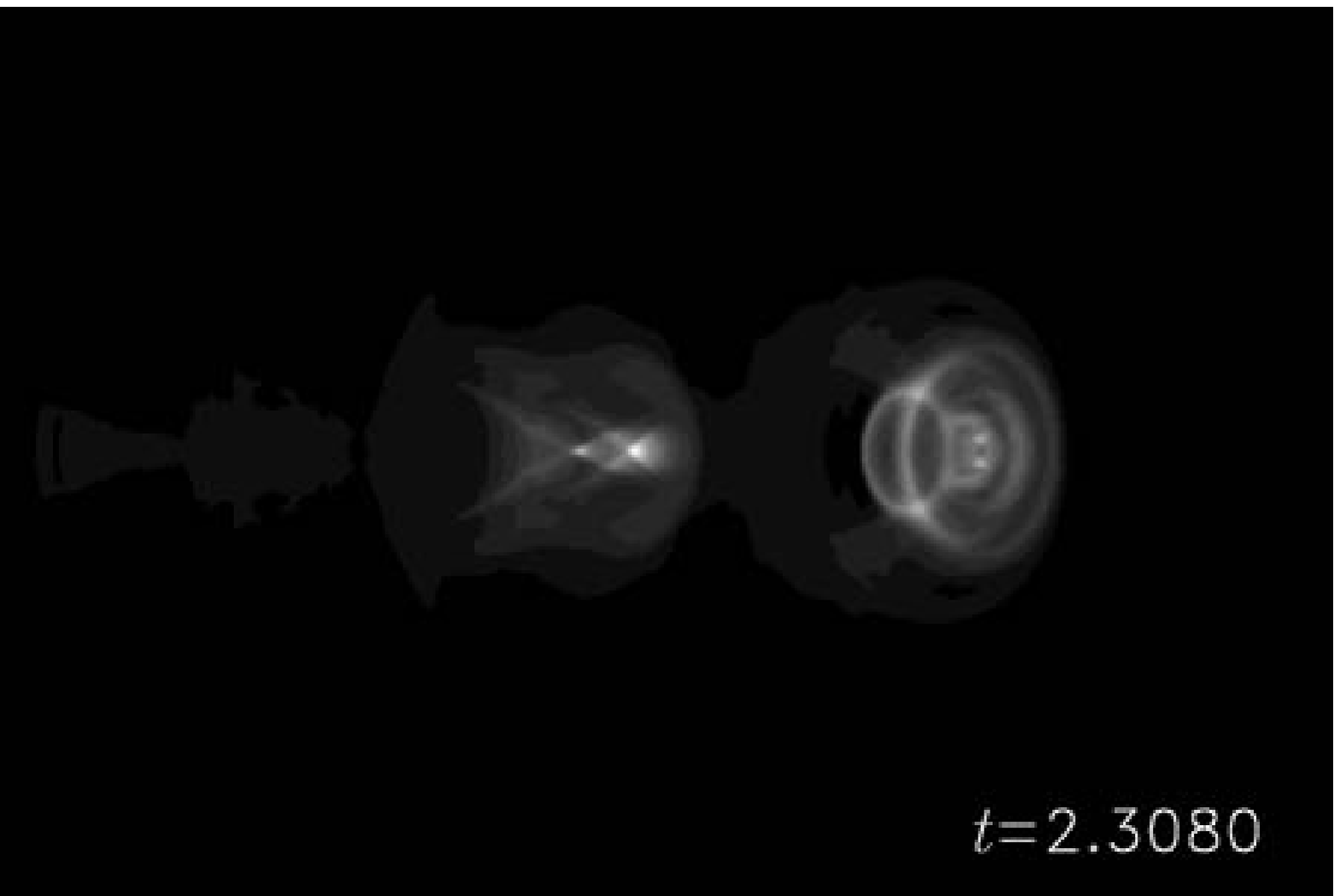}
&\includegraphics[width=4.5cm]{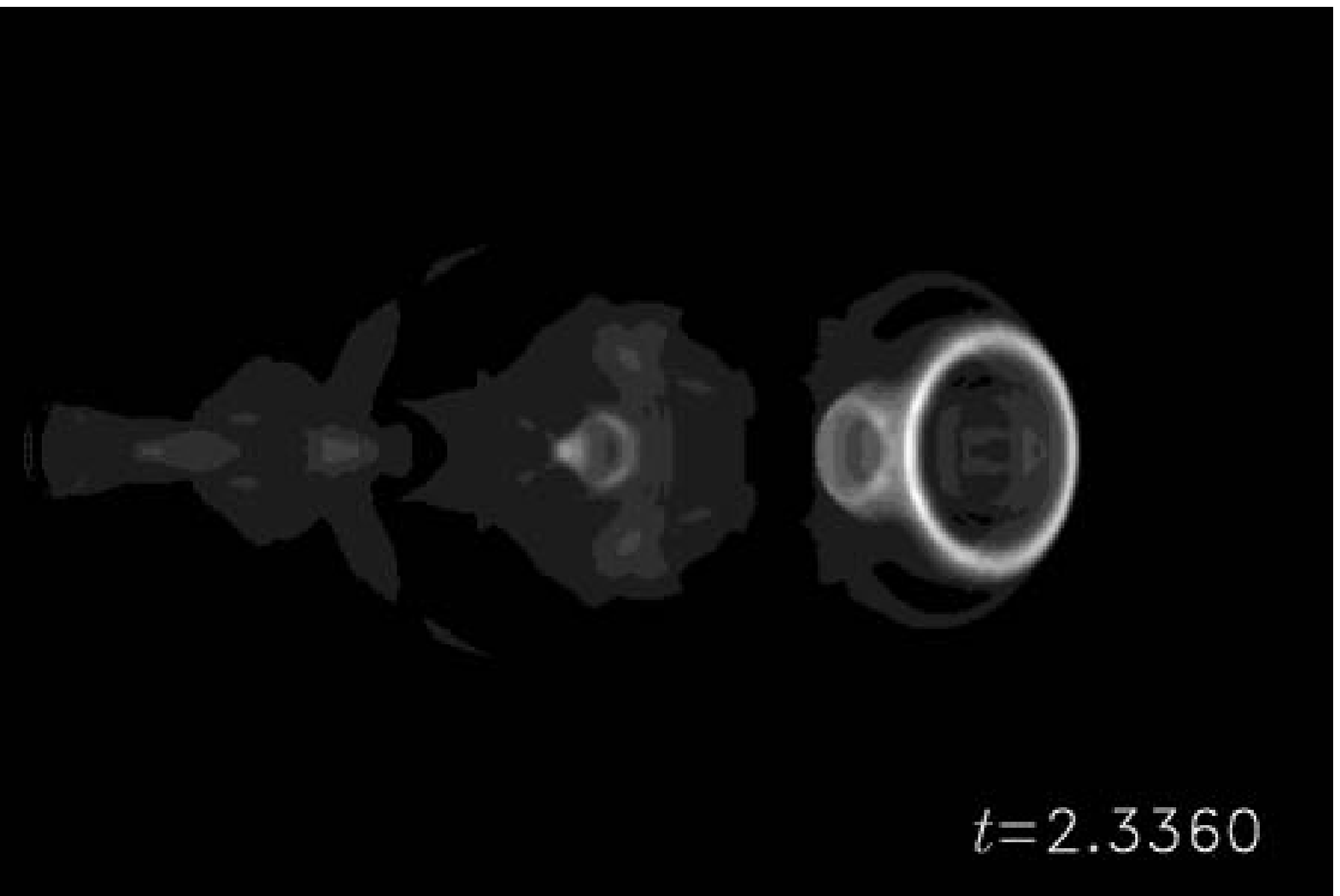}
\end{array}
$
\caption{
Selected images rendered with the jet
at an orientation of $\theta=45^\circ$
to the line of sight,
from the simulation with jet parameters $(\eta,M)=(10^{-2},5)$
and an open left boundary.  
Compared to $(\eta,M)= (10^{-4},5)$, 
this sequence shows a greater concentration of bright features 
towards the head of the lobe. 
There are few instances of equally bright well-separated rings.  
}
\label{f:pageant.pxit-2m5}
\end{center}
\end{figure}

\begin{figure}
\begin{center}
$
\begin{array}{ccc}
 \includegraphics[width=4.5cm]{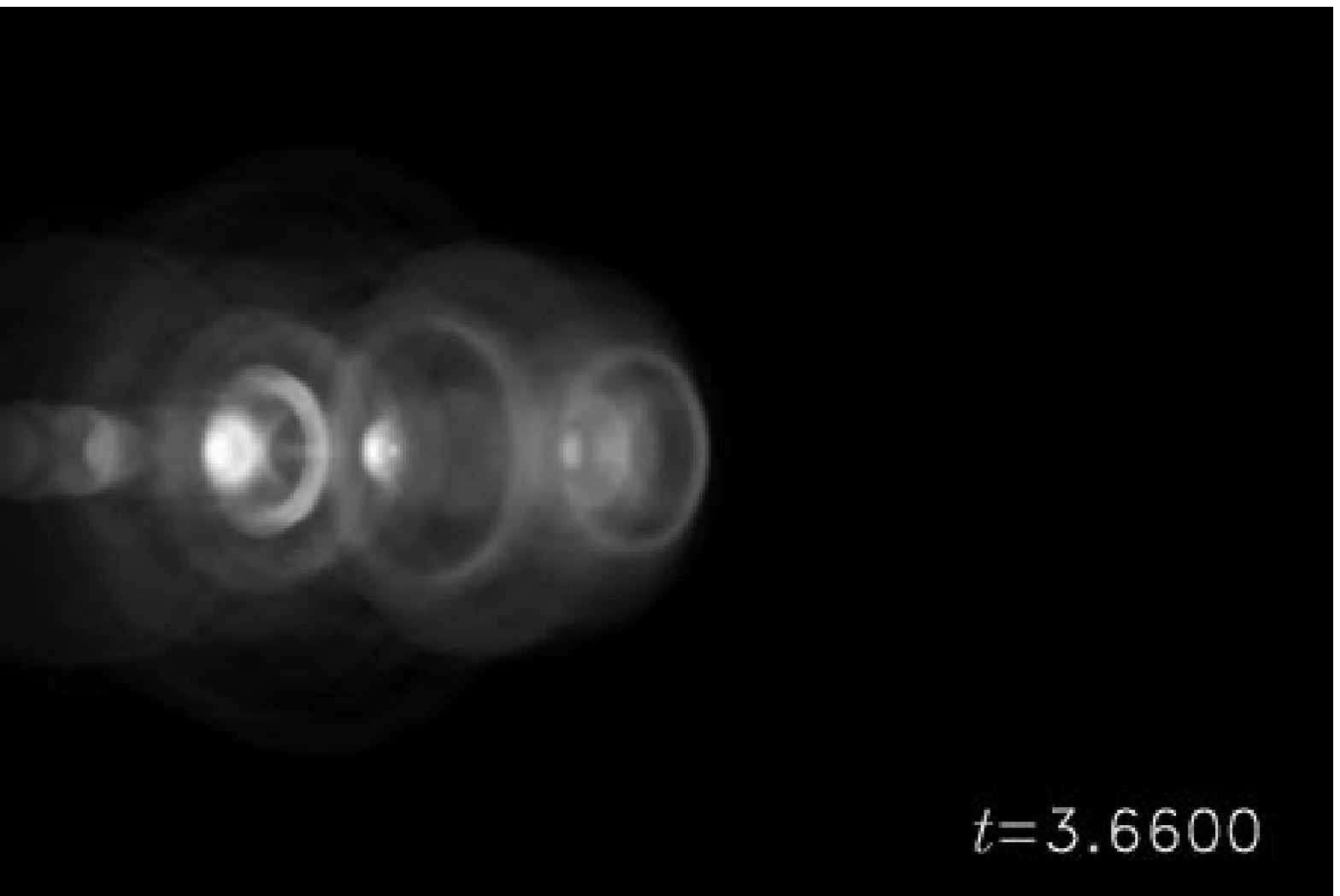}
&\includegraphics[width=4.5cm]{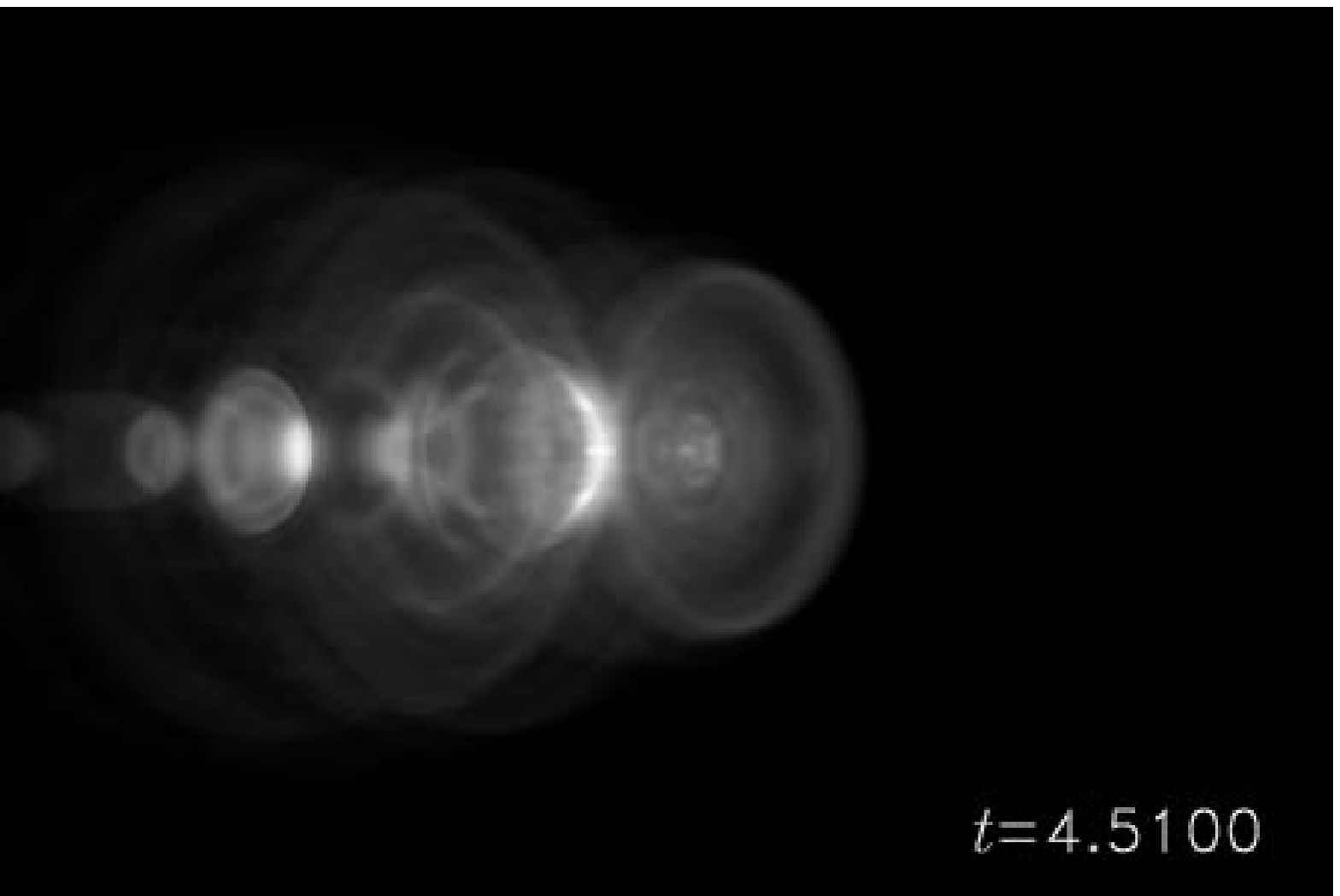}
&\includegraphics[width=4.5cm]{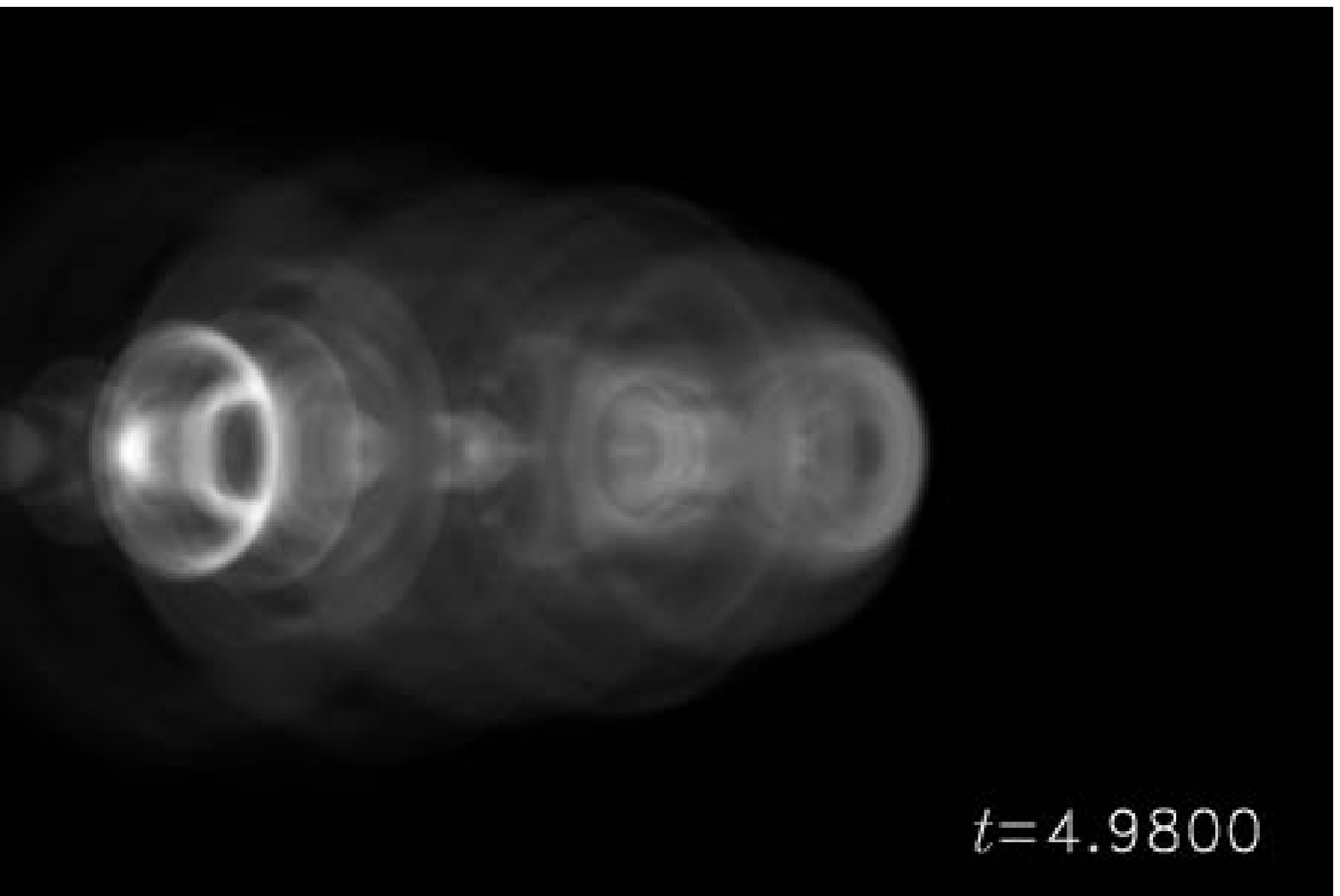}
\\
 \includegraphics[width=4.5cm]{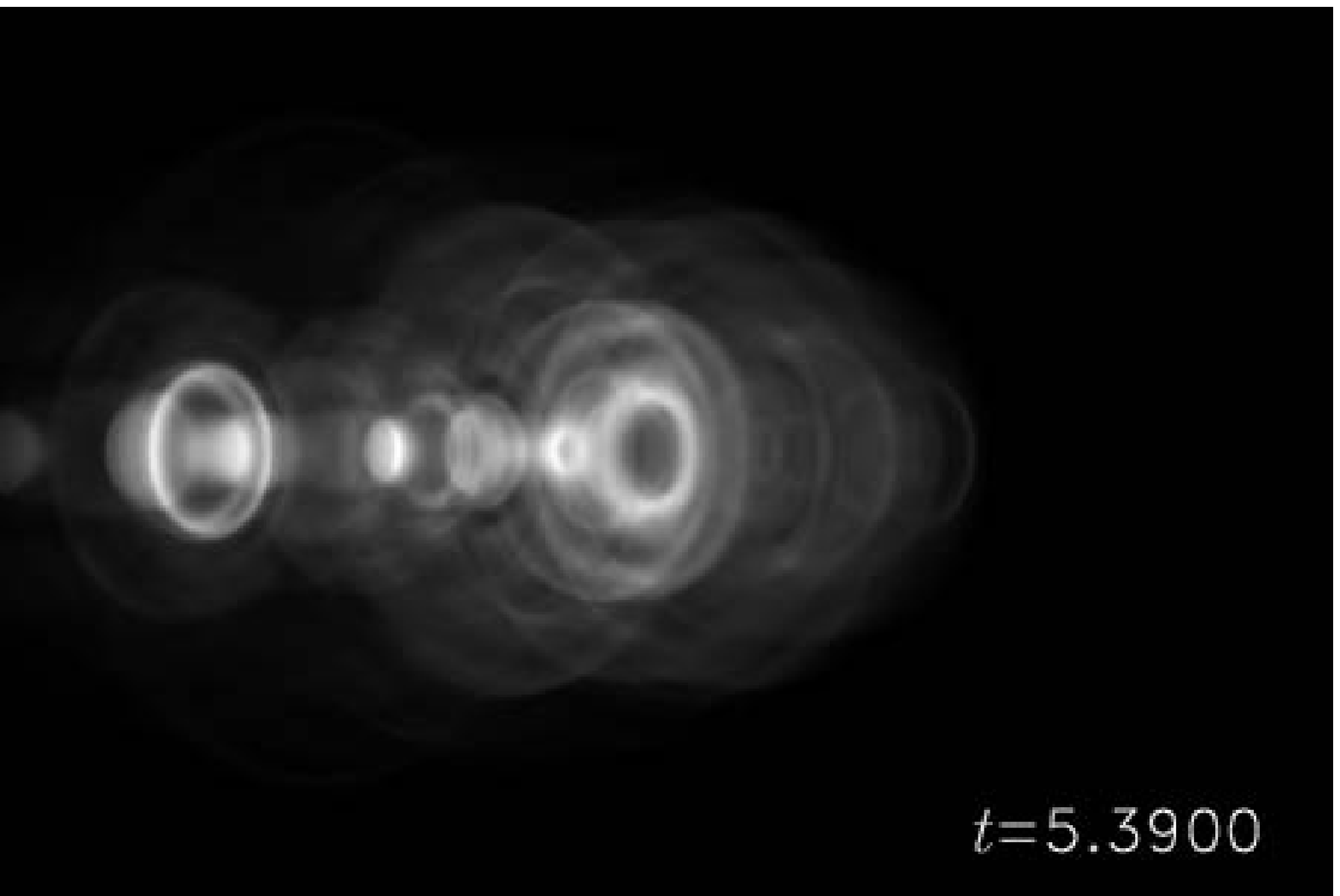}
&\includegraphics[width=4.5cm]{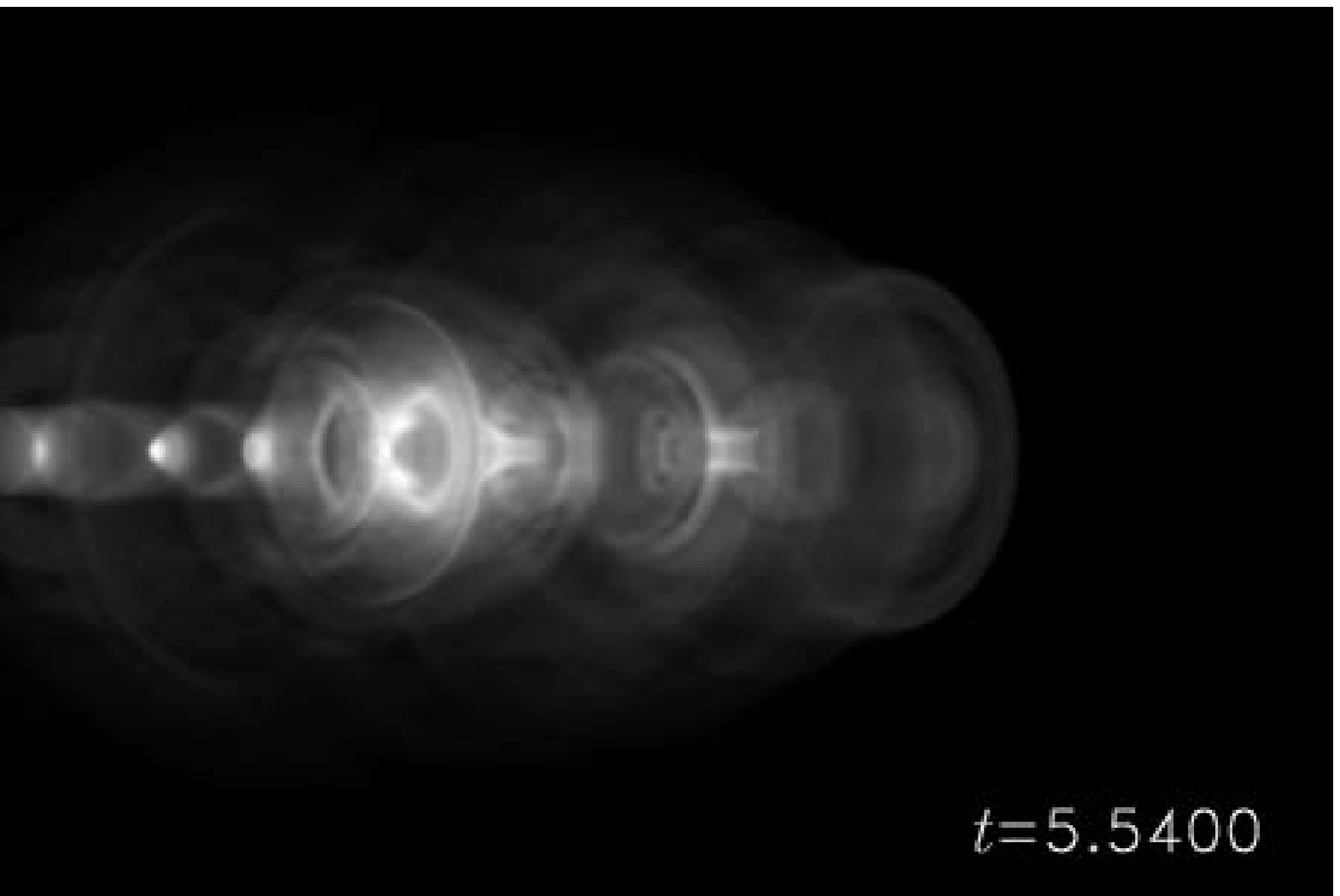}
&\includegraphics[width=4.5cm]{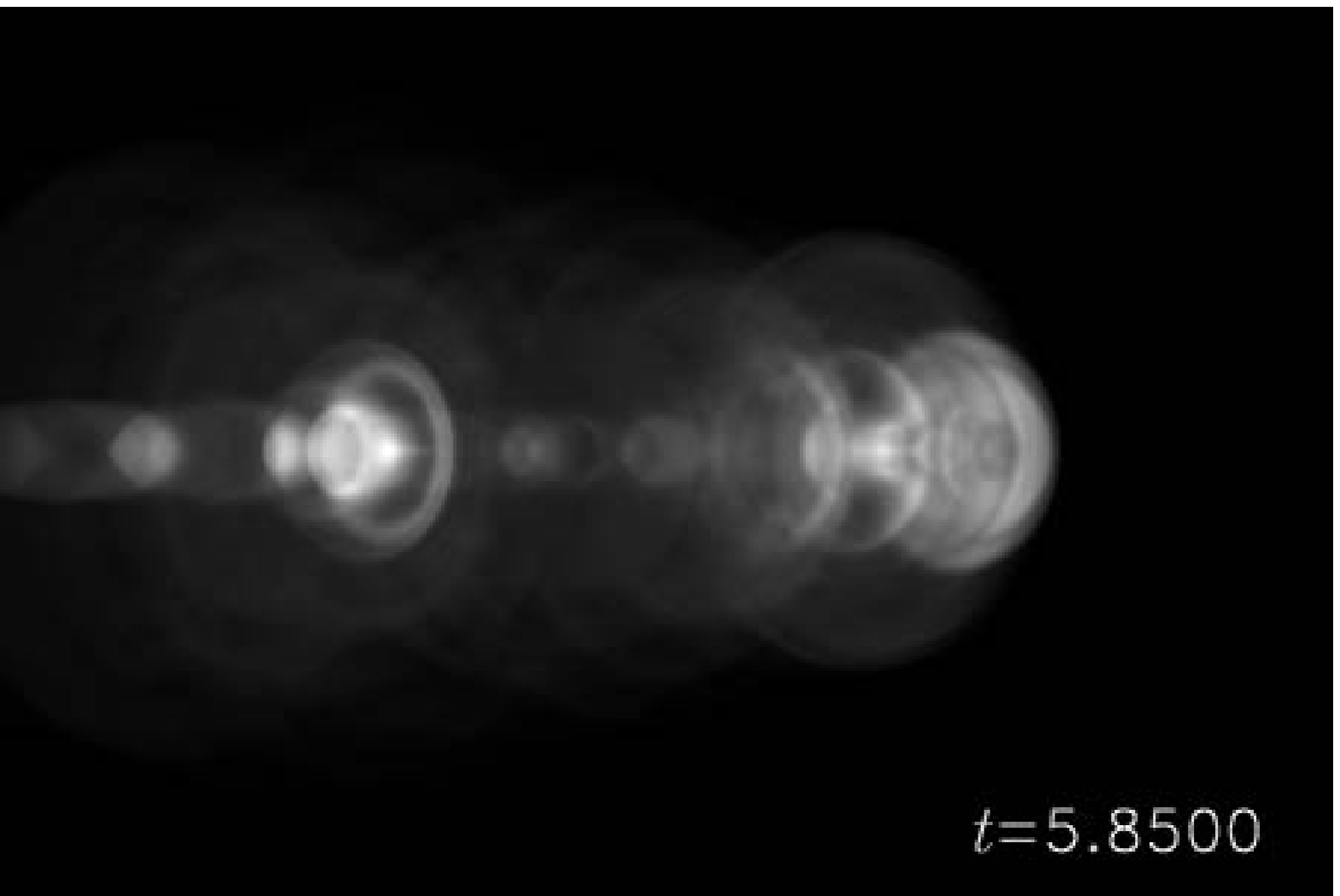}
\\
 \includegraphics[width=4.5cm]{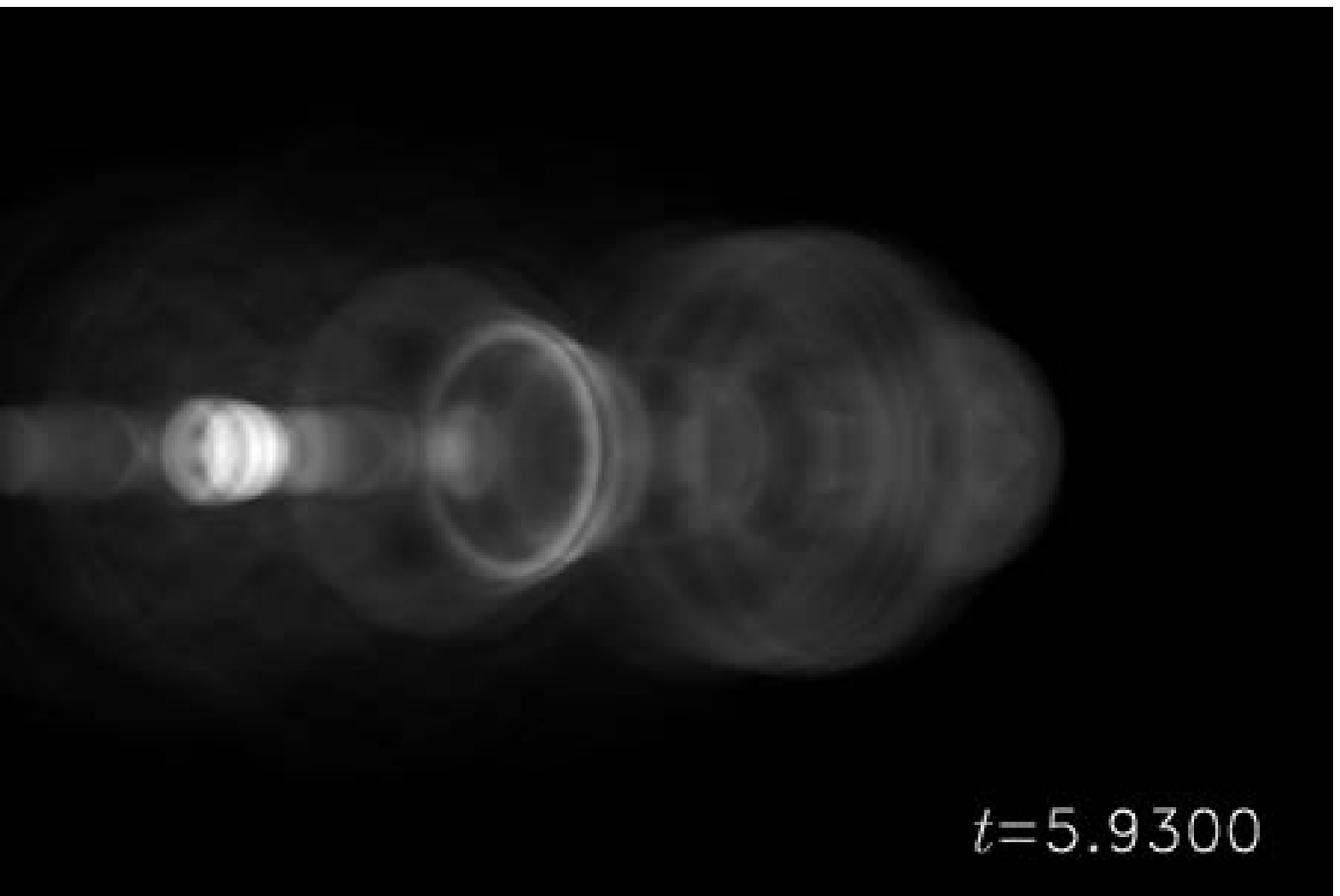}
&\includegraphics[width=4.5cm]{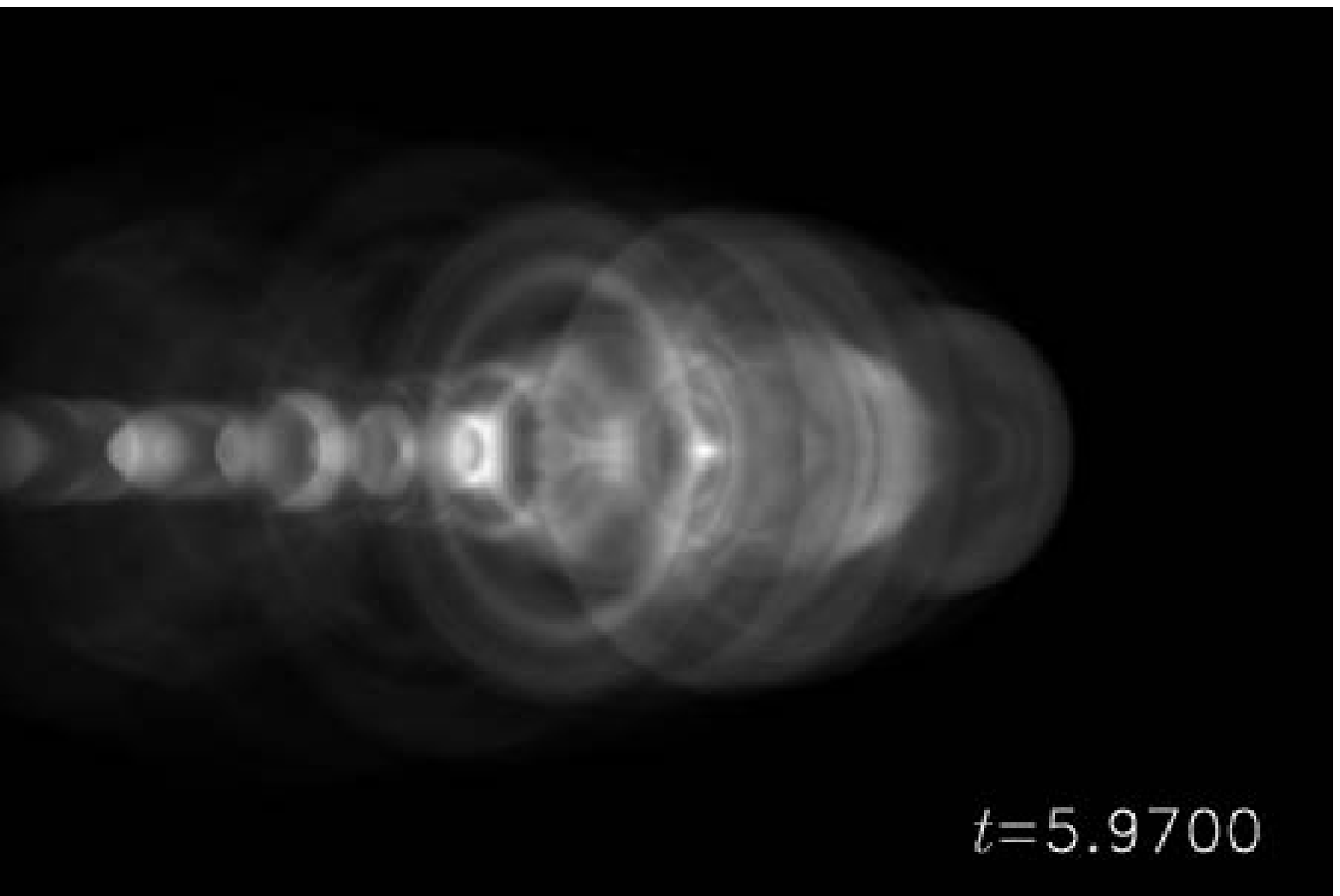}
&\includegraphics[width=4.5cm]{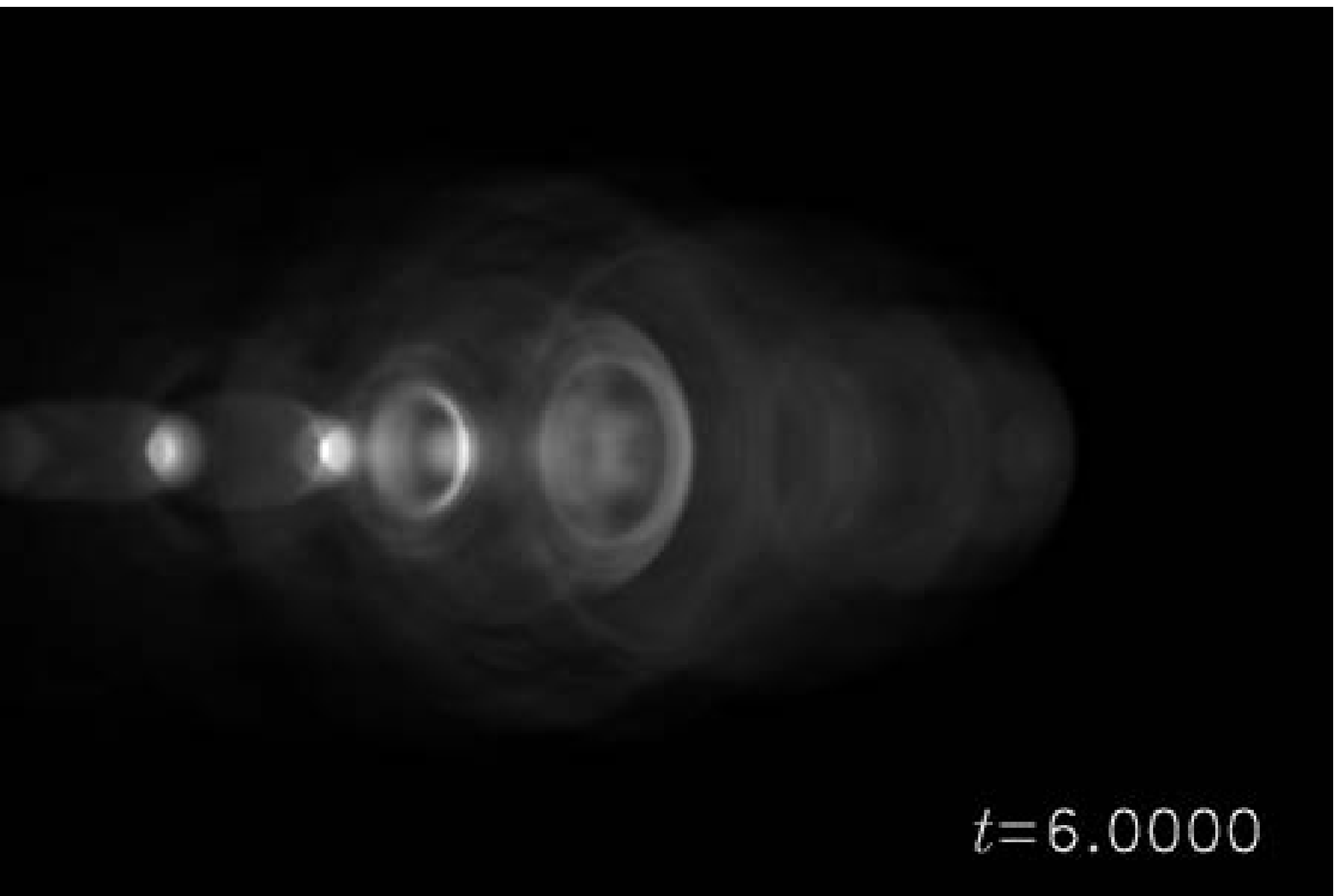}
\end{array}
$
\caption{
Selected images rendered with the jet
at an orientation of $\theta=45^\circ$,
from the simulation with jet parameters $(\eta,M)=(10^{-4},2)$
and an open left boundary. 
The lower brightness contrast of the rings, jet knots and hot-spots 
with respect to the rest of the lobe, 
compared to the  higher density, higher Mach number simulations, 
is apparent.
}
\label{f:pageant.pxit-4m2}
\end{center}
\end{figure}

This work was supported by
an Australian Research Council Large  Grant, A69905341
and grants of computing
time from the ANU Supercomputer Facility.
We are grateful to Drs~N.~Gizani, M.~Garrett and P.~Leahy
for permission to publish their radio image (Figure~\ref{'fig.radio'}).

\clearpage


\clearpage

\begin{table}
\caption{
Classical evaluations of the
velocity, mass flux, force  and power of a jet
with radius
$r_{\rm jet}=0.5 \> \mathrm{kpc}$
for our choices of the parameters $(\eta,M)$.
These are non-relativistic evaluations, assuming cosmic abundances,
an electron number density of
$n_0=7.8\times10^{-3} \> \mathrm{cm}^{-3}$
and a temperature of
$kT_\mathrm{ism}=2.45 \> \mathrm{keV}$
in the undisturbed ambient medium.
}
\label{table.jet.properties}
\begin{center}
$\begin{array}{rrllllll}
\eta&M
&\begin{array}{c}{{v_{\rm j}}\over{(T/T_\mathrm{ism})^{1/2} }}
\\({\rm cm} \>.\> {\rm s}^{-1})
\end{array}
&\begin{array}{c}{{\dot{M}_{\rm jet}}\over{(n/n_0) (T/T_\mathrm{ism})^{1/2} }}
\\(M_\odot \>.\> {\rm yr}^{-1})
\end{array}
&\begin{array}{c}{{F_\mathrm{jet}}\over{(n/n_0) (T/T_\mathrm{ism}) }}
\\({\rm dyn})
\end{array}
&\begin{array}{c}
{{L_{\rm jet}}\over{(n/n_0) (T/T_\mathrm{ism})^{3/2} }}
\\({\mathrm{erg}} \>.\> {\mathrm{s}}^{-1})
\end{array}
\\\hline
\\
10^{-2}&2&1.59\times10^{~9}
&2.87\times10^{-2}&3.31\times10^{33}&4.00\times10^{42}
\\
10^{-3}&2&5.03\times10^{~9}
&9.08\times10^{-3}&3.31\times10^{33}&1.26\times10^{43}
\\
10^{-4}&2&1.59\times10^{10}
&2.87\times10^{-3}&3.31\times10^{33}&4.00\times10^{43}
\\
10^{-2}&5&3.97\times10^{~9}
&7.18\times10^{-2}&1.84\times10^{34}&4.00\times10^{43}
\\
10^{-3}&5&1.26\times10^{10}
&2.27\times10^{-2}&1.84\times10^{34}&1.26\times10^{44}
\\
10^{-4}&5&3.97\times10^{10}
&7.18\times10^{-3}&1.84\times10^{34}&4.00\times10^{44}
\\
10^{-2}&10&7.95\times10^{~9}
&1.44\times10^{-1}&7.23\times10^{34}&2.94\times10^{44}
\\
10^{-3}&10&2.51\times10^{10}
&4.54\times10^{-2}&7.23\times10^{34}&9.30\times10^{44}
\\
10^{-4}&10&7.95\times10^{10}
&1.44\times10^{-2}&7.23\times10^{34}&2.94\times10^{45}
\\
10^{-2}&50&3.97\times10^{10}
&7.18\times10^{-1}&1.80\times10^{36}&3.57\times10^{46}
\\
10^{-3}&50&1.26\times10^{10}
&2.27\times10^{-1}&1.80\times10^{36}&1.13\times10^{47}
\\
10^{-4}&50&3.97\times10^{11}
&7.18\times10^{-2}&1.80\times10^{36}&3.57\times10^{47}
\\
\\\hline
\end{array}$
\end{center}
\end{table}

\begin{table}
\caption{
Typical temperatures in shocked thermal gas
at the front of the bow shock ($T_\mathrm{b}/T_\mathrm{ism}$)
and
on the flank where the shock is approximately parallel to the jet
($T_\mathrm{f}/T_\mathrm{ism}$),
for various choices of the system parameters.
}
\label{table.temperature.ratios}
\begin{center}$
\begin{array}{cc}
\begin{array}{crcc}
\eta&M
&T_\mathrm{b}/T_\mathrm{ism}
&T_\mathrm{f}/T_\mathrm{ism}
\\\hline
\\
10^{-2}&2&1.08&1.02
\\
10^{-3}&2&1.08&1.02
\\
10^{-4}&2&1.11&1.03
\\
\\
10^{-2}&5&1.71&1.18
\\
10^{-3}&5&1.63&1.13
\\
10^{-4}&5&1.69&1.17
\\
\\\hline
\end{array}
&
\begin{array}{crcc}
\eta&M
&T_\mathrm{b}/T_\mathrm{ism}
&T_\mathrm{f}/T_\mathrm{ism}
\\\hline
\\
10^{-2}&10&3.28&1.54
\\
10^{-3}&10&3.64&1.76
\\
10^{-4}&10&2.97&1.45
\\
\\
10^{-2}&50&20.6&4.45
\\
10^{-3}&50&27.1&6.53
\\
10^{-4}&50&13.9&6.71
\\
\\\hline
\end{array}
\end{array}
$\end{center}
\end{table}

\end{document}